\documentclass{article}

\usepackage{jheppub}

\usepackage{bbm}
\usepackage{booktabs}
\usepackage{braket}
\usepackage{comment}
\usepackage{dcolumn}
\usepackage{graphicx}
\usepackage[utf8]{inputenc}
\usepackage{multirow}
\usepackage{rotating}
\usepackage{siunitx}
\usepackage{slashed}
\usepackage[normalem]{ulem}
\usepackage{xcolor}
\usepackage{xspace}
\usepackage{hyperref}
\usepackage[nameinlink,capitalize]{cleveref}
\usepackage{tikz}
\usepackage[compat=1.1.0]{tikz-feynman}
\usepackage{tabularx}

\newcommand{\eg}{\textit{e.g.}\xspace}
\newcommand{\ie}{\textit{i.e.}\xspace}

\newcommand{\eq}[1]{\begin{equation} #1 \end{equation}}
\newcommand{\eqa}[1]{\begin{eqnarray} #1 \end{eqnarray}}
\newcolumntype{C}{>{\centering\arraybackslash}X}
\newcommand{\BqToDqL}{\ensuremath{\bar{B}_{(s)}^0 \to D_{(s)}^{(*)+} P^-}\xspace}
\newcommand{\BqToDqP}{\ensuremath{\bar{B}_{(s)}^0 \to D_{(s)}^+ P^-}\xspace}
\newcommand{\BqToDqV}{\ensuremath{\bar{B}_{(s)}^0 \to D_{(s)}^{* +} P^-}\xspace}
\newcommand{\TeV}{\text{TeV}}
\newcommand{\GFermi}{\ensuremath{G_{\text{F}}}}
\newcommand{\wcbern}[2][]{\mathcal{C}_{#2}^{#1}}
\newcommand{\fsofd}{\frac{f_s}{f_d}}
\newcommand{\EOS}{\texttt{EOS}\xspace}
\DeclareMathOperator{\Tr}{Tr}
\DeclareMathOperator{\Li}{Li_2}
\newcommand{\vecx}{\ensuremath{\Vec{x}}}
\newcommand{\vecnu}{\ensuremath{\Vec{\nu}}}
\newcommand{\vecth}{\ensuremath{\Vec{\vartheta}}}

\begin{document}

\setcounter{tocdepth}{2}

\title{Towards a Global Analysis of the $b\to c\bar{u} q$ Puzzle}

\author[a,b]{Stefan Meiser,}
\author[c]{Danny van Dyk,}
\author[a,b]{Javier Virto}

\emailAdd{smeiser@icc.ub.edu}
\emailAdd{danny.van.dyk@gmail.com}
\emailAdd{jvirto@icc.ub.edu}
\affiliation[a]{Departament de Física Quàntica i Astrofísica, Universitat de Barcelona,\\
Martí i Franquès 1, E08028 Barcelona, Catalonia}
\affiliation[b]{
Institut de Ciències del Cosmos (ICCUB), Universitat de Barcelona,\\
Martí i Franquès 1, E08028 Barcelona, Catalonia}
\affiliation[c]{Institute for Particle Physics Phenomenology and Department of Physics,\\
Durham University, Durham~DH1~3LE, UK}

\abstract{%
We study the nonleptonic decays $\bar{B}_s^0 \to D_s^{(*)+} \pi^-$ and $\bar{B}^0 \to D^{(*)+} K^-$
within the Weak Effective Theory (WET) up to mass-dimension six.
We revisit the calculation of the hadronic matrix elements within QCD Factorization including the full set of WET operators.
We recalculate the two-particle contributions to the hard-scattering kernels at next-to-leading order in $\alpha_s$, confirming recent results in the literature.
We also calculate the three-particle contributions at leading order in $\alpha_s$, clarifying the procedure, refining the SM results in the literature, and providing for the first time the complete set of contributions within the WET.
We use these results to perform a global phenomenological study of the effective couplings,
putting bounds on the size of the WET Wilson coefficients in four distinct fit models.
The fits include constraints from the nonleptonic $B$-meson decay width, which we calculate at the leading order for the full set of WET operators for the first time.
This study is the first one to account for simultaneous variation of up to six effective couplings.
We identify two distinct modes in all fit models and discuss how future measurements can be used to distinguish between them.
}

\begin{flushright}
    EOS-2024-04\\
    IPPP/24/71
\end{flushright}
\vspace*{-3\baselineskip}

\maketitle


\section{Introduction}
\label{sec:intro}

The study of $B$-meson decays is one of the most important sources of knowledge we have on the flavour structure of the Standard Model (SM) and its possible extensions~\cite{Isidori:2010kg}.
One type of such decays are those with only hadrons in the final state, the ``nonleptonic'' $B$-decays, which have been essential in the study of the CKM matrix and CP violation~\cite{Antonelli:2009ws,Belle-II:2018jsg}. A great amount of experimental results have been produced in the last few decades, boosted during the $B$-factory era~\cite{BaBar:2014omp}, and pushed even further by the hadronic machines at Tevatron and the LHC~\cite{Kuhr:2013hd,LHCb:2012myk}. 

However, rigorous and precise theoretical predictions for such $B$-decay observables are rather difficult to make. This is due to non-perturbative QCD effects at the level of the amplitudes, in the calculation of matrix elements of dipole and four-quark operators in the Weak Effective Theory (WET)~\cite{Aebischer:2017gaw}, between the $B$-meson initial state and the all-hadronic final state. In this regard, a significant step forward was achieved when the concept of factorization in the heavy-quark limit was made formal~\cite{Beneke:1999br,Bauer:2000yr}, already a quarter of a century ago. This approach, often called QCD Factorization~(QCDF) has been applied to two-body $B$-decays into two light mesons~\cite{Beneke:1999br,Beneke:2000ry,Beneke:2001ev,Beneke:2003zv,Bauer:2004tj} (such as $B\to \pi\pi$) or a heavy and a light meson~\cite{Beneke:2000ry,Bauer:2001cu} (such as $B\to D\pi$), and to three-body decays such as $B\to \pi\pi\pi$~\cite{Krankl:2015fha} or $B\to D\pi\pi$~\cite{Huber:2020pqb}.

Within QCDF, the simplest of the nonleptonic decays are those two-body $B$ decays to a heavy-light final state, such as $B^0\to D^- K^+$ or $B_s\to D_s^{*-} \pi^+$, which do not receive contributions from annihilation topologies, and where the spectator degrees of freedom in the $B$-meson end up in the heavy meson in the final state.
These will be the \emph{decays of interest} in our analysis.
They are a subset of what in the older traditional nomenclature are known as ``class-I decays''~\cite{Neubert:1997uc}.
We will abbreviate the decays of interest as \BqToDqL from this point forward.
In this case QCDF establishes the following important result~\cite{Beneke:2000ry}, often referred to as a ``factorization formula'':
\begin{equation}
    \langle Q_i \rangle \equiv
    \langle D^{(*)} P | Q_i | B \rangle = 
    \sum_j F_j^{B\to D^{(*)}}(m_P^2)
    \int_0^1 du\,T_{ij}(u,\mu)\Phi_P(u,\mu) + {\cal O}(\Lambda_\text{QCD}/m_b)\ .
\end{equation}
Here $Q_i$ is a four-quark operator in the WET (see below),  $F_j^{B\to D}$ are local $B\to D$ form factors, $\Phi_P$ is the light-cone distribution amplitude of the light pseudoscalar meson and $T_{ij}(u,\mu)$ are perturbative ``hard-scattering kernels''. This factorization formula is simpler than the corresponding one for heavy-to-light decays, where in general one encounters two such form factor terms and a hard-spectator scattering term dependent on the light-cone distribution of the $B$-meson~\cite{Beneke:1999br,Beneke:2000wa}. Therefore, theoretical predictions for class-I decays are numerically more precise and reliable than those for heavy-to-light decays.

Within the SM, only two effective operators contribute to their amplitudes: the two `VLL' operators with different colour structure
\begin{equation}
\begin{aligned}
    Q_{1}
        & = [\bar c \gamma_\mu P_L T^A b] [\bar q \gamma_\mu P_L T^A u]\ ,
        &
    Q_{2}
        & = [\bar c \gamma_\mu P_L b] [\bar q \gamma_\mu P_L u]\ .
\end{aligned}
\end{equation}
The corresponding matrix elements (\ie the hard-scattering kernels $T_{1j}$ and $T_{2j}$) are known to two-loop order, that is, at NNLO in QCD~\cite{Huber:2016xod}, and a first exploration of a QED factorization formula for the decays of interest has been investigated in Ref.~\cite{Beneke:2021jhp}.
Beyond the SM, the general set of effective operators up to mass dimension six with the same flavour structure contains 20 operators in total. Thus, a general analysis of class-I decays beyond the SM requires to calculate the corresponding matrix elements for all 20 operators.
Such a general analysis needs little motivation in the current moment, but it is further called upon in view of recent findings that branching ratio measurements seem to be in tension with theoretical predictions~\cite{Huber:2016xod,Bordone:2020gao}.
Our approach is complementary to others in the literature, which include SMEFT-based~\cite{Bordone:2021cca,Atkinson:2024hqp}, model-based~\cite{Iguro:2020ndk}, and symmetry-based~\cite{Fleischer:2021cct,Fleischer:2021cwb} BSM studies.
Other approaches to resolve the puzzle include the computation of the decay amplitude beyond the framework of QCDF~\cite{Piscopo:2023opf}.

A recent analysis~\cite{Cai:2021mlt} has considered class-I decays within this general set of effective operators, finding that BSM effects in the SM-like operators or in scalar operators
can explain the observed tensions. However, due to the lack of complementary constraints, Ref.~\cite{Cai:2021mlt} investigates exclusively scenarios with BSM contributions to only one or two Wilson coefficients at a time.
In this work, we determine to what extent the Wilson coefficients of the ``$qbcu$'' sectors of the Weak Effective Theory
can be simultaneously constrained from data. As we show below, this is a much more ambitious and complicated type of
analysis. 
To remain within the most predictive set of observables,
we limit our investigation to decays which receive no contributions from annihilation topologies: $\bar B^0\to D^{(*)+}K^-$ and $\bar B_s\to D_s^{(*)+}\pi^-$.
Because of the absence of annihilation topologies, these modes are expected to be more reliably predicted within QCDF.
Moreover, we calculate for the first time the three-particle hard-scattering kernels to the QCD factorization
formulas of the decays at hand at LO in $\alpha_s$ and for the full basis of WET operators.
Incidental to that calculation, we check the analytic expressions for the QCDF formulas
for the two-particle contributions to the hard scattering kernels up to NLO in $\alpha_s$ for the full WET basis.

The sole use of the exclusive nonleptonic decay modes is, however, not sufficient to bound the
WET parameter space. To achieve that, we additionally consider the inclusive nonleptonic
decay width of the $B$ meson, which we calculate at the leading order for the full set of ``$qbcu$'' WET operators.
We thus obtain a qualitative picture of the WET parameter space and the type of solutions
to the extant puzzle in the decays of interest.

The structure of the article is as follows. We begin defining the relevant sectors
of the Weak Effective Theory in \cref{sec:wet} and discuss the two different operator bases used
in our analysis. We continue in \cref{sec:class-I} with a detailed discussion of the decays of interest within the
QCDF approach, and a presentation of our new results at the three-particle
level. \Cref{sec:lifetime} is dedicated to our calculation of the
inclusive nonleptonic $B$-meson decay width for the full set of ``$qbcu$'' WET operators
at leading power in $1/m_b$ and at LO in $\alpha_s$.
The setup, methods, and results of our phenomenological analysis are presented in \cref{sec:pheno}.
We conclude in \cref{sec:conclusion}.
A series of appendices provides complementary information on: the WET operator bases used and the changes of basis, the two- and three-particle LCDAs for the light mesons, parity transformations of the various matrix elements, a collection of the leading-order two-particle contributions to the hard scattering kernels in QCDF, and the approach used to separate the various modes in the posterior distributions.


\section{Weak Effective Theory for $b\to c \bar{u} q$ processes}
\label{sec:wet}

The analysis presented here uses the Weak Effective Theory (WET) to predict a variety of exclusive and inclusive $B$ decays
both in the Standard Model (SM) and beyond. Following Refs.~\cite{Aebischer:2017gaw,Aebischer:2017ugx}, the WET can be partitioned into independent
``sectors'' of operators which do not mix with each other under renormalization up to order $\GFermi \sim g^2/M_W^2$.
The sectors relevant to this work are the $dbcu$ and $sbcu$ sectors, each comprised of 20 four-quark operators.
The effective Lagrangian reads
\begin{equation}
\label{eq:Lagrangian}
    \mathcal{L}^{qbcu}
        = -\frac{4 \GFermi}{\sqrt{2}} V_{cb} V_{uq}^* \sum_{i=1}^{20} C_i(\mu)\,Q_i(\mu) + \text{h.c.}\,,
    \qquad
    q \in \lbrace d, s \rbrace\,,
\end{equation}
where $V$ represents the Cabibbo-Kobyashi-Maskawa quark mixing matrix, $\lbrace Q_i \rbrace$ represents
a basis of local field operators of dimension six, and $\lbrace C_i\rbrace$ are their Wilson coefficients defined at the
renormalization scale $\mu$.
Throughout this paper, we will describe each sector by using two different sets of bases.\\

The first basis is a Fierz transformed version of the ``BMU basis''~\cite{Buras:2000if}.
Its operators are labelled~$\mathcal{Q}_i^{X}$, with $i$ enumerating the operators for a given label $X$,
and $X$ labelling the Dirac structures and chiralities of the quark fields.
Our choice of basis is comprised of the following ten operators
\begin{align}
    \mathcal{Q}_1^{VLL} &= \left[ \bar{c}_\alpha \gamma_\mu P_L b_\beta \right] \left[ \bar{q}_\beta \gamma^\mu P_L u_\alpha \right] \, , & 
    \mathcal{Q}_2^{VLL} &= \left[ \bar{c}_\alpha \gamma_\mu P_L  b_\alpha \right] \left[ \bar{q}_\beta \gamma^\mu P_L u_\beta \right]  \, , 
    \notag \\
    \mathcal{Q}_1^{SLR} &= \left[ \bar{c}_\alpha P_L b_\beta \right] \left[ \bar{q}_\beta P_R u_\alpha \right]\, , &
    \mathcal{Q}_2^{SLR} &= \left[ \bar{c}_\alpha P_L b_\alpha \right] \left[ \bar{q}_\beta P_R u_\beta \right] \, , 
    \notag \\
    \mathcal{Q}_1^{VRL} &= \left[ \bar{c}_\alpha \gamma_\mu P_R b_\beta \right] \left[ \bar{q}_\beta \gamma^\mu P_L u_\alpha \right]\, , & 
    \mathcal{Q}_2^{VRL} &= \left[ \bar{c}_\alpha \gamma_\mu P_R b_\alpha \right] \left[ \bar{q}_\beta \gamma^\mu P_L u_\beta \right] \, , 
    \label{eq:FierzedBMU}
    \\
    \mathcal{Q}_1^{SRR} &= \left[ \bar{c}_\alpha P_R b_\beta \right] \left[ \bar{q}_\beta P_R u_\alpha \right]\, , &
    \mathcal{Q}_2^{SRR} &= \left[ \bar{c}_\alpha P_R b_\alpha \right] \left[ \bar{q}_\beta P_R u_\beta \right] \, , 
    \notag \\
    \mathcal{Q}_3^{SRR} &= \left[ \bar{c}_\alpha \sigma_{\mu \nu} P_R b_\beta \right] \left[ \bar{q}_\beta \sigma^{\mu \nu} P_R u_\alpha \right]\, , &
    \mathcal{Q}_{4}^{SRR} &= \left[ \bar{c}_\alpha \sigma_{\mu \nu} P_R b_\alpha \right] \left[ \bar{q}_\beta \sigma^{\mu \nu} P_R u_\beta \right] \, , 
    \notag 
\end{align}
as well as their chirality-flipped counterparts, with $P_L\leftrightarrow P_R$. Here $\alpha, \, \beta$ are colour indices and we use $\sigma_{\mu \nu} = \frac{i}{2} [\gamma_\mu, \gamma_\nu]$.
Our ordering of the operators does not reflect the ordering in Ref.~\cite{Buras:2000if}.\footnote{%
    We choose this particular ordering to ensure that a basis transformation at tree-level from the JMS basis~\cite{Jenkins:2017jig}
    to ours can be achieved in a convenient form.
}
This basis is used in the analytic calculations of the exclusive decays of interest \BqToDqL.
Our rationale for using this basis is that previous computations of these processes are available
in the literature, which serve as a reference point and comparison for our own calculations; see \cref{sec:class-I}.\\

The second basis used in this work is the ``Bern basis''~\cite{Aebischer:2017gaw}.
Its operators are labelled $\mathcal{O}_i$, where $i$ enumerates all operators.
It is comprised of the ten operators
\begin{align}
\mathcal{O}_1^{qbcu} &= \left[ \bar{q} P_R \gamma_\mu b \right] \left[ \bar{c} \gamma^\mu u \right] \, , & 
\mathcal{O}_2^{qbcu} &= \left[ \bar{q} P_R \gamma_\mu T^A b \right] \left[ \bar{c} \gamma^\mu T^A u \right]  \, , \notag 
\\
\mathcal{O}_3^{qbcu} &= \left[ \bar{q} P_R \gamma_{\mu \nu \rho} b \right] \left[ \bar{c} \gamma^{\mu \nu \rho} u \right]\, , &
\mathcal{O}_4^{qbcu} &= \left[ \bar{q} P_R \gamma_{\mu \nu \rho} T^A b \right] \left[ \bar{c} \gamma^{\mu \nu \rho} T^A u \right] \, , \notag 
\\
\mathcal{O}_5^{qbcu} &= \left[ \bar{q} P_R b \right] \left[ \bar{c} u \right]\, , & 
\mathcal{O}_6^{qbcu} &= \left[ \bar{q} P_R T^A b \right] \left[ \bar{c} T^A u \right] \, , 
\\
\mathcal{O}_7^{qbcu} &= \left[ \bar{q} P_R \sigma_{\mu \nu} b \right] \left[ \bar{c} \sigma^{\mu \nu} u \right]\, , &
\mathcal{O}_8^{qbcu} &= \left[ \bar{q} P_R \sigma_{\mu \nu} T^A b \right] \left[ \bar{c} \sigma^{\mu \nu} T^A u \right] \, , \notag 
\\
\mathcal{O}_9^{qbcu} &= \left[ \bar{q} P_R \gamma_{\mu \nu \rho \sigma} b \right] \left[ \bar{c} \gamma^{\mu \nu \rho \sigma} u \right]\, , &
\mathcal{O}_{10}^{qbcu} &= \left[ \bar{q} P_R \gamma_{\mu \nu \rho \sigma} T^A b \right] \left[ \bar{c} \gamma^{\mu \nu \rho \sigma} T^A u \right] \, , \notag 
\end{align}
together with the ten parity-flipped operators $\mathcal{O}_{i'}^{qbcu}= \mathcal{O}_{i}^{qbcu}|_{P_R \to P_L}$. 
The notation used for the products of gamma matrices is $\gamma^{\mu_1 \mu_2 ... \mu_n} \equiv \gamma^{\mu_1} \gamma^{\mu_2} \dots \gamma^{\mu_n}$.
We use the Bern basis for the calculation of the $B$-meson lifetime in \cref{sec:lifetime} as well as for the phenomenological analysis presented in \cref{sec:pheno}.
In contrast to the BMU basis, where only two operators have non-vanishing SM matching conditions, the Bern basis features four such operators: $\mathcal{O}_{1 \dots 4}^{qbcu}$.
This tells us that only two independent linear combinations of these four operators appear in the SM, corresponding to the more traditional operators $\mathcal{Q}_{1}^{VLL}$ and $\mathcal{Q}_{2}^{VLL}$ of the BMU basis.
The remaining 16 operators $\mathcal{O}_{5 \dots 10}^{qbcu}$ and $\mathcal{O}_{1' \dots 10'}^{qbcu}$ are purely BSM operators.
Furthermore, the Anomalous Dimension Matrix (ADM) splits up into four blocks of operators that mix with each other~\cite{Aebischer:2017gaw}. The first block mixes $\mathcal{O}_1^{qbcu}$ through $\mathcal{O}_4^{qbcu}$, while the second block mixes $\mathcal{O}_5^{qbcu}$ through $\mathcal{O}_{10}^{qbcu}$. The same holds true for the primed operators. 
This block structure holds at least up to NLL.\\

For both bases, we regularize the one-loop integrals using naive dimensional regularization (NDR)
with anti-commuting $\gamma_5$ in $d = 4 - 2 \epsilon$ dimensions.
For the renormalization of the physical operators we employ the $\overline{\text{MS}}$-scheme.
(We also use the $\overline{\text{MS}}$-scheme for all quark masses that occur in our calculations.)
For the renormalization of the evanescent operators we follow the Buras-Weisz prescription \cite{Buras:1989xd,Dugan:1990df,Herrlich:1994kh}, which ensures that matrix elements of evanescent operators vanish to all orders. Furthermore, we use the method of Greek projections~\cite{Tracas:1982gp,Buras:2000if}
--- in the form presented in~Ref.~\cite{Dekens:2019ept} --- to find the evanescent operators necessary for our calculations. A full list of the evanescent operators relevant for this article can be found in \cref{app:bases:evanescent}. This fully specifies the renormalization scheme.\\

The SM matching conditions in the BMU basis are known up to two loops. 
In our notation, the one-loop matching conditions read~\cite{Gorbahn:2004my,Buras:1998raa,Egner:2024azu}:
\begin{equation}
\label{eq:preliminaries:matching-conditions-BMU}
C_1^{VLL}(\mu)
= 3 \frac{\alpha_s}{4 \pi} \left(\log{\frac{\mu^2}{m_W^2}} + \frac{11}{6} \right)\ ,
\qquad
C_2^{VLL}(\mu)
= 1 - \frac{3}{N_c} \frac{\alpha_s}{4 \pi} \left(\log{\frac{\mu^2}{m_W^2}} + \frac{11}{6}\right)\ ,
\end{equation}
where $N_c$ is the number of colours.
We have verified the matching conditions by explicit calculation.
Performing the change of basis to the Bern basis as described in \cref{app:bases:change-of-basis}, we obtain the four non-vanishing matching conditions
\begin{equation}
\label{eq:preliminaries:matching-conditions-bern}
\begin{aligned}
    \wcbern[qbcu]{1}(\mu)
        & = -\frac{1}{9} - \frac{\alpha_s}{4 \pi} \left(\frac{8}{9} \log{\frac{\mu^2}{m_W^2}} - \frac{52}{27}\right) \,, &
    \wcbern[qbcu]{2}(\mu)
        & = -\frac{2}{3} + \frac{\alpha_s}{4 \pi} \left(\frac{2}{3} \log \frac{\mu^2}{m_W^2} - \frac{85}{9}\right)   \,, \\
    \wcbern[qbcu]{3}(\mu)
        & = +\frac{1}{36} + \frac{\alpha_s}{4 \pi} \left(\frac{2}{9} \log \frac{\mu^2}{m_W^2} - \frac{1}{27} \right)  \,,  &
    \wcbern[qbcu]{4}(\mu)
        & = +\frac{1}{6} - \frac{\alpha_s}{4 \pi} \left(\frac{1}{6} \log \frac{\mu^2}{m_W^2} - \frac{19}{36}\right)   \,.
\end{aligned}
\end{equation}
Note that in the conventions of~Ref.~\cite{Aebischer:2017gaw} the CKM factors are put into the Wilson coefficients. We use a different convention, as in~\cref{eq:Lagrangian}, and factor them out. 
Since the characteristic energy of the processes that we study is of the order of $m_b$, one needs to evaluate the Wilson coefficients at a scale $\mu_b \simeq m_b$. In order to avoid large logarithms, one has to resum them using the renormalization group equation
\begin{align}
    \mu \frac{d}{d\mu} C_i = {\gamma}_{ji} \, C_j \, ,
\end{align}
where $\hat{\gamma}$ is the Anomalous Dimension Matrix (ADM), with the perturbative expansion
\begin{align}
    \hat{\gamma} = \frac{\alpha_s}{4 \pi} \, \hat{\gamma}^{(0)} + \left(\frac{\alpha_s}{4 \pi}\right)^2 \, \hat{\gamma}^{(1)} + \mathcal{O}(\alpha_s^3) \ .
\end{align}
Here $\hat{\gamma}^{(0)}$ arises at one-loop, and $\hat{\gamma}^{(1)}$ at two-loops.
In the BMU basis $\hat{\gamma}^{(0)}$ and $\hat{\gamma}^{(1)}$ are known~\cite{Buras:2000if}.
The one-loop ADM $\hat{\gamma}^{(0)}$ in the Bern basis is given in~Ref.~\cite{Aebischer:2017gaw},
while the two-loop ADM $\hat{\gamma}^{(1)}$ has not been explicitly presented.
We derive the two-loop ADM in the Bern basis for $N_c = 3$ and $n_f = 5$
by performing a change of basis, as described in \cref{app:bases:change-of-basis}. The full $20\times 20$ ADM is comprised of two identical $10\times 10$ diagonal blocks, each given by
\begin{align}
\label{eq:preliminaries:adms-bern}
\hat{\gamma}^{(1)} = \left(
\begin{array}{cccccccccc}
 \frac{44}{9} & -\frac{899}{3} & -\frac{32}{9} & \frac{245}{12} & 0 & \
0 & 0 & 0 & 0 & 0 
\\
 -\frac{646}{27} & -\frac{2072}{9} & -\frac{115}{54} & \frac{739}{72} \
& 0 & 0 & 0 & 0 & 0 & 0 
\\
 -\frac{6848}{9} & -1344 & \frac{524}{9} & 178 & 0 & 0 & 0 & 0 & 0 & 0 
\\
0  & -\frac{8468}{9} & -\frac{172}{9} & \frac{367}{18} & 0 & 0 & 0 & \
0 & 0 & 0 
\\
0  & 0 & 0 & 0 & -\frac{1832}{9} & -\frac{64}{3} & -\frac{104}{9} & -\
\frac{296}{9} & -\frac{7}{18} & \frac{11}{48} 
\\
0  & 0 & 0 & 0 & -\frac{128}{27} & \frac{608}{9} & -\frac{52}{81} & -\
\frac{1783}{108} & \frac{11}{216} & \frac{59}{144} 
\\
0  & 0 & 0 & 0 & -\frac{9488}{27} & \frac{7108}{9} & \frac{3052}{9} & \
-\frac{31}{9} & \frac{521}{27} & -\frac{217}{36} 
\\
0  & 0 & 0 & 0 & -\frac{25528}{81} & \frac{896}{3} & -\frac{6974}{81} \
& -\frac{4727}{27} & \frac{863}{162} & \frac{38}{9} 
\\
0  & 0 & 0 & 0 & -\frac{26368}{27} & -\frac{249088}{9} & \
-\frac{91456}{9} & \frac{68192}{9} & -\frac{8912}{27} & \frac{8143}{9} 
\\
0  & 0 & 0 & 0 & \frac{510976}{81} & -\frac{14080}{9} & \
\frac{1600}{81} & \frac{46960}{27} & -\frac{11794}{81} & -\frac{41}{6} 
\\
\end{array}
\right) \, . 
\end{align}
Note that the one-loop matching coefficients as well as the two-loop ADM depend on the evanescent operators used, and thus are scheme dependent. Our choice of scheme is described in \cref{app:bases:evanescent}.

\section{Class-I decays in the Weak Effective Theory}
\label{sec:class-I}

The factorization formula for class-I decays,
\begin{equation}
    \langle Q_i \rangle \equiv
    \langle D^{(*)} P | Q_i | B \rangle = 
    \sum_j F_j^{B\to D^{(*)}}(m_P^2)
    \int_0^1 du\,T_{ij}(u,\mu)\Phi_P(u,\mu) + {\cal O}(\Lambda_\text{QCD}/m_b)\ ,
\end{equation}
establishes the notion of
the hard(-collinear) scattering kernels $T_{ij}(u, \mu)$ and the soft matrix elements (\ie, the hadronic
form factors $F_j$ and the light-meson light-cone distribution amplitudes).
The purpose of this section is to provide results for the kernels $T_{ij}$ at the two-
and three-particle level for the full basis of WET operators.
Although the calculation of the decay amplitudes for class-I decays at the level of two-particle Fock states
for the full basis of WET operators is well documented in the literature~\cite{Beneke:2000ry,Cai:2021mlt},
the calculation of three-particle Fock states is incomplete.
We begin by reiterating the calculation of two-particle contributions in \cref{sec:class-I:hard-scattering-2pt}
and setting the notation before discussing the calculation of three-particle contributions in \cref{sec:class-I:hard-scattering-3pt}.

\subsection{Two-particle contributions to the hard-scattering kernels at one loop}
\label{sec:class-I:hard-scattering-2pt}

\begin{figure}
    \centering
    \begin{tikzpicture}[baseline=(b.base)]
    \begin{feynman}
    \vertex (a) at (0,0);
    \vertex (b) at (0.45,0);
    \vertex (c) at (0.35, 0.77);
    \vertex (f1) at (1.5,0);
    \vertex (i1) at (-1.5,0);
    \vertex (i2) at (-1.5, -0.7);
    \vertex (f2) at (1.5, -0.7);
    \vertex (f3) at (-0.5, 1.1);
    \vertex (f4) at (0.5, 1.1);

    \diagram* {
      (i1) -- [fermion] (a) -- (b) -- [fermion] (f1),
      (f2) -- [fermion] (i2),
      (f3) -- [fermion] (a),
      (a) -- [fermion] (f4),
      (b) -- [gluon] (c),
    };
  \end{feynman}
  \end{tikzpicture} \qquad
  \begin{tikzpicture}[baseline=(b.base)]
  \begin{feynman}
    \vertex (a) at (-0.45,0);
    \vertex (b) at (0,0); 
    \vertex (c) at (-0.35, 0.77);
    \vertex (f1) at (1.5,0);
    \vertex (i1) at (-1.5,0);
    \vertex (i2) at (-1.5, -0.7);
    \vertex (f2) at (1.5, -0.7);
    \vertex (f3) at (-0.5, 1.1);
    \vertex (f4) at (0.5, 1.1);

    \diagram* {
      (i1) -- [fermion] (a) -- (b) -- [fermion] (f1),
      (f2) -- [fermion] (i2),
      (f3) -- [fermion] (b),
      (b) -- [fermion] (f4),
      (a) -- [gluon] (c),
    };
  \end{feynman}
  \end{tikzpicture} \qquad 
  \begin{tikzpicture}[baseline=(b.base)]
    \begin{feynman}
    \vertex (a) at (0,0);
    \vertex (b) at (0.45,0);
    \vertex (c) at (-0.35, 0.77);
    \vertex (f1) at (1.5,0);
    \vertex (i1) at (-1.5,0);
    \vertex (i2) at (-1.5, -0.7);
    \vertex (f2) at (1.5, -0.7);
    \vertex (f3) at (-0.5, 1.1);
    \vertex (f4) at (0.5, 1.1);

    \diagram* {
      (i1) -- [fermion] (a) -- (b) -- [fermion] (f1),
      (f2) -- [fermion] (i2),
      (f3) -- [fermion] (a),
      (a) -- [fermion] (f4),
      (b) -- [gluon] (c),
    };
  \end{feynman}
  \end{tikzpicture} \qquad
  \begin{tikzpicture}[baseline=(b.base)]
  \begin{feynman}
    \vertex (a) at (-0.45,0);
    \vertex (b) at (0,0); 
    \vertex (c) at (0.35, 0.77);
    \vertex (f1) at (1.5,0);
    \vertex (i1) at (-1.5,0);
    \vertex (i2) at (-1.5, -0.7);
    \vertex (f2) at (1.5, -0.7);
    \vertex (f3) at (-0.5, 1.1);
    \vertex (f4) at (0.5, 1.1);

    \diagram* {
      (i1) -- [fermion] (a) -- (b) -- [fermion] (f1),
      (f2) -- [fermion] (i2),
      (f3) -- [fermion] (b),
      (b) -- [fermion] (f4),
      (a) -- [gluon] (c),
    };
  \end{feynman}
  \end{tikzpicture} 
 
\caption{Vertex corrections that contribute to the hard-scattering kernels at the one-loop level.
These are commonly labelled ``non-factorizable'' in the BBNS nomenclature~\cite{Beneke:2000ry}.}
\label{fig:class-I:hard-scattering-2pt}
\end{figure}
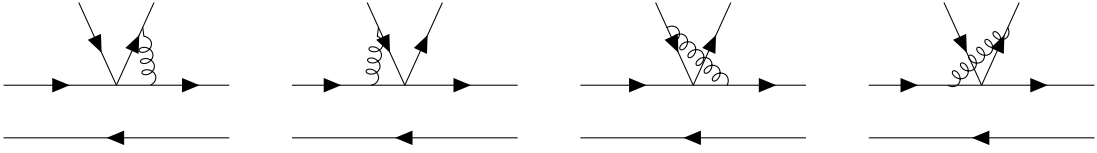

We find it convenient to work with the Fierz-transformed BMU basis given in~\cref{eq:FierzedBMU}.
The usual procedure to calculate the two-particle contributions to the hard-scattering kernels at one loop is as follows.
For each WET operator $\mathcal{Q}_i \sim [\bar{c} \Gamma_1 b] [\bar{q} \Gamma_2 u]$ we take the following steps:
\begin{enumerate}
    \item The partonic $b(p_b)\to c(p_c) q(l_1)\bar u(l_2)$ amplitudes are calculated to the desired order in $\alpha_s$. The relevant one-loop diagrams are shown in \cref{fig:class-I:hard-scattering-2pt}. They yield the following amplitude
    \begin{align}
        A_{\text{2-part}} = -ig_s^2 F_c^{\text{2-part}} \int \frac{d^4k}{(2\pi)^4} \bar{u}(p_c) A_{1\nu}(k) u(p_b) \frac{1}{k^2} \bar{u}(l_1) A_2^\nu(l_1,l_2,k) v(l_2)\, ,
    \end{align}
    with
    \begin{align}
        A_1^\nu(k)
            & = \gamma^\nu \frac{\slashed{p}_c -\slashed{k} + m_c}{2 p_c \cdot k - k^2} \Gamma_1 - \Gamma_1 \frac{\slashed{p}_b + \slashed{k} + m_b}{2 p_b \cdot k + k^2}\gamma^\nu \,,
        \label{eq:A1}
        \\[2mm]
        A_2^\nu(l_1, l_2, k)
            & = \Gamma_2 \frac{\slashed{l}_2 + \slashed{k}}{2 l_2 \cdot k + k^2} \gamma^\nu - \gamma^\nu \frac{\slashed{l}_1 + \slashed{k}}{2 l_1 \cdot k + k^2} \Gamma_2 \, .
    \end{align}
    The colour structures are accounted for by the factor $F_c^{\text{2-part}}$. Upon projecting onto meson states, it turns out to be $F_c^{\text{2-part}} = 0$ for the singlet operator $[\bar{c}_\alpha \Gamma_1 b_\alpha] [\bar{q}_\beta \Gamma_2 u_\beta]$, and $F_c^{\text{2-part}} = C_F/N_c$ for the operator with crossed colour indices $[\bar{c}_\alpha \Gamma_1 b_\beta] [\bar{q}_\beta \Gamma_2 u_\alpha]$.
    
    The loop integration is carried out and the result is simplified using equations of motion, neglecting explicitly all light quark masses. The Dirac structures multiplying $1/\epsilon$ terms in the amplitude are decomposed into physical and evanescent structures, according to the definition of the renormalization scheme (see~\cref{app:bases:evanescent}). Then the one-loop counterterms are added in order to renormalize the amplitude.
    At this point all IR divergences should cancel, which is a manifestation of factorization at the one-loop level.

\item In order to perform the collinear expansion, one needs to specify the kinematics. The two light quarks carry light-like, collinear momenta
\eq{
l_1 = u q + l_\perp + \frac{\vec{l}_\perp^2}{4 u E} n_- \ , 
\qquad 
l_2 = \bar{u}q - l_\perp + \frac{\vec{l}_\perp^2}{4 \bar{u} E} n_- \ ,
}
where $q = E n_+$ is the light-like light meson momentum with energy $E = (m_b^2 - m_c^2)/(2 m_b)$. The light-like directions are defined as $n_\pm = (1, 0, 0, \pm 1)$. The perpendicular component $l_\perp$ is orthogonal to the two light-like directions ($n_\pm \cdot l_\perp = 0)$. Here we can safely neglect the light meson mass $m_P$ since it scales (at most) as $\Lambda_{\text{QCD}}$. We use the notation $\bar{u} \equiv (1-u)$.
In the following we will use a general decomposition of four-vectors into their $n_\pm$ and perpendicular components:
\eq{
v = \frac{v_+}{\sqrt{2}} n_+ + \frac{v_-}{\sqrt{2}} n_- + v_\perp \, .
}

We are interested in contributions up to twist-3. Investigating the structure of the momentum-space representation of the light-meson projection \cref{eq:PseudoScalarMomentum}, it is clear that only terms up to $l_\perp$ can contribute. Hence, for this calculation, we suppress the $n_-$ components of the light quark momenta: 
\begin{align}
l_1 = u q + l_\perp + {\cal O}(l_\perp^2) \ ,
\qquad l_2 = \bar{u} q - l_\perp + {\cal O}(l_\perp^2)\ .
\end{align}
The collinear expansion now requires to separate the amplitude into $\mathcal{O}(l_\perp^0)$ and $\mathcal{O}(l_\perp)$ terms, 
\begin{align}
     A_{\text{2-part}} = A^{(0)} + l_\perp^\mu A_\mu^{(1)} + {\cal O}(l_\perp^2)\ ,
\end{align}
where $A^{(0)}$ are the terms that do not explicitly multiply a factor $l_\perp$ and $A_\mu^{(1)}$ are the terms that do multiply a factor $l_\perp$. This separation is ambiguous because by the use of equations of motion it is possible to move terms from $A^{(0)}$ to $A^{(1)}$ and vice-versa. However the final result will not depend on this freedom, and it is by itself unambiguous.

\item It is now straightforward to apply a projection onto the hadronic external states by means of the following substitution
\begin{align}
    \bar{u}(p_c) \Gamma u(p_b) \, \bar{u}(l_1) \Gamma' v(l_2) \to \int_0^1 du \, \bra{D^{(*)+}_{(s)}}\bar{c}\Gamma b \ket{\bar{B}_{(s)}} \, \Tr \left[M_P \Gamma' \right]\bigg|_{l_\perp = 0} \,  ,
\end{align}
with~\cite{Beneke:2000wa}
\begin{align}
    M_P =  \frac{if_P}{4} \left \{\slashed{q} \gamma_5 \phi_{P}(u) - \mu_P \gamma_5 \left(\phi_P^p(u)
    - i \sigma_{\mu \nu} n_+^\mu n_-^\nu \frac{(1-2u)}{2} \, \phi_P^p(u)
    + i \sigma_{\mu \nu} q^\mu u \bar{u} \, \phi_P^p(u) \frac{\partial}{\partial l_{\perp \nu}}  \right) \right\} \, .
\end{align}
These traces can be evaluated in four dimensions.
The functions $\phi_P$ and $\phi_P^p$ are two-particle Light-Cone Distribution Amplitudes (LCDAs), as defined in~\cref{app:lcda:2pt}.
We also projected onto the heavy meson states, which in our case can be done easily by exchanging partonic external states with mesonic states.
Note that setting $l_\perp = 0$ after acting with the projector on the amplitude ensures that only terms up to $\mathcal{O}(l_\perp)$ in the partonic amplitude can contribute, since higher powers of $l_\perp$ would be set to zero in this collinear limit.

\end{enumerate}

This procedure allows us to compute the hard-scattering kernels at one-loop order for a pseudoscalar light meson in the final state. The corresponding kernels for a light vector meson in the final state can be obtained easily from the ones computed here.

\subsection{Three-particle contributions to the hard-scattering kernels}
\label{sec:class-I:hard-scattering-3pt}

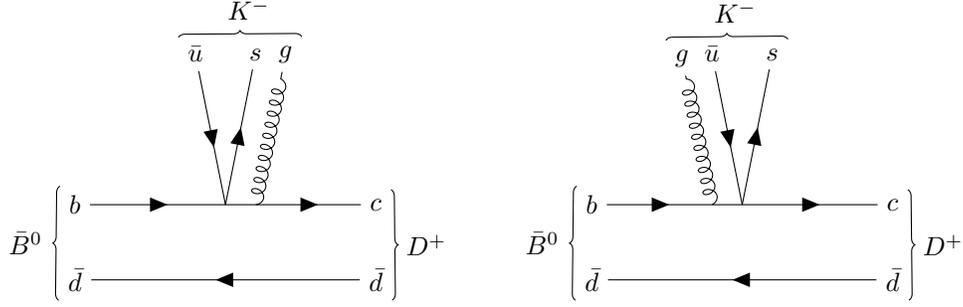
\begin{figure}[t]
    \centering
    \begin{tikzpicture}[baseline=(b.base)]
    \begin{feynman}
    \vertex (a) at (0,0);
    \vertex (b) at (0.4,0); 
    \vertex (f1) at (2,0) {\(c\)};
    \vertex (i1) at (-2,0) {\(b\)};
    \vertex (i2) at (-2, -1) {\(\bar{d}\)};
    \vertex (f2) at (2, -1) {\(\bar{d}\)};
    \vertex (f3) at (-0.4, 2) {\(\bar{u}\)};
    \vertex (f4) at (0.4, 2) {\(s\)};
    \vertex (f5) at (0.8, 2) {\(g\)};

    \diagram* {
      (i1) -- [fermion] (a) -- (b) -- [fermion] (f1),
      (f2) -- [fermion] (i2),
      (f3) -- [fermion] (a),
      (a) -- [fermion] (f4),
      (b) -- [gluon] (f5),
    };
    
    \draw[decoration={brace}, decorate] (i2.south west)-- (i1.north west) node[pos=0.5, left, xshift=-3pt] {\(\bar{B}^0\)};
    \draw[decoration={brace}, decorate] (f1.north east)-- (f2.south east) node[pos=0.5, right, xshift=2pt] {\(D^+\)};
    \draw[decoration={brace}, decorate] (f3.north west)-- (f5.north east) node[pos=0.57, above, yshift=3pt] {\(K^-\)};
    \end{feynman}
    \end{tikzpicture} \qquad
    \begin{tikzpicture}[baseline=(b.base)]
    \begin{feynman}
    \vertex (a) at (-0.4,0);
    \vertex (b) at (0,0); 
    \vertex (f1) at (2,0) {\(c\)};
    \vertex (i1) at (-2,0) {\(b\)};
    \vertex (i2) at (-2, -1) {\(\bar{d}\)};
    \vertex (f2) at (2, -1) {\(\bar{d}\)};
    \vertex (f3) at (-0.4, 2) {\(\bar{u}\)};
    \vertex (f4) at (0.4, 2) {\(s\)};
    \vertex (f5) at (-0.8, 1.95) {\(g\)};

    \diagram* {
      (i1) -- [fermion] (a) -- (b) -- [fermion] (f1),
      (f2) -- [fermion] (i2),
      (f3) -- [fermion] (b),
      (b) -- [fermion] (f4),
      (a) -- [gluon] (f5),
    };
    
    \draw[decoration={brace}, decorate] (i2.south west)-- (i1.north west) node[pos=0.5, left, xshift=-3pt] {\(\bar{B}^0\)};
    \draw[decoration={brace}, decorate] (f1.north east)-- (f2.south east) node[pos=0.5, right, xshift=2pt] {\(D^+\)};
    \draw[decoration={brace}, decorate] (f5.north west)-- (f4.north east) node[pos=0.57, above, yshift=3pt] {\(K^-\)};
    \end{feynman}
    \end{tikzpicture}
\caption{The $q\bar{q}g$ contribution to the amplitude for the decay $\bar{B}^0 \to D^+ K^-$. The contribution for the decay $\bar{B}^0_s \to D^+_s \pi^-$ can be obtained by changing the spectator quark to a strange quark and the strange quark of the final state kaon to a down quark.}
\label{fig:class-I:hard-scattering-3pt:qqg}
\end{figure}

The three-particle contributions to the hard-scattering kernels are obtained by calculating the partonic amplitude to a four-body final state $b(p_b)\to c(p_c) q(l_1)\bar u(l_2)g(p_g)$, where the partons $q\bar u g$ will become part of the final-state light meson in the hadronic process. The kinematics here are different than in the two-parton case since the light meson momentum now has to be split between the three partons. All three partons carry collinear and light-like momenta with the dominant component in the direction of the dominant component of the light meson momentum $q$. Note that for the insertion of vector operators we work in the limit $q^2 = 0$. For the insertion of tensor operators however, the matrix elements vanish in the  $q^2 = 0$ limit and thus we set $q^2 = m_P^2$ and keep the leading $\mathcal{O}(m_P^2)$ term.

The tree-level contributions to the partonic amplitude are shown in~\cref{fig:class-I:hard-scattering-3pt:qqg}, which give~\footnote{We use $D_\mu = \partial_\mu - i g_s A_\mu$, fixing the sign of the quark-quark-gluon vertex. The same convention should be used in the definition of the light-meson three-particle LCDAs.}
\begin{align}
    A_{\text{3-part}} = g_s F_c^{\text{3-part}}\, \bar{u}(p_c) A_1^\nu(-p_g)u(p_b)\,  \Bar{u}(l_1) \Gamma_2 v(l_2) \, \epsilon^{*}_\nu(p_g) \, ,
\end{align}
and $A_1^\nu$ is defined in~\cref{eq:A1}.
Again $\Gamma_{1,2}$ refer to the Dirac structure of the operator under consideration,
and the colour factor will turn out to be $F_c^{\text{3-part}} = 0$ for the colour singlet operator $[\bar{c}_\alpha \Gamma_1 b_\alpha] [\bar{q}_\beta \Gamma_2 u_\beta]$, and $F_c^{\text{3-part}} = 1/N_c$ for the operator with crossed colour indices $[\bar{c}_\alpha \Gamma_1 b_\beta] [\bar{q}_\beta \Gamma_2 u_\alpha]$.

Following~Ref.~\cite{Descotes-Genon:2002crx}, the aim is to write this amplitude as a convolution of a hard-scattering kernel that only depends on hard scales, and a ``partonic'' LCDA.
We rewrite the amplitude as
\begin{align}
    A_{\text{3-part}} 
    &= F_c^{\text{3-part}} \int \frac{d^4k d^4z}{(2 \pi)^4} e^{-i k \cdot z} \, \bar{u}(p_c) A_1^\nu(-k) u(p_b) \,  \bra{\bar{u}(l_2) q(l_1) g(p_g)} \Bar{q}(0)\,\Gamma_2 \,g_s A_\nu(z) \, u(0) \ket{0} \, .
\end{align}
Now it is convenient to use Fock-Schwinger gauge in order to express the gluon field in terms of the gluon field-strength tensor
\begin{align}
    A_\nu(z) = \int_0^1 dv \,v\,z^\mu G_{\mu \nu}(vz) \ ,
\end{align}
and to rewrite $z$ as a derivative with respect to $k$ and use integration by parts to let it act on $A_1^\nu$,
\begin{equation}
A_{\text{3-part}} =
-i F_c^{\text{3-part}} \int \frac{d^4k\,d^4z}{(2 \pi)^4} \, \widetilde{A}_1^{\mu\nu}(k)  
\int_0^1 dv \, v \, e^{-i k \cdot z}  \bra{\bar{u}(l_2) q(l_1) g(p_g)} \Bar{q}(0)\, \Gamma_2\, g_s G_{\mu\nu}(vz) \, u(0) \ket{0} 
\ ,
\label{eq:A3part}
\end{equation}
where we have defined
\begin{equation}
\widetilde{A}_1^{\mu\nu}(k)  \equiv  
\bar{u}(p_c) \frac{\partial}{\partial k_\mu} A_1^\nu(-k) u(p_b)   \, .
\end{equation}
The key step now is to expand the amplitude in $k_\perp$,
\begin{equation}
\widetilde A_1^{\mu\nu}(k) = \widetilde A^{(0) \mu\nu}_1(k_+) + k_{\perp \rho}\, \widetilde A^{(1)\mu\nu\rho}_1(k_+) + \cdots \ ,
\label{eq:A1exp}
\end{equation}
and define a partonic distribution amplitude
\begin{align}
\phi_{\mu \nu}(v z) \equiv \bra{\bar{u}(l_2) q(l_1) g(p_g)} \Bar{q}(0)\, \Gamma_2\,g_s\,G_{\mu \nu}(vz) \, u(0) \ket{0} \, .
\label{eq:partDA}
\end{align}
Putting~\cref{eq:A1exp,eq:partDA} into~\cref{eq:A3part}, one gets
\begin{align}
A_{\text{3-part}} = &
-i F_c^{\text{3-part}} \int \frac{dk_+}{2 \pi} \, \widetilde{A}_1^{(0)\mu\nu}(k_+) \,  \int dz_- \, e^{-i k_+ z_-}  \, \int_0^1 dv \, v \,   \phi_{\mu\nu}(vz) \big |_{z_+ , z_\perp = 0}
\notag\\[2mm]
&
-F_c^{\text{3-part}} \int \frac{dk_+}{2 \pi} \, \widetilde{A}_1^{(1)\mu\nu\rho}(k_+) \, \int dz_- \, e^{-i k_+ z_-} \, \int_0^1 dv \, v \,\left(\frac{\partial}{\partial z_\perp^\rho} \phi_{\mu\nu}(vz) \right)\bigg |_{z_+ , z_\perp = 0}\ .
\end{align}
The functions $\widetilde{A}_1^{(0)\mu\nu}(k_+)$ and $\widetilde{A}_1^{(1)\mu\nu\rho}(k_+)$ do not depend on the light-meson degrees of freedom, the information of which is contained in the distribution amplitudes $\phi_{\mu\nu}(vz)$, evaluated on the light cone (for $z_+ = 0$ and $z_\perp = 0$). Thus this establishes the desired factorization formula.
At this point one can exchange the partonic external states with the hadronic states that we are actually interested in. 
The functions $\widetilde{A}_1^{(0)}$ and $\widetilde{A}_1^{(1)}$ contain now the $B \to D^{(*)}$ hadronic form factors, as defined in~Ref.~\cite{Bordone:2019vic} except for a factor of ``$i$'' in the definition of the $B\to D^*$ form factors. Our phase convention matches the one of BBNS~\cite{Beneke:2000ry}. The vacuum to light meson matrix element $\phi$ is expressed in terms of hadronic three-particle LCDAs as in~\cref{app:lcda:3pt}.

Note that the expansion of $\widetilde{A}_1$ in~\cref{eq:A1exp} seems to be arbitrary since one integrates over all~$k$. However, higher orders in $k_\perp$ correspond to higher $z_\perp$ derivatives acting on $\phi$. Considering the definition of the twist-3 and twist-4 three-particle LCDAs in~\cref{app:lcda:3pt}, one can see that higher $z_\perp$-derivatives vanish up to twist-4.

\subsection{Analytic results}

In this subsection we present the explicit expressions for the hard-scattering kernels up to one loop for the two-particle contributions as well as the tree-level kernels for the three-particle contributions.

\subsubsection{Two-particle contributions at one loop}
\label{sec:class-I:results:2pt}

Here we present the full set of one-loop hard-scattering kernels. Although in the rest of the paper we only consider pseudoscalar light mesons in the final state, we present also the results for light vector mesons in the final state since they are easily obtained from the former.

\paragraph{Vector sector}
We start by presenting our result for the SM operator $\mathcal{Q}_1^{VLL}$:
\begin{align}
    \bra{D^{(*)+}_{(s)}L^-} \mathcal{Q}_1^{VLL} \ket{\bar{B}^0_{(s)}}^{\text{1-loop}} =  \pm \frac{i f_L}{4} \int_0^1 du\, \phi_L(u) \, \left[J_V \, T^{VLL}(u,z_c) - J_A \, T^{VLL}(u,-z_c) \right] \,  ,
\end{align}
where $z_c = m_c/m_b$ is the scale invariant ratio of the heavy quark masses.
The hard-scattering kernel is given as
\begin{align}
    T^{VLL}(u,z_c) = \frac{\alpha_s}{4 \pi} \frac{C_F}{N_c} \left[-6 \log{\frac{\mu^2}{m_b^2}} - 18 + F^{VLL}(u,z_c) \right]\, ,
\end{align}
where
\begin{align}
    F^{VLL}(u,z_c) = \left(3+2 \log{\frac{u}{\bar{u}}}\right)\log{z_c^2} + f^{VLL}(u,z_c) + f^{VLL}(\bar{u},1/z_c)\, , \label{FVLLDef}
\end{align}
and
\begin{align}
    f^{VLL}(u,z_c) =
    &- \frac{u(1-z_c^2)[3(1-u(1-z_c^2))+z_c]}{\left[1-u(1-z_c^2)\right]^2} \log{[u(1-z_c^2)]}- \frac{z_c}{1-u(1-z_c^2)} \notag \\
    &+ 2\left\{\frac{\log{\left[u(1-z_c^2)\right]}}{1-u(1-z_c^2)}-\Li\left[1-u(1-z_c^2)\right] - \log^2{\left[u(1-z_c^2)\right]} - \{u \to \bar{u}\} \right\} \, .
\end{align}
The dilogarithm is defined as
\begin{align}
    \Li(x) = -\int_0^x \frac{\log{(1-t)}}{t}\, dt \, .
\end{align}
The upper (lower) sign corresponds to a light pseudoscalar (vector) meson in the final state. 
The non-perturbative quantities $f_L$ and $\phi_L$ are the decay constant of the light meson and the twist-2 LCDA of the light-meson.
For $L$ pseudoscalar, it is $\phi_L = \phi_P$ as defined in \cref{eq:LCDAs:2pt:psd:twist2}, while for $L = \rho, \, K^*$, the LCDA $\phi_L = \phi_V$ is defined in~Ref.~\cite{Cai:2021mlt}.
We use the following notation for the matrix elements of the heavy (axial-)vector current:
\begin{align}
    J_{V (A)} := \bra{D_{(s)}^{(*)+}} \bar{c} \slashed{q} (\gamma_5) b \ket{\bar{B}^0_{(s)}} \, ,
\end{align}
which can be further parametrized in terms of invariant form factors (with the conventions of~Ref.~\cite{Bordone:2019vic} up to a factor of ``$i$'' in the definition of $A_0$).
The perturbative kernel that we obtain coincides with the results given in Refs.~\cite{Beneke:2000ry, Cai:2021mlt}\footnote{Note that for the comparison with Ref.~\cite{Beneke:2000ry}, one has to divide our result by two since they used a different colour structure.}.

Furthermore, in the limit $z_c \to 0$, we get the loop function
\begin{align}
    \lim_{z_c \to 0} F^{VLL}(u,z_c) =& 3 \left(\frac{1-2u}{1-u} \log{u} -i\pi \right) \notag\\
    &+ \left[\Li(u) - \log^2{u} + \frac{2 \log{u}}{1-u} - (3+2i\pi) \log{u} - \{u \to \bar{u}\} \right]\, ,
\end{align}
which is the result for the vertex-corrections for charmless $B$-decays as presented in Refs.~\cite{Beneke:2001ev, Beneke:2003zv}.

For the $(V-A) \otimes (V+A)$ structure, we get
\begin{align}
    \bra{D^{(*)+}_{(s)}L^-}\mathcal{Q}_1^{VLR}\ket{\bar{B}^0_{(s)}}^{\text{1-loop}} &= -\frac{i f_L}{4}  \int_0^1 du\, \phi_L(u) \, \left[ J_V \,T^{VLR}(u,z_c) - J_A \, T^{VLR}(u,-z_c) \right] \, , 
\end{align}
with
\begin{align}
    T^{VLR}(u,z_c) = \frac{\alpha_s}{4 \pi} \frac{C_F}{N_c} \left[ 6 \log \frac{\mu^2}{m_b^2} + 6 + F^{VLR}(u,z_c) \right] \, ,
\end{align}
and
\begin{align}
    F^{VLR}(u,z_c) =- \left(3+2 \log{\frac{\bar{u}}{u}}\right)\log{z_c^2} - f^{VLL}(\bar{u},z_c) - f^{VLL}(u,1/z_c)\,. \label{FVLRDef}\
\end{align}
This aligns with the results presented in Ref.~\cite{Cai:2021mlt}. Note that $F^{VLR}(u,z_c) = -F^{VLL}(\bar{u},z_c)$. Hence, the limit $z_c\to 0$ is
\begin{align}
    \lim_{z_c \to 0} F^{VLR}(u,z_c) = -\lim_{z_c \to 0} F^{VLL}(\bar{u},z_c)\, ,
\end{align}
which again correctly reproduces the case of charmless $B$-decays in Refs.~\cite{Beneke:2001ev, Beneke:2003zv}.

\paragraph{Scalar and Tensor sector}
For the $(1-\gamma_5) \otimes (1-\gamma_5)$ structure, the result reads
\begin{align}
    \bra{D^{(*)+}_{(s)}L^-} \mathcal{Q}_1^{SLL} \ket{\bar{B}^0_{(s)}}^{\text{1-loop}} &= 
    \frac{i f_L}{4} \mu_L \int_0^1 du \, \phi_{3;L}(u) \, \left[J_S \,T^{SLL}(u,z_c) - J_P \,T^{SLL}(u,-z_c) \right]\,.
\end{align}
The hard-scattering kernel $T^{SLL}$ is defined as
\begin{align}
    T^{SLL}(u,z_c) = \frac{\alpha_s}{4 \pi} \frac{C_F}{N_c} \left[-\frac{4(u-\bar{u})(1-z_c)}{1+z_c} \log{\frac{\mu^2}{m_b^2}} + F^{SLL}(u,z_c) \right]\, ,
\end{align}
with
\begin{align}
    F^{SLL}(u,z_c) = 2\left[ \frac{(u - \bar{u})(1-z_c)}{1+z_c} + \log{\frac{u}{\bar{u}}} \right] \log{z_c^2} + f^{SLL}(u,z_c) + f^{SLL}(\bar{u}, 1/z_c) \, ,
\end{align}
and
\begin{align}
    f^{SLL}(u,z_c) =& -2 \bigg\{\frac{u(1-z_c)\left[u(1-z_c) + 2z_c \right] -1}{1 - u(1-z_c^2)} \log\left[u(1-z_c^2)\right] + \frac{5u}{1+z_c} + \log^2{\left[u(1-z_c^2)\right]} \notag \\
    &\qquad+ \Li \left[1-u(1-z_c^2)\right] \bigg\} - \{ u \rightarrow \bar{u} \}\, .
\end{align}
Here, we introduced the following notation for the heavy (pseudo-) scalar current:
\begin{align}
    J_{S (P)} := \bra{D_{(s)}^{(*)+}} \bar{c} (\gamma_5) b \ket{\bar{B}^0_{(s)}} \, .
\end{align}
The parameter $\mu_L$ differs between pseudoscalar and vector light mesons. For pseudoscalar $L$, we have $\mu_L(\mu) = \mu_p(\mu) = m_P^2/(m_u(\mu) + m_q(\mu))$, with the $\overline{\text{MS}}$ quark masses $m_i$ where $i = u, \, q$. For $L = \rho, \, K^*$, we have $\mu_L(\mu) = \mu_v(\mu) = m_V f_V^\perp(\mu)/f_V$ with the scale dependent transverse decay constant $f_V^\perp(\mu)$.
The twist-3 LCDA $\phi_{3;L}$ in the pseudoscalar case reduces to $\phi_P^p(u) = 1$ in the Wandzura-Wilczek limit as explained in \cref{app:lcda:2pt}. In the vector case, it is given by $\phi_{3;L} = \phi_v$ as defined in~Ref.~\cite{Cai:2021mlt}.

For the other scalar structure $(1-\gamma_5) \otimes (1+\gamma_5)$, the matrix element reads
\begin{align}
    \bra{D^{(*)+}_{(s)}L^-} \mathcal{Q}_1^{SLR} \ket{\bar{B}^0_{(s)}}^{\text{1-loop}} &= 
    \mp \frac{i f_L}{4} \mu_L \int_0^1 du \, \phi_{3;L}(u) \, \left[ J_S \,T^{SLR}(u,z_c) - J_P \, T^{SLR}(u,-z_c) \right] \,.
\end{align}
The hard-scattering kernel $T^{SLR}$ is given by
\begin{align}
    T^{SLR}(u,z_c) = \frac{\alpha_s}{4 \pi} \frac{C_F}{N_c}F^{SLR}(u,z_c) \, ,
\end{align}
with the loop function
\begin{align}
    F^{SLR}(u,z_c) = 2\log{\frac{u}{\bar{u}}} \log{z_c^2} - 6 + f^{SLR}(u,z_c) + f^{SLR}(\bar{u}, 1/z_c) \, ,
\end{align}
and 
\begin{align}
    f^{SLR}(u,z_c) &= \bigg \{\frac{u^2(z_c-1)^2(3z_c^2 + 4z_c + 2) - 2}{\left[1-u(1-z_c^2)\right]^2} \log{\left[u(1-z_c^2)\right]} + \frac{z_c^2}{(1+z_c)^2 \left[1-u(1-z_c^2)\right]} \notag \\
    &+ 2 \left[\frac{2\log{\left[u(1-z_c^2)\right]}}{1-u(1-z_c^2)} - \log^2{\left[u(1-z_c^2)\right]} - \Li\left[1- u(1-z_c^2)\right] \right] \bigg \}  - \{u \rightarrow \bar{u}\} \,.
\end{align}
We again investigate the $z_c\to 0$ limit, which yields
\begin{align}
    \lim_{z_c \to 0} F^{SLR}(u,z_c) = -6 + 2 \Li(u) - \log^2{u} - (1+2\pi i) \log{u} - \{u \to \bar{u}\} \, .
\end{align}
This agrees with the result for charmless $B$-decays presented in Ref.~\cite{Beneke:2003zv}.

Finally, for the $\sigma_{\mu \nu} (1-\gamma_5) \otimes \sigma^{\mu \nu} (1-\gamma_5)$ structure, we get
\begin{align}
    \bra{D^{(*)+}_{(s)}L^-} \mathcal{Q}_3^{SLL} \ket{\bar{B}^0_{(s)}}^{\text{1-loop}} &= 
    - \frac{i f_L}{4} \mu_L \int_0^1 du \, \phi_{3;L}(u) \left[ \, J_S \,T^{TLL}(u,z_c) - J_P\, T^{TLL}(u,-z_c) \right ]\,.
\end{align}
The hard-scattering kernel $T^{TLL}$ can be written as
\begin{align}
    T^{TLL}(u,z_c) = \frac{\alpha_s}{4 \pi} \frac{C_F}{N_c} \left [-48 \log{\frac{\mu^2}{m_b^2}} + F^{TLL}(u,z_c)\right] \, ,
\end{align}
with the loop function
\begin{align}
    F^{TLL}(u,z_c) = 8 \left [3+ \frac{(u- \bar{u})(1-z_c)}{z_c+1}\log{\frac{u}{\bar{u}}} \right] \log{z_c^2} + f^{TLL}(u,z_c) + f^{TLL}(\bar{u}, 1/z_c)\, ,
\end{align}
and
\begin{align}
    f^{TLL}(u,z_c) =& -\frac{8(4u+3)}{1+z_c} + \frac{8(1-z_c)}{1+z_c} \bigg \{ \frac{u [(u-2)z_c^2 - 2z_c + 2 - u] - 1}{1-u(1-z_c^2)} \log{\left[u(1-z_c^2)\right]} \nonumber\\
    &+ (1-2u) \Big[\log^2{\left[u(1-z_c^2)\right]} + \Li[1-u(1-z_c^2)]\Big] + \{ u \rightarrow \bar{u} \} \bigg \} \, .
\end{align}
Note the extra minus sign in comparison to Ref.~\cite{Cai:2021mlt}. This is because of the additional $i$ in our definition of $\sigma_{\mu \nu}$. The hard-scattering kernels of the parity flipped operators can be easily obtained from the ones given above as explained in \cref{app:parity}.
Our results for the hard-scattering kernels are in full agreement with the ones given in Ref.~\cite{Cai:2021mlt}. Additionally, we calculated the hard-scattering kernels in the case $m_c = 0$ independently and found agreement with the full hard-scattering kernels in the $z_c \to 0$ limit. Note that in order to get the sign of the imaginary parts of the hard-scattering kernels right, one has to set $z_c^2 \to z_c^2 - i \epsilon$, with $\epsilon > 0$ infinitesimal. Furthermore, the constant finite terms in the hard-scattering kernels depend on the choice of evanescent operators. Our choice is described in \cref{app:bases:evanescent} and coincides with the ones used in Ref.~\cite{Cai:2021mlt}.
The corresponding LO QCDF results are collected in \cref{app:LO}.

\subsubsection{Three-particle contributions at tree level}
\label{sec:class-I:results:3pt}

Here we present the contributions from additional collinear gluons in the final state of the light meson as shown in \cref{fig:class-I:hard-scattering-3pt:qqg}.  
We will again give the results for the full basis of BSM operators. 

We start with the results for the vector operators:
\begin{align}
    \bra{D^{(*)+}_{(s)} P^-}\mathcal{Q}_1^{VLL}\ket{\bar{B}^0_{(s)}}_{q\bar{q}g} &= 
    +i \frac{f_P}{m_b^2}  \int \mathcal{D} \underline{u} \, [J_V - J_A]  \, T^{VLL}_{q\bar{q}g}(u_3) \, \Phi_{4; P}(\underline{u}) \, , \\
    \bra{D^{(*)+}_{(s)} P^-}\mathcal{Q}_1^{VLR}\ket{\bar{B}^0_{(s)}}_{q\bar{q}g} &= 
    -i \frac{f_P}{m_b^2}  \int \mathcal{D} \underline{u} \, [J_V - J_A] \, T^{VLL}_{q\bar{q}g}(u_3) \, \Phi_{4; P}(\underline{u}) \, , \\
    \bra{D^{(*)+}_{(s)} P^-}\mathcal{Q}_1^{VRL}\ket{\bar{B}^0_{(s)}}_{q\bar{q}g} &= 
    +i \frac{f_P}{m_b^2}  \int \mathcal{D} \underline{u} \, [J_V + J_A] \, T^{VLL}_{q\bar{q}g}(u_3) \, \Phi_{4; P}(\underline{u}) \, , \\
    \bra{D^{(*)+}_{(s)} P^-}\mathcal{Q}_1^{VRR}\ket{\bar{B}^0_{(s)}}_{q\bar{q}g} &= 
    -i \frac{f_P}{m_b^2}  \int \mathcal{D} \underline{u} \, [J_V + J_A] \, T^{VLL}_{q\bar{q}g}(u_3) \, \Phi_{4; P}(\underline{u}) \, ,
\end{align}
where $\int \mathcal{D} \underline{u} \equiv \int_0^1 du_1 du_2 du_3\, \delta(1-u_1-u_2-u_3)$. Here, $\underline{u} = \{u_1, u_2, u_3\}$ denote the momentum fractions of the quark, antiquark, and the gluon of the light meson momentum $q$. The twist-4 LCDA $\Phi_{4;P}$ is defined in \cref{eq:lcda:3pt:tw4}.
The hard-scattering kernel is given by
\begin{align}
    T^{VLL}_{q\bar{q}g}(u_3, z_c) = \frac{1}{N_c}\frac{1}{u_3 \, (1 - z_c^2)} \, .
\end{align}
Our result for the SM operator $Q_1^{V LL}$ in the limit $m_c \to 0$ differs from the result in Eq.~(68) of Ref.~\cite{Beneke:2000ry}.
Accounting for different choices of operator basis, we find that Eq.~(68) of Ref.~\cite{Beneke:2000ry} should be multiplied by a factor of $1/N_c$.

For the scalar operators there is no non-vanishing three-particle contribution since the scalar operators cannot generate any non-vanishing vacuum-to-light meson transition for three-particle light meson states. For the tensor operators, however, we obtain the following
\begin{align}
    \bra{D^{(*)+}_{(s)} P^-} \mathcal{Q}_3^{SLL} \ket{\bar{B}^0_{(s)}}_{q\bar{q}g} =
        & +i f_{3 P} \frac{m_P^2}{m_b^2} \int \mathcal{D} \underline{u} \, \left[J_S \,T^{TLL}_{q\bar{q}g}(u_3, z_c) - J_P \, T^{TLL}_{q\bar{q}g}(u_3, -z_c)\right] \Phi_{3; P}(\underline{u})
    \nonumber\\
        &+ \mathcal{O}(q^4) \,, \\
    \bra{D^{(*)+}_{(s)} P^-} \mathcal{Q}_3^{SRR} \ket{\bar{B}^0_{(s)}}_{q\bar{q}g} =& -i f_{3 P} \frac{m_P^2}{m_b^2} \int \mathcal{D} \underline{u} \, \left[J_S \,T^{TLL}_{q\bar{q}g}(u_3, z_c) + J_P \, T^{TLL}_{q\bar{q}g}(u_3, -z_c)\right] \Phi_{3; P}(\underline{u})
    \nonumber\\
        &+ \mathcal{O}(q^4) \,,
\end{align}
with the hard-scattering kernel
\begin{align}
    T^{TLL}_{q\bar{q}g}(u_3, z_c) = \frac{2}{N_c} \frac{1}{u_3 \, (1+z_c)^2} \, ,
\end{align}
and the twist-3 LCDA $\Phi_{3; P}$ defined in \cref{eq:lcda:3pt:tw3}.
These contributions are zero for $q^2 = 0$, so we keep the leading term with $q^2 = m_P^2$.

\subsection{Branching ratios}
\label{sec:class-I:observables}

We write decay amplitudes for our decays of interest in the standard way as \footnote{For the $\bar{B}_{(s)} \to D_{(s)}^{*+} P^-$ amplitude we differ from the expression in Ref.~\cite{Beneke:2000ry} by a factor of minus one. This is because we insert this minus sign into the definition of $a_1$ such that $a_1(D^*P) \sim -1$ in the SM.}
\begin{align}
    \mathcal{A}(\bar{B}_{(s)} \to D_{(s)}^{+} P^-) &= i\frac{G_F}{\sqrt{2}} \, V_{cb} \, V_{uq}^*  \, a_1(D_{(s)} P) \, f_P \, f_0^{B_{(s)} \to D_ {(s)}}(m_P^2) \, (m_{B_{(s)}}^2 - m_{D_{(s)}}^2) \, , \\ 
    \mathcal{A}(\bar{B}_{(s)} \to D_{(s)}^{*+} P^-) &= i\frac{G_F}{\sqrt{2}} \, V_{cb} \, V_{uq}^* \, a_1(D^{*}_{(s)} P) \, f_P \, A_0^{B_{(s)} \to D_{(s)}^*}(m_P^2) \, \sqrt{\lambda(m_{B_{(s)}}^2, m_{D_{(s)}^*}^2, m_P^2)}  \, ,
\end{align}
with the Källén function $\lambda(a,b,c) = a^2 + b^2 + c^2 - 2 ab - 2 ac - 2bc$. Note that in our definition of $a_1$ we include BSM effects as well as three-particle contributions.
It is convenient to split up $a_1$ as
\eq{
a_1(D^{(*)}P) = a_1^{(0)}(D^{(*)}P) + \frac{\alpha_s}{4 \pi} \, a_1^{(1)}(D^{(*)}P) + a_1^{q\bar{q}g}(D^{(*)}P) \, .
}
Here, $a_1^{(n)}(D^{(*)}P)$ is the two-particle $n$-loop contribution, while $a_1^{q\bar{q}g}(D^{(*)}P)$ is the tree-level three-particle contribution.
The explicit expressions for $a_1$ are given by
\eqa{
a_1^{(0)}(DP) & = &  
\sum_{i,j} \, v_i \, M^{(0)}_{ij}(DP) \, (\wcbern[qbcu]{j} - \wcbern[qbcu]{j'}) \, ,
}
with the coefficient matrix
\begin{align}
\hat{M}^{(0)}(DP) & = 
\begin{pmatrix}
\frac{1}{3} & \frac{4}{9} & \frac{16}{3} & \frac{64}{9} & -\frac{1}{6} & -\frac{2}{9} & 0 & 0 & -\frac{8}{3} & -\frac{32}{9} \\
\frac{2}{3} & \frac{8}{9} & \frac{8}{3} & \frac{32}{9} & \frac{1}{6} & \frac{2}{9} & 2 & \frac{8}{3} & -\frac{64}{3} & -\frac{256}{9}
\end{pmatrix} \, ,
\end{align}
and the prefactor vector
$\vec{v} = \left( 1, \, \mu_p/(m_b - m_c) \right)$ .

The NLO contribution can be cast in an analogous form
\eq{
a_1^{(1)}(DP) = 
\sum_{i,j} \, v_i \, M^{(1)}_{ij}(DP) \, (\wcbern[qbcu]{j} - \wcbern[qbcu]{j'}) \, ,
}
with
\eqa{
\left(\hat{M}^{(1)}(DP)\right)^T = 
\begin{pmatrix}
 H_P^{VLL} - \frac{16}{9}                   & 2 H_P^{SLR}(z_c)+\frac{32}{9}                                  \\
 -\frac{1}{6} H_P^{VLL}(z_c)-\frac{28}{27}  & \frac{56}{27}-\frac{1}{3} H_P^{SLR}(z_c)                       \\
 16 H_P^{VLL}(z_c)+\frac{320}{9}            & 8 H_P^{SLR}(z_c)-\frac{256}{9}                                 \\
 -\frac{8}{3} H_P^{VLL}(z_c)-\frac{304}{27} & \frac{1280}{27}-\frac{4}{3} H_P^{SLR}(z_c)                     \\
 -\frac{1}{2} H_P^{VLR}(z_c)-\frac{4}{3}    & \frac{1}{2} H_P^{SLL}(z_c)-\frac{1}{8}H_P^{TLL}(z_c)+\frac{40}{9}  \\
 \frac{1}{12}H_P^{VLR}(z_c)-\frac{7}{9}     & -\frac{1}{12}H_P^{SLL}(z_c)+\frac{1}{48} H_P^{TLL}(z_c)+\frac{43}{27} \\
 -\frac{32}{9}                            & 6 H_P^{SLL}(z_c)+\frac{1}{2}H_P^{TLL}(z_c)-\frac{224}{9}                 \\
 -\frac{56}{27}                           & -H_P^{SLL}(z_c)-\frac{1}{12}H_P^{TLL}(z_c)-\frac{140}{27}                \\
 \frac{1280}{9}-8 H_P^{VLR}(z_c)            & -64 H_P^{SLL}(z_c)-16 H_P^{TLL}(z_c)+\frac{1024}{3}                      \\
 \frac{4}{3}H_P^{VLR}(z_c)-\frac{1216}{27}  & \frac{32}{3} H_P^{SLL}(z_c)+\frac{8}{3} H_P^{TLL}(z_c)+\frac{2176}{9} \\
\end{pmatrix} \, .
}
Here $H_P^{X}$ is the convolution integral of the hard-scattering kernel for the Dirac structure $X$ with the corresponding LCDA:
\eq{
H^X_P(z_c) = \frac{4 \pi}{\alpha_s}\int_0^1 du \, T^X(u, z_c) \, \phi_P^X(u) \, , 
}
where $\phi_P^X$ is given by
\eq{
\phi_P^X(u) =
\begin{cases}
    \phi_P(u),       & \text{if } X \in \{VLL, \, VLR\}         \\
    \phi_P^p(u), & \text{if } X \in \{SLL, \, SLR, \, TLL\}
\end{cases} \, .
}
We normalized the $H^X_P$ by $\frac{\alpha_s}{4 \pi}$ such that $a_1^{(1)}$ is $\mathcal{O}(\alpha_s^0)$.

We also write the three-particle contributions in a similar form,
\eq{
a_1^{q\bar{q}g}(DP) = 
\sum_{i, j} \, \widetilde{v}_i \, M^{q\bar{q}g}_{ij}(DP) \, (\wcbern[qbcu]{j} - \wcbern[qbcu]{j'}) \, ,
}
with
\eq{
\widetilde{v} = \left(\frac{1}{m_b^2}, \frac{f_{3L}}{f_L} \frac{m_L^2}{m_b^2} \frac{1}{(m_b - m_c)}\right)\ ,
}
and
\eqa{
(\hat M^{q\bar{q}g}(DP))^T =
\left(
\begin{array}{cc}
 4 H^{VLL}_{P,q\bar{q}g}(z_c)             & 0                                      \\
 -\frac{2}{3} H^{VLL}_{P,q\bar{q}g}(z_c)  & 0                                      \\
 64 H^{VLL}_{P,q\bar{q}g}(z_c)            & 0                                      \\
 -\frac{32}{3} H^{VLL}_{P,q\bar{q}g}(z_c) & 0                                      \\
 -2 H^{VLL}_{P,q\bar{q}g}(z_c)            & \frac{1}{2} H^{TLL}_{P,q\bar{q}g}(z_c)   \\
 \frac{1}{3} H^{VLL}_{P,q\bar{q}g}(z_c)   & -\frac{1}{12} H^{TLL}_{P,q\bar{q}g}(z_c) \\
 0                                      & -2 H^{TLL}_{P,q\bar{q}g}(z_c)            \\
 0                                      & \frac{1}{3} H^{TLL}_{P,q\bar{q}g}(z_c)   \\
 -32 H^{VLL}_{P,q\bar{q}g}(z_c)           & 64 H^{TLL}_{P,q\bar{q}g}(z_c)            \\
 \frac{16}{3} H^{VLL}_{P,q\bar{q}g}(z_c)  & -\frac{32}{3} H^{TLL}_{P,q\bar{q}g}(z_c) \\
\end{array}
\right)
\, .
}
The three-particle convolutions $H^X_{P,q\bar{q}g}$ are given by:
\eq{
    H^X_{P,q\bar{q}g}(z_c) = \int \mathcal{D}\underline{u} \,  
    \begin{cases}
    T^{VLL}_{q\bar{q}g}(u_3, z_c) \, \Phi_{4;P}(\underline{u}),     & \text{if } X = VLL \\
    T^{TTL}_{q\bar{q}g}(u_3, z_c) \, \Phi_{3;P}(\underline{u}),     & \text{if } X = TLL
    \end{cases} \, .
}
The corresponding expressions for $a_1(D^*P)$ can be written in a similar form. The vectors $v$ and $\widetilde{v}$ have to be modified by exchanging $m_b - m_c$ with $m_b + m_c$ and the primed Wilson coefficients get an extra minus sign. The matrices read
\begin{align}
\hat{M}^{(0)}(D^*P) & = 
\left(
\begin{array}{cccccccccc}
 -\frac{1}{3} & -\frac{4}{9} & -\frac{16}{3} & -\frac{64}{9} & -\frac{1}{6} & -\frac{2}{9} & 0 & 0 & -\frac{8}{3} & -\frac{32}{9} \\
 \frac{2}{3} & \frac{8}{9} & \frac{8}{3} & \frac{32}{9} & -\frac{1}{6} & -\frac{2}{9} & -2 & -\frac{8}{3} & \frac{64}{3} & \frac{256}{9} \\
\end{array}
\right) \, ,
\end{align}
\begin{align}
\left(\hat{M}^{(1)}(D^*P)\right)^T = 
\left(
\begin{array}{cc}
 -H_P^{VLL}(-z_c) + \frac{16}{9}             & 2 H_P^{SLR}(-z_c)+\frac{32}{9}                                           \\
 \frac{1}{6}H_P^{VLL}(-z_c)+\frac{28}{27}    & -\frac{1}{3}H_P^{SLR}(-z_c) + \frac{56}{27}                              \\
 -16 H_P^{VLL}(-z_c)-\frac{320}{9}           & 8 H_P^{SLR}(-z_c)-\frac{256}{9}                                          \\
 \frac{8}{3} H_P^{VLL}(-z_c)+\frac{304}{27}  & -\frac{4}{3} H_P^{SLR}(-z_c) + \frac{1280}{27}                           \\
 -\frac{1}{2}H_P^{VLR}(-z_c)-\frac{4}{3}     & -\frac{1}{2}H_P^{SLL}(-z_c)+\frac{1}{8}H_P^{TLL}(-z_c)-\frac{40}{9}      \\
 \frac{1}{12}H_P^{VLR}(-z_c)-\frac{7}{9}     & \frac{1}{12}H_P^{SLL}(-z_c)-\frac{1}{48}H_P^{TLL}(-z_c)-\frac{43}{27}     \\
 -\frac{32}{9}                             & -6 H_P^{SLL}(-z_c)-\frac{1}{2}H_P^{TLL}(-z_c)+\frac{224}{9}               \\
 -\frac{56}{27}                            & H_P^{SLL}(-z_c)+\frac{1}{12}H_P^{TLL}(-z_c)+\frac{140}{27}              \\
 \frac{1280}{9}-8 H_P^{VLR}(-z_c)            & 64 H_P^{SLL}(-z_c)+16 H_P^{TLL}(-z_c)-\frac{1024}{3}                      \\
 \frac{4}{3}H_P^{VLR}(-z_c)-\frac{1216}{27}  & -\frac{32}{3}H_P^{SLL}(-z_c)-\frac{8}{3}H_P^{TLL}(-z_c) -\frac{2176}{9}   \\
\end{array}
\right) \, ,
\end{align}
\begin{align}
(\hat{M}^{q\bar{q}g}(D^*P))^T = 
\left(
\begin{array}{cc}
 -4 H^{VLL}_{P,q\bar{q}g}(-z_c)           & 0                                      \\
 \frac{2}{3}H^{VLL}_{P,q\bar{q}g}(-z_c)   & 0                                      \\
 -64 H^{VLL}_{P,q\bar{q}g}(-z_c)          & 0                                      \\
 \frac{32}{3} H^{VLL}_{P,q\bar{q}g}(-z_c) & 0                                      \\
 -2 H^{VLL}_{P,q\bar{q}g}(-z_c)           & -\frac{1}{2}H^{TLL}_{P,q\bar{q}g}(-z_c)  \\
 \frac{1}{3}H^{VLL}_{P,q\bar{q}g}(-z_c)   & \frac{1}{12}H^{TLL}_{P,q\bar{q}g}(-z_c)  \\
 0                                      & 2 H^{TLL}_{P,q\bar{q}g}(-z_c)            \\
 0                                      & -\frac{1}{3}H^{TLL}_{P,q\bar{q}g}(-z_c)  \\
 -32 H^{VLL}_{P,q\bar{q}g}(-z_c)          & -64 H^{TLL}_{P,q\bar{q}g}(-z_c)          \\
 \frac{16}{3}H^{VLL}_{P,q\bar{q}g}(-z_c)  & \frac{32}{3} H^{TLL}_{P,q\bar{q}g}(-z_c) \\
\end{array}
\right) \, .
\end{align} 
Finally, the decay rates are obtained by performing the phase space integration, 
\eqa{
\Gamma\big(\bar{B}^0_{(s)} \to D^+_{(s)} P^-\big) 
& = &  \frac{G_F^2 \, (m_{B_{(s)}}^2 - m_{D_{(s)}}^2)^2 \, \lambda^{1/2} (m_{B_{(s)}}^2, m_{D_{(s)}}^2, m_P^2)}{32 \pi \, m_{B_{(s)}}^3}  \notag \\
& & \times \left|V_{cb}V_{uq}^*\right|^2 \, \left|a_1\left(D_{(s)} P\right)\right|^2 \, f_P^2 \,  \left(f_0^{B_{(s)} \to D_{(s)}} \left(m_P^2\right) \right)^2 \, , 
\\[5mm]
\Gamma\big(\bar{B}^0_{(s)} \to D^{*+}_{(s)} P^-\big)
& = &
\frac{G_F^2 \, \lambda^{3/2}(m_{B_{(s)}}^2, m_{D_{(s)}^*}^{2}, m_P^2)}{32 \pi \, m_{B_{(s)}}^3} \,  \left|V_{cb}V_{uq}^*\right|^2 \, \left|a_1\left(D_{(s)}^{*
} P\right)\right|^2 \, f_P^2 \, \left(A_0^{B_{(s)} \to D_{(s)}^*} \left(m_P^2\right) \right)^2 \, .  \notag \\
}
%


\section{The $B$-meson lifetime in the Weak Effective Theory}
\label{sec:lifetime}

\begin{figure}
    \centering
    \begin{tikzpicture}
        \begin{feynman}
        \tikzstyle{vertexbox} = [draw, rectangle, minimum size=1.5mm, inner sep=0mm, fill=black]
        \vertex (i1) {\(b\)};
        \vertex[right=2cm of i1] (v1);
        \vertex[right=3cm of v1] (v2);
        \vertex[right=2cm of v2] (o1) {\(b\)};

        \node[vertexbox] at (v1) {};
        \node[vertexbox] at (v2) {};
        \node[above left=0.2cm of v1] {$\mathcal{O}_i$};
        \node[above right=0.2cm of v2] {$\mathcal{O}_j^{\dagger}$};
        \node at ($ (v1)!0.5!(v2) + (0.2,0.2) $) {\(c\)};
        \node at ($ (v1)!0.5!(v2) + (0.2,+1.3) $) {\(q\)};
        \node at ($ (v1)!0.5!(v2) + (0.2,-1.3) $) {\(\bar{u}\)};
        \diagram* {
            (i1) -- [fermion] (v1),                
            (v1) -- [fermion] (v2),  

            (v1) -- [fermion, out=70, in=110] (v2),

            (v2) -- [fermion, out=-110, in=-70] (v1),

            (v2) -- [fermion] (o1),                
        };
        \end{feynman}
        \draw[dashed, thick] (3.3, -1.3) -- (3.3, 1.3);
    \end{tikzpicture}
    \caption{Tree-level contribution to the $(i,j)$ interference to the partonic rate $\Gamma(b\to c\bar u q)$.}
    \label{fig:lifetime}
\end{figure}
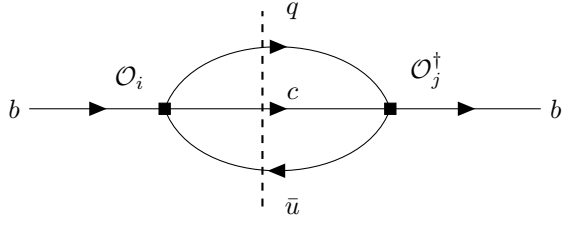

Apart from the constraints coming from exclusive branching ratios, we also constrain the WET coefficients using the total lifetime of the $B$-meson.
The lifetime constraint serves the purpose of providing an absolute bound on the WET parameters, \ie,
to ensure a finite volume of the parameter space.
Hence, for our purposes, it is sufficient to compute the leading-order expression for the lifetime,
without taking into account either corrections beyond the heavy-quark limit or perturbative QCD or QED corrections.
In the heavy quark limit, the inclusive decay rate $\Gamma(B\to X_{c\bar u q})$
is equal to the partonic decay rate, 
\begin{equation}
    \Gamma(B\to X_{c\bar u q})
        = \Gamma(b\to X_{c\bar u q}^\text{part}) + {\cal O}(1/m_b)
        = \underbrace{\Gamma(b\to c\bar u q)}_{\Gamma_q} + {\cal O}(1/m_b,\alpha_s,\alpha) \, .
\end{equation}
The leading three-particle inclusive partonic rate is given by
\begin{equation}
    \Gamma_q = \frac1{2N_c}\frac1{2m_b} \int \ \sum|{\cal M}|^2 dPS_3 \ ,
\end{equation}
where the sum over colour and spin of the squared matrix element at tree level is obtained directly by evaluating the cut diagram shown in \cref{fig:lifetime}.
We write the result of this computation in the form
\begin{equation}
    \Gamma_q
        = \Gamma_0 \, |V_{uq}^*|^2 \, \sum_{i,j} \Big(\wcbern[qbcu *]{i} \wcbern[qbcu]{j}+\wcbern[qbcu *]{i'} \wcbern[qbcu]{j'}\Big)\,G_{ij} \, ,
\end{equation}
with
\begin{equation}
\Gamma_0 \equiv \frac{G_F^2 \, m_b^5 \, |V_{cb}|^2}{192 \, \pi^3} \, .
\end{equation}
The coefficients $G_{ij}$ fulfil a number of symmetry relations. It is easy to see that only insertions of either two singlet or two octet operators yield non-zero contributions due to the colour structure of the diagram and hence we have $G_{ij} = 0$ if one of the indices is even and the other one is odd. Furthermore, the contributions from two singlet operator insertions are equal to the two octet operator insertions up to an overall colour factor, such that $G_{i+1 \, j+1} = \frac{2}{9} \, G_{ij}$ for $i,j$ odd. 
Additionally, it is clear that due to the structure of the Dirac traces coming from the cut diagram \cref{fig:lifetime}, the chirality-flipped operators yield the same contributions as their non-flipped counterparts.
Finally, $G_{ij}= G_{ji}$. 
Taking into account these considerations, the whole matrix can be constructed from the entries with $i,j$ odd and $i \leq j$:
\begin{align}
    G_{11}
        & = \frac{1}{10} G_{13} = \frac{1}{136} G_{33} = 4 G_{55} = \frac{1}{10} G_{59} = \frac{1}{6} G_{77} = -\frac{1}{96} G_{79} = \frac{1}{2080} G_{99} = 6 \, f(z_c) \, , 
        \nonumber\\
    G_{15}
        & = \frac{1}{6} G_{17} = -\frac{1}{56} G_{19} = \frac{1}{10} G_{35} = \frac{1}{24} G_{37} = -\frac{1}{128}  G_{39} = 6 \, g(z_c) \,, \\
    G_{57} &= 0 \, ,
     \nonumber
\end{align}
where the two functions $f$ and $g$ are given as 
\begin{align}
    f(z_c)
        & = 1 - 8 z_c^2 + 8 z_c^6 - z_c^8 - 24 z_c^4 \log(z_c) \, , \\[1mm]
    g(z_c)
        & = z_c \, \left(1 + 9 z_c^2 - 9 z_c^4 - z_c^6 + 12 (z_c^2 + z_c^4) \log(z_c) \right) \, .
\end{align}
The SM part of our results reproduces the LO result in Ref.~\cite{Egner:2024azu}.

\section{Phenomenological analysis}
\label{sec:pheno}

We now investigate the implications of the experimental data on the non-leptonic decays, and the corresponding $b\to c \bar{u} q$ puzzle, by carrying out a phenomenological analysis within the WET.
The analysis makes use of the available data on \BqToDqP and \BqToDqV decays as well as measurements of the $B$-meson lifetime.
As mentioned in~\cref{sec:wet}, we perform the analysis within the Bern basis of WET operators.

\subsection{Setup}
\label{sec:pheno:setup}

\subsubsection{Statistical framework}
\label{sec:pheno:setup:framework}

We perform a Bayesian analysis of the available data on \BqToDqL decays. Central to any Bayesian analysis
is the so-called posterior probability density function (PDF):
\begin{equation}
    \label{eq:pheno:setup:posterior}
    P(\vecx\,|\, D, M) \propto P(D \,|\, \vecx, M)\, P_0(\vecx \,|\, M)\,.    
\end{equation}
As part of the analysis, we maximize this posterior PDF with respect to $\vecx$,
or we draw random samples $\vecx \sim P(\vecx\,|\, D,M)$ to determine the allowed parameter space and produce posterior predictions.
The parameters $\vecx = (\vecth, \vecnu)$ split into the parameters of interest $\vecth$ (here: the WET
Wilson coefficients) and nuisance parameters $\vecnu$ (here: hadronic parameters).
The posterior PDF factorizes into the likelihood function $P(D \,|\, \vecx, M)$, which accounts for 
the experimental and theoretical constraints following from the labelled data set $D$,
and the prior PDF $P_0(\vecx \,|\, M)$, which accounts for our a-priori knowledge of the parameters.
The fit model $M$ governs what parameters are varied and how their priors are chosen.
We discuss the data set and the fit models entering our analysis in \cref{sec:pheno:setup:exp} and \cref{sec:pheno:setup:models}, respectively.\\

To compare a pair of fit models $M_1$ and $M_2$, we compute the normalization of the posterior
for both models given a common dataset $D$:
\begin{equation}
    \label{eq:pheno:setup:evidence}
    P(D \,|\, M_i) \equiv \int d\vecx\, P(D \,|\, \vecx, M_i)\, P_0(\vecx \,|\, M_i)\,.
\end{equation}
This quantity is known as the marginal posterior of $M_i$ or the evidence of $M_i$.
Ratios of evidences $K(M_1, M_2) \equiv P(D \,|\, M_1) / P(D \,|\, M_2)$ are called Bayes factors
and provide information about the efficiency of the two models in describing their common dataset $D$.\\
Following Jeffreys' interpretation~\cite{Jeffreys:1939xee}, $M_1$ is preferred over the model $M_2$ if $K(M_1, M_2) > 1$.
The levels of preference range from ``barely worth mentioning'' ($1 < K(M_1, M_2) \leq 3$), substantial ($3 < K(M_1, M_2) \leq 10$),
strong ($10 < K(M_1, M_2) \leq 100$), and decisive ($100 < K(M_1, M_2)$).
For $K(M_1, M_2) < 1$, the above interpretation favours $M_2$ over $M_1$, and the level is determined from $1/K$ instead.\\

In the course of our analysis we are confronted with posterior densities that feature well-separated
local modes for the parameters of interest.
To determine the relevance of these local modes, we proceed as follows. First, we distinguish
between any pair of local modes in an unambiguous way. In our analysis, this can be achieved
either through rotating the parameter space or by creating a characteristic scalar function $f$.
We use the latter approach, by defining a hyperplane in the vector space of $\vecth$ and
further defining $f = \vec{n} \cdot \vecx + \text{const}$.
This hyperplane (and its normal vector $\vec{n}$) is chosen such that one of the modes yields $f < 0$,
while the other mode yields $f > 0$.
The procedure is iterated if more than one hyperplane is needed to cleanly separate the modes.
Second, we determine the \emph{local evidence} of a mode:
\begin{equation}
    P(D \,|\, M_i)_{m} \equiv \int d\vecx\, P(D \,|\, \vecx, M_i)\, P_0(\vecx \,|\, M_i) \, \prod_{i} \mathbbm{1}_{f_i^m}(\vecth)\,,
\end{equation}
where $m$ indicates the mode of interest, $\mathbbm{1}$ represents a suitable indicator function,
and $i$ iterates over all characteristic functions as collected in \cref{app:modes}.
The local evidence is therefore obtained by constraining the integration volume to contain only the mode $m$.
Naturally, we recover the global evidence by adding up all local evidences, \ie,
\begin{equation}
    P(D \,|\, M_i) = \sum_m P(D \,|\, M_i)_{m}\,.
\end{equation}
To streamline our discussion of the phenomenological results, we dismiss all modes with local evidence below $2\%$ of the global evidence.

\subsubsection{Experimental data}
\label{sec:pheno:setup:exp}

\begin{table}[p]
    \centering
    \renewcommand*{\arraystretch}{1.1}
    \resizebox{\textwidth}{!}{%
    \begin{tabular}{@{}cccc@{}}
        \toprule
        Measurement
            & Central value \& Uncertainty
            & Reference
            & Notes
            \\
        \midrule
        \multicolumn{3}{c}{B factory analyses}\\
        \midrule
            $\mathcal{B}(B_s^0 \to D_s^- \pi^+)$ 
            &  $(3.6 \pm 0.5 \pm 0.5) \cdot 10^{-3} $
            & \cite{Belle:2008ezn}
            &
            \\
            $\displaystyle\frac{\mathcal{B}(B^0 \to D^- K^+)}
            {\mathcal{B}(B^0 \to D^- \pi^+)}$ 
            &  $(6.8 \pm 1.7) \cdot 10^{-2} $
            & \cite{Belle:2001ccu}
            &
            \\
            $f_{00} \mathcal{B}(B^0 \to D^- \pi^+) 
            \mathcal{B}(D^- \to K^+ \pi^- \pi^-)$
            & $(1.21 \pm 0.05) \cdot 10^{-4}$
            & \cite{BaBar:2006rof, CLEO:2002iij}
            &
            \\
            $\mathcal{B}(B^0 \to D^- \pi^+) \mathcal{B}(D^- \to K^+ \pi^- \pi^-)$
            &   $(2.88 \pm 0.29) \cdot 10^{-4}$
            & \cite{BaBar:2006zod} 
            & $^\dagger$
            \\
            $\displaystyle\frac{\mathcal{B}(B_s^0 \to D_s^ {*-} \pi^+)}{\mathcal{B}(B_s^0 \to D_s^- \pi^+)}$
            & $0.66 \pm 0.16$
            & \cite{Belle:2010ldr}
            &
            \\
            $f_{00} \mathcal{B}(B^0 \to D^{*-} \pi^+)$
            & $(1.40 \pm 0.07) \cdot 10^{-3}$ 
            & \cite{BaBar:2006rof, CLEO:1997vmd}
            &
            \\
            $\displaystyle\frac{\mathcal{B}(B^0 \to D^{*-} \pi^+)}{\mathcal{B}(B^0 \to D^- \pi^+)}$
            & $(0.99 \pm 0.14)$
            & \cite{BaBar:2006zod}
            &
            \\
        \midrule
        \multicolumn{3}{c}{Tevatron analyses}\\
        \midrule
            $\displaystyle
            \fsofd\big|_\text{Tevatron}
            \frac{\mathcal{B}(B_s^0\to D_s^- \pi^+)}{\mathcal{B}(B^0\to D^- \pi^+)}
            \frac{\mathcal{B}(D_s^-\to \phi \pi^-)}{\mathcal{B}(D^-\to K^+\pi^-\pi^-)}
            $
            & $(6.7 \pm 0.5) \cdot 10^{-2}$
            & \cite{CDF:2006hob}
            &
            \\
        
        \midrule
        \multicolumn{3}{c}{LHC analyses}\\
        \midrule
            $\displaystyle
            \fsofd\big|_\text{LHCb(7 TeV)}
            \frac{\mathcal{B}(B_s^0\to D_s^-\pi^+)}{\mathcal{B}(B^0\to D^- \pi^+)}
            \frac{\mathcal{B}(D_s^-\to K^+K^-\pi^-)}{\mathcal{B}(D^-\to K^+\pi^-\pi^-)}
            $
            & $(17.4 \pm 0.7) \cdot 10^{-2}$
            & \cite{LHCb:2012wdi}
            &
            \\
            $\displaystyle
            \fsofd\big|_\text{LHCb(7 TeV)}
            \frac{\mathcal{B}(B_s^0 \to D_s^- \pi^+)}
            {\mathcal{B}(B^0 \to D^- K^+)}
            \frac{\mathcal{B}(D_s^- \to K^+ K^- \pi^-)}
            {\mathcal{B}(D^- \to K^+ \pi^- \pi^+)}
            $
            & $2.08 \pm 0.08$
            & \cite{LHCb:2013vfg} 
            & $^*$
            \\
            $
            \frac{\mathcal{B}(B^0 \to D^- K^+)}{\mathcal{B}(B^0 \to D^- \pi^+)}
            $
            & $(8.22 \pm 0.28) \cdot 10^{-2}$
            & \cite{LHCb:2013vfg} 
            & $^*$
            \\
            
        \midrule
        \multicolumn{3}{c}{PDG averages}\\
        \midrule
            $\mathcal{B}(D_s^- \to \phi \pi^-) \mathcal{B}(\phi \to K^+ K^-)$
            & $(2.27 \pm 0.08) \cdot 10^{-2}$
            & \cite{ParticleDataGroup:2018ovx}
            &
            \\
            $\mathcal{B}(D_s^- \to K^+ K^- \pi^-)$
            & $(5.45 \pm 0.17) \cdot 10^{-2}$
            & \cite{ParticleDataGroup:2018ovx}
            &
            \\
            $\mathcal{B}(D^- \to K^+ \pi^- \pi^-)$
            & $(9.38 \pm 0.16) \cdot 10^{-2}$
            & \cite{ParticleDataGroup:2018ovx}
            &
            \\
        
        \midrule
        \multicolumn{3}{c}{Other}\\
        \midrule
            $\displaystyle\frac{\mathcal{B}(B^0 \to D^{*-} K^+)}
            {\mathcal{B}(B^0 \to D^{*-} \pi^+)}$
            & $(7.75 \pm 0.30) \cdot 10^{-2}$
            & \cite{Belle:2001ccu, BaBar:2005osj, LHCb:2013fqe}
            &
            \\
            $f_{00}$
            & $0.488 \pm 0.010$
            & \cite{Jung:2015yma, Belle:2002lms, BaBar:2005uwr}
            &
            \\
        
        \bottomrule
    \end{tabular}
    }
    \caption{%
        Summary of the definition of the auxiliary observables entering the likelihood and their
        respective statistical constraints.
        The measurements marked with $^*$ are correlated with each other with a correlation coefficient of $-0.56$. 
        In the reference marked with $^\dagger$, both the $D^- \to K^+ \pi^- \pi^-$ and $D^- \to K_s \pi^-$ channels are considered, but the $D^- \to K^+ \pi^- \pi^-$ channel dominates.
    }
    \label{tab:pheno:setup:exp:likelihood}
\end{table}

\begin{figure}[t]
    \centering
    \includegraphics[width=0.8\textwidth]{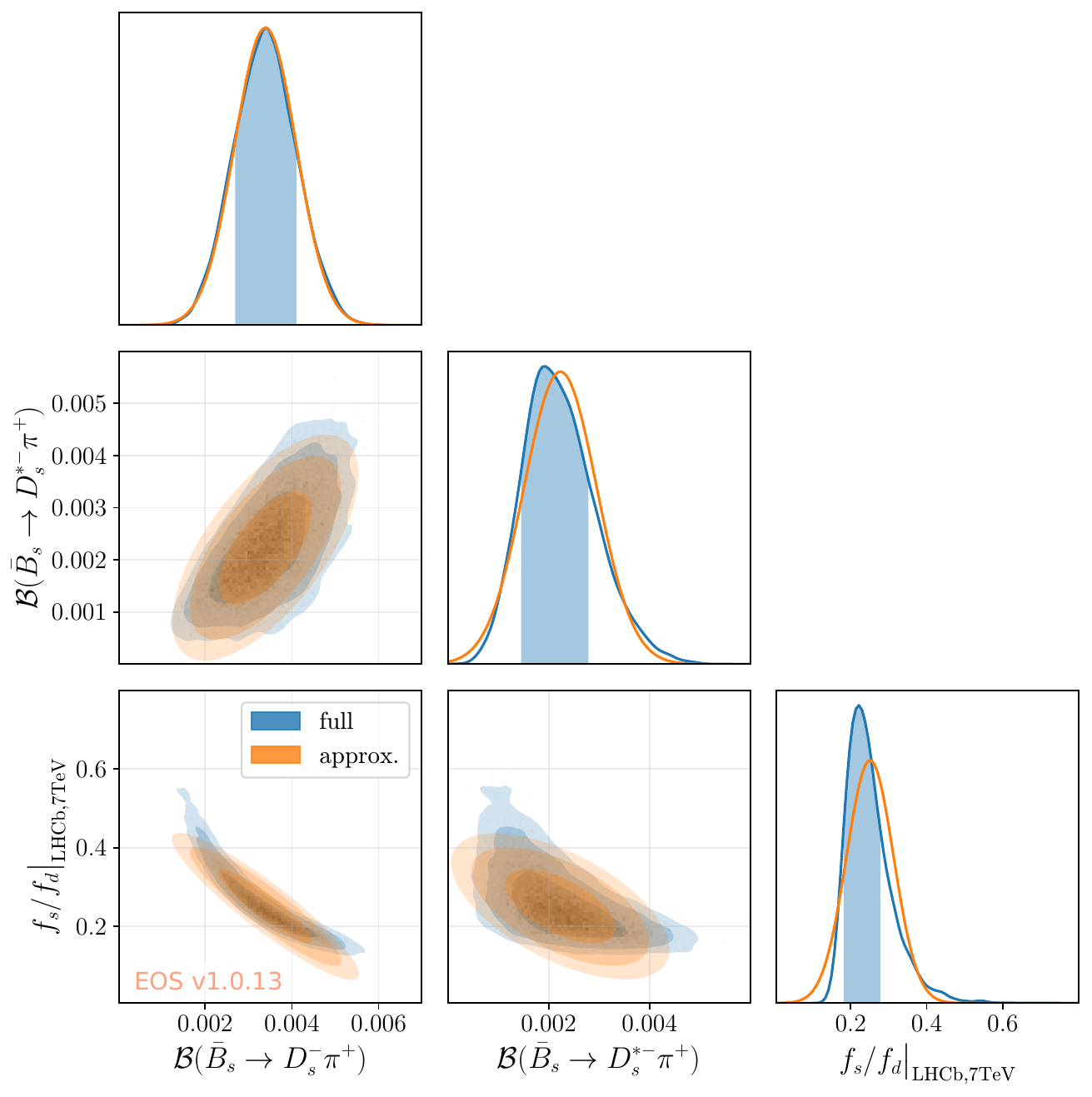}
    \caption{%
        Selection of one-dimensional and two-dimensional marginal likelihoods arising from the global likelihood
        as described in \cref{tab:pheno:setup:exp:likelihood}. We choose to display
        the branching ratios $\mathcal{B}(\bar{B}_s\to D_s^{(*)} \pi)$ and the hadronisation fraction
        $f_s/f_d$ for the LHCb phase at $7\,\TeV$ centre-of-mass energy.
        Our choice highlights the non-Gaussian nature of the full likelihood and
        illustrates the effects of our Gaussian approximation.
    }
    \label{fig:pheno:setup:exp:likelihood:marginals}
\end{figure}

As discussed in Ref.~\cite{Bordone:2020gao}, an accurate description of the experimental constraints on
the branching ratios for the \BqToDqP and \BqToDqV decays of interest is difficult,
and an indiscriminate use of the PDG world
averages for the various branching ratios~\cite{ParticleDataGroup:2022pth} would lead to inconsistent results.
This is due to the fact that previous experimental analyses have relied on
\begin{itemize}
    \item contemporary absolute measurements of the branching ratios for the subsequent $D_{(s)}$ decay,
        \eg, $D_s \to \phi \pi$;
    \item contemporary absolute measurements of the branching ratios of a normalisation channel,
        \eg, $\bar{B}^0\to D^+ \pi^-$.
    \item contemporary determinations of the $B_s$-meson fragmentation fraction $f_s$,
        typically normalised to the $B_d$-meson fragmentation fraction $f_d$;
    \item an ad-hoc assumption on the $p_T$ dependence of $f_s$.
\end{itemize}
Following Ref.~\cite{Bordone:2020gao}, we construct our experimental likelihood
for the branching ratios of \BqToDqP and \BqToDqV decays from the measurements of auxiliary observables.
Absolute branching ratios of the subsequent $D_{(s)}$ decay, the normalisation channels, and
the hadronisation fraction are included as nuisance parameters.
All auxiliary observables and nuisance parameters as well as their statistical constraints are listed in
\cref{tab:pheno:setup:exp:likelihood}.
A machine-readable description of the likelihood for use with the \EOS software is available as the file \texttt{likelihood.yaml}
within the supplementary material~\cite{EOS-DATA-2024-03}.\\
For the purpose of our proof-of-concept analysis, we approximate
this likelihood with a multivariate Gaussian likelihood, which is available in \EOS under the qualified name
\begin{center}
    \verb|B_(s)->D_(s)^(*)::BR@MvDV:2024A|\,.
\end{center}

The effect of the Gaussian approximation is illustrated in \cref{fig:pheno:setup:exp:likelihood:marginals}.
Approximating the full likelihood avoids using a total of eight experimental nuisance parameters.
This step considerably simplifies and speeds up our analysis, in particular the sampling processes.
The approximative likelihood of the branching ratios for the decays of interest contributes a total of four observations to the analysis.

In addition to the exclusive decay modes discussed above, we further impose a penalty function on the
theory prediction for the total decay width $\Gamma(\vecx)$ of the $B^-$ meson, assuming that this
decay width minus the contributions from the semileptonic $B^- \to X \ell^- \nu$  channels is completely saturated by decays mediated by the $qbcu$ sector.
As discussed in \cref{sec:lifetime}, our theory prediction only accounts for tree-level contributions
at leading order in $\alpha_s$ and leading power in $1/m_b$.
Our choice of the penalty function contributes to the overall likelihood function as
\begin{equation}
    \label{eq:pheno:penalty-function}
    -2 \log P(\text{$B^-$ lifetime}\,|\, \vecx, M) = \begin{cases}
            \left(\frac{\Gamma(\vecx) - \mu}{\sigma}\right)^2    &  \text{if $\Gamma(\vecx) > \mu$}\\
            0                                                    &  \text{otherwise}
        \end{cases}
\end{equation}
where we use
\begin{equation}
\begin{aligned}
    \mu
        & = \Gamma(B^-)_\text{PDG} - \Gamma(B^- \to X \ell^-\nu)_\text{PDG} = 0.4763\,\unit{\ps^{-1}}\,,\\
    \sigma
        & = \sigma(\Gamma(B^-))_\text{PDG} = 0.0015\,\unit{\ps^{-1}}\,.
\end{aligned}
\end{equation}
This penalty function provides a very weak constraint on the size of the $qbcu$-sector coefficients, and
its dominant effect is to limit the volume of the WET parameter space.
The constraint could be strengthened by dedicated inclusive measurements; see~\cref{sec:pheno:outlook} for further discussion.
We do not count this penalty function as an observation for the purpose of determining the total
degrees of freedom or a $p$ value.

\subsubsection{Parameters of interest and their priors}
\label{sec:pheno:setup:models}

We analyse the available data using five fit models, which we label SM and \mbox{WET-1} through \mbox{WET-4}.
These models differ only in terms of the parameters of interest $\vecth$ but share a common set of (hadronic) nuisance parameters $\vecnu$.
All of our fit models assume the Wilson coefficients to be real and flavour universal between the $dbcu$ and $sbcu$ sectors. This reduces the maximum number of independent parameters of interest to $20$.
The five fit models are the following:

\begin{description}
    \item[\textbf{SM}] This model lacks any parameters of interest whatsoever. The WET Wilson coefficients are fixed to their SM values.
    Hence, it serves as the \emph{null hypothesis} for the purpose of model comparisons.

    \item[\textbf{SM+PC}] This model has two parameters of interest, which
    account for power corrections to the QCDF results in the SM via
    \begin{equation}
    \begin{aligned}
        a_1(D_{(s)} P)
            & \to a_1(D_{(s)} P) \times (1 + \delta_P)\,, &
        a_1(D_{(s)}^* P)
            & \to a_1(D_{(s)}^* P) \times (1 + \delta_V)\,.
    \end{aligned}
    \end{equation}
    As for the SM model, the WET Wilson coefficients are fixed to their SM values.
    The newly introduced parameters $\delta_P$ and $\delta_V$ are used universally for both quark flavours $q=d,s$.
    Their priors are chosen to be uniform distributions
    \begin{equation}
    \begin{aligned}
        \delta_P
            & \in [-0.5\%, 0]\,, &
        \delta_V
            & \in [-0.3\%, 0]\,.
    \end{aligned}
    \end{equation}
    These intervals correspond to the larger of the estimates provided in Ref.~\cite{Bordone:2020gao}
    for the as-of-yet unknown soft-gluon contributions.
    Note that the estimate in Ref.~\cite{Bordone:2020gao} includes the sign of this contribution,
    and that allowing the parameters to take the opposite sign would only increase the tension with the data.

    \item[\textbf{SM+PC'}] This model corresponds to SM+PC, albeit with much larger prior intervals
    for the power correction parameters:
    \begin{equation}
        \begin{aligned}
        \delta_P
            & \in [-30\%, 0]\,, &
        \delta_V
            & \in [-30\%, 0]\,.
    \end{aligned}
    \end{equation}
    These intervals correspond to the size of power corrections needed to explain the anomalies through power corrections
    alone, as discussed in the literature~\cite{Bordone:2020gao}.

    \item[\textbf{WET-1}] This model has four parameters of interest, corresponding to the WET Wilson coefficients $\wcbern[qbcu]{1}$ through $\wcbern[qbcu]{4}$.
    We do not just vary the BSM part of the Wilson coefficients but the whole Wilson coefficients.
    Hence, there is no fixed SM contribution in this scenario and thus the likelihood is symmetric under $\wcbern[qbcu]{i} \to -\wcbern[qbcu]{i}$.
    The priors are chosen as independent uniform distribution on the support
    \begin{equation}
    \begin{aligned}
        \wcbern[qbcu]{1} & \in [-3.0, 3.0] \,, &
        \wcbern[qbcu]{2} & \in [-4.0, 4.0] \,, \\
        \wcbern[qbcu]{3} & \in [-0.3, 0.3] \,, &
        \wcbern[qbcu]{4} & \in [-0.4, 0.4] \,. \\
    \end{aligned}
    \end{equation}
    All other WET Wilson coefficient are set to zero.
    Hence, this fit model corresponds to floating only the SM-like WET Wilson coefficients.
    
    \item[\textbf{WET-2}] This model has six parameters of interest, corresponding to the WET Wilson coefficients $\wcbern[qbcu]{5}$ through $\wcbern[qbcu]{10}$.
    Their priors are chosen as independent uniform distribution on the support
    \begin{equation}
    \begin{aligned}
        \wcbern[qbcu]{5} & \in [-3.5, 4.0] \,, &
        \wcbern[qbcu]{6} & \in [-6.5, 6.5] \,, \\
        \wcbern[qbcu]{7} & \in [-1.5, 1.5] \,, &
        \wcbern[qbcu]{8} & \in [-3.0, 2.0] \,, \\
        \wcbern[qbcu]{9} & \in [-0.1, 0.1] \,, &
        \wcbern[qbcu]{10} & \in [-0.2, 0.2] \,. \\
    \end{aligned}
    \end{equation}
    The Wilson coefficients $\wcbern[qbcu]{1}$ though $\wcbern[qbcu]{4}$ are set to their SM values.
    Hence, this fit model corresponds to floating only the six non-SM Wilson coefficients of the unprimed operators while setting the SM-like Wilson coefficients to their SM values. The SM contribution breaks the $\wcbern[qbcu]{i} \to -\wcbern[qbcu]{i}$ symmetry that was present in the WET-1 scenario.
    
   \item[\textbf{WET-3}] This model has four parameters of interest, corresponding to the WET Wilson coefficients $\wcbern[qbcu]{1'}$ through $\wcbern[qbcu]{4'}$.
    Their priors are chosen as independent uniform distribution on the support
    \begin{equation}
    \begin{aligned}
        \wcbern[qbcu]{1'} & \in [-2.5, 2.5] \,, &
        \wcbern[qbcu]{2'} & \in [-2.5, 5.0] \,, \\
        \wcbern[qbcu]{3'} & \in [-0.25, 0.25] \,, &
        \wcbern[qbcu]{4'} & \in [-0.25, 0.25] \,. \\
    \end{aligned}
    \end{equation}
    All other BSM WET Wilson coefficients are set to zero while the four SM WET Wilson coefficients $\wcbern[qbcu]{1 \dots 4}$ are set to their SM value.

     \item[\textbf{WET-4}] This model has six parameters of interest, corresponding to the WET Wilson coefficients $\wcbern[qbcu]{5'}$ through $\wcbern[qbcu]{10'}$.
    Their priors are chosen as independent uniform distribution on the support
    \begin{equation}
    \begin{aligned}
        \wcbern[qbcu]{5'} & \in [-4.0, 3.5] \,, &
        \wcbern[qbcu]{6'} & \in [-10.0, 4.5] \,, \\
        \wcbern[qbcu]{7'} & \in [-1.5, 1.5] \,, &
        \wcbern[qbcu]{8'} & \in [-2.0, 3.0] \,, \\
        \wcbern[qbcu]{9'} & \in [-0.1, 0.1] \,, &
        \wcbern[qbcu]{10'} & \in [-0.2, 0.2] \,. \\
    \end{aligned}
    \end{equation}
    The Wilson coefficients $\wcbern[qbcu]{1}$ through $\wcbern[qbcu]{4}$ are set to their SM values while all the other BSM Wilson coefficients are set to zero.
\end{description}
Our choice of fit models is motivated by the structure of the SM contribution and the anomalous mass dimension matrix
in the Bern basis of operators.

Within the SM, the coefficients $\wcbern[qbcu]{1}$ through $\wcbern[qbcu]{4}$ take non-zero values, while all other coefficients vanish; see \cref{eq:preliminaries:matching-conditions-bern}.
Under change of the renormalization scale, these four SM-like coefficients mix with each other but do not induce contributions
in the remaining coefficients. BSM physics inducing effects amongst the SM-like coefficients only is therefore captured by
model WET-1. If, on the other hand, BSM physics does not produce SM-like effects, we would not see interference with the
SM-like operators. This scenario motivates fit model WET-2. The remaining models WET-3 and WET-4 arise from BSM physics
that induces effects with the opposite chirality.

\subsubsection{Hadronic nuisance parameters and their priors}
\label{sec:pheno:setup:nuisances}

\begin{table}
\renewcommand{\arraystretch}{1.1}
\setlength{\tabcolsep}{15pt}
\centering
\begin{tabular}{@{}ccc@{}}
    \toprule
    Parameter & Value $\pm$ uncertainty & Comments    \\
    \midrule
    $f^{B_s \to D_s}_0(m_\pi^2)$
    & $0.669 \pm 0.011$
    & Gaussian
    \\
    $f^{B \to D}_0(m_K^2)$
    & $0.675 \pm 0.011$
    & Gaussian
    \\
    $A^{B_s \to D_s^*}_0(m_\pi^2)$
    & $0.688 \pm 0.056$
    & Gaussian
    \\
    $A^{B \to D^*}_0(m_K^2)$
    & $0.704 \pm 0.035$
    & Gaussian
    \\
   \bottomrule
\end{tabular}
\caption{%
    Summary of the 1D marginal priors for the hadronic nuisance parameters that are varied in our analysis.
    For the full covariance matrix we refer to the ancillary material~\cite{EOS-DATA-2024-03}.
}
\label{tab:pheno:setup:nuisances}
\end{table}

For this proof-of-concept of a Bayesian analysis of the available data on \BqToDqL decays,
we aim to include only the major sources of theoretical uncertainties. In our case,
these are the hadronic $B_{q} \to D_{q}^{(*)}$ form factors with $q=d,s$. Their central values and uncertainties are collected in \cref{tab:pheno:setup:nuisances}.
Further theoretical uncertainties
that emerge from the hadronic matrix elements of the light meson (decay constants and LCDA parameters) are not accounted for.
We find that these uncertainties are small compared to those of the hadronic form factors
and compared to the effects incurred by the approximation to the experimental likelihood
as discussed in \cref{sec:pheno:setup:exp}.
We also fix all the other input parameters such as masses, couplings, CKM elements and $B$-meson lifetimes to their central values, which are collected in~\cref{tab:pheno:setup:pars}.

\begin{table}
\renewcommand{\arraystretch}{1.1}
\setlength{\tabcolsep}{30pt}
\centering
\resizebox{!}{.5\textheight}{%
\begin{tabular}{@{}lccc@{}}
    \toprule
    Parameter & Value & Units  & Source    \\
    \midrule
    \multicolumn{4}{c}{Strong coupling and quark masses}\\
    \midrule
    $\alpha_s(M_z)$            & $0.1180$  & \textendash &  \cite{ParticleDataGroup:2024cfk}\\
    $M_Z$                      & $91.1880$ & GeV         &  \cite{ParticleDataGroup:2024cfk} \\
    $M_W$                      & $80.3692$ & GeV         &  \cite{ParticleDataGroup:2024cfk} \\
    $m_b(m_b)$     & $4.18$    & GeV         &  \cite{ParticleDataGroup:2024cfk} \\
    $m_c(m_c)$     & $1.273$   & GeV         &  \cite{ParticleDataGroup:2024cfk} \\
    $m_s(2\,\text{GeV})$ & $0.0935$  & GeV         &  \cite{ParticleDataGroup:2024cfk} \\
    $m_d(2\,\text{GeV})$ & $0.0047$  & GeV         &  \cite{ParticleDataGroup:2024cfk} \\
    $m_u(2\,\text{GeV})$ & $0.00216$ & GeV         &  \cite{ParticleDataGroup:2024cfk} \\
    \midrule
    \multicolumn{4}{c}{Hadronic parameters of the $\pi$ and $K$ mesons}\\
    \midrule
    $m_\pi$                        & $0.13957039$ & GeV       & \cite{ParticleDataGroup:2024cfk} \\
    $f_\pi$                        & $0.1302$   & GeV         & \cite{FlavourLatticeAveragingGroupFLAG:2021npn,RBC:2014ntl,Follana:2007uv,MILC:2010hzw} \\
    $\alpha_2^\pi(1\,\text{GeV})$  & $0.157$    & \textendash & \cite{RQCD:2019osh}${}^\dagger$ \\
    $\alpha_4^\pi(1\,\text{GeV})$  & $0.06$     & \textendash & \cite{LatticeParton:2022zqc}${}^\dagger$ \\
    $f_{3\pi}(1\,\text{GeV})$      & $0.0045$   & GeV$^2$     & \cite{Ball:2006wn} \\
    $\omega_{3\pi}(1\,\text{GeV})$ & $-1.5$     & \textendash & \cite{Ball:2006wn} \\
    $m_K$                          & $0.493677$ & GeV         &  \cite{ParticleDataGroup:2024cfk} \\
    $f_K$                          & $0.1557$   & GeV         & \cite{FlavourLatticeAveragingGroupFLAG:2021npn,FermilabLattice:2014tsy,Dowdall:2013rya,Carrasco:2014poa} \\
    $\alpha_1^{K^-}(1\,\text{GeV})$    & $-0.06336$ & \textendash & \cite{RQCD:2019osh}${}^\dagger$ \\
    $\alpha_2^{K^-}(1\,\text{GeV})$    & $0.142191$ & \textendash & \cite{RQCD:2019osh}${}^\dagger$ \\
    $\alpha_3^{K^-}(1\,\text{GeV})$    & $-0.06219$ & \textendash & \cite{LatticeParton:2022zqc}${}^\dagger$ \\
    $\alpha_4^{K^-}(1\,\text{GeV})$    & $0.11196$  & \textendash & \cite{LatticeParton:2022zqc}${}^\dagger$ \\
    $f_{3 K}(1\,\text{GeV})$       & $0.0045$   & GeV$^2$     & \cite{Ball:2006wn} \\
    $\omega_{3 K}(1\,\text{GeV})$  & $-1.2$     & \textendash & \cite{Ball:2006wn} \\ 
    \midrule
    \multicolumn{4}{c}{Masses and lifetimes of the $B_{(s)}$ and $D_{(s)}^{(*)}$ mesons}\\
    \midrule
    $m_B$                          & $5.27972$ & GeV        & \cite{ParticleDataGroup:2024cfk} \\
    $m_{B_s}$                      & $5.36693$ & GeV        & \cite{ParticleDataGroup:2024cfk} \\
    $m_D$                          & $1.86966$ & GeV        & \cite{ParticleDataGroup:2024cfk} \\
    $m_{D^*}$                      & $2.01026$ & GeV        & \cite{ParticleDataGroup:2024cfk} \\
    $m_{D_s}$                      & $1.96835$ & GeV        & \cite{ParticleDataGroup:2024cfk} \\
    $m_{D_s^*}$                    & $2.1122$  & GeV        & \cite{ParticleDataGroup:2024cfk} \\
    $\tau_{B}$                     & $1.517$   & ps         & \cite{ParticleDataGroup:2024cfk} \\
    $\tau_{B_s}$                   & $1.520$   & ps         & \cite{ParticleDataGroup:2024cfk}\\
    \midrule
    \multicolumn{4}{c}{CKM matrix elements}\\
    \midrule
    $|V_{cb}|$                       & $0.0411$  & \textendash & \cite{ParticleDataGroup:2024cfk} \\ 
    $|V_{ud}|$                       & $0.97367$ & \textendash & \cite{ParticleDataGroup:2024cfk} \\ 
    $|V_{us}|$                       & $0.22431$ & \textendash & \cite{ParticleDataGroup:2024cfk} \\ 
   \bottomrule
\end{tabular}
}
\renewcommand{\arraystretch}{1.0}
\caption{
Numerical values for the parameters that are kept fixed in the analysis.
${}^\dagger$The value has been evolved down from $2~\text{GeV}$ to $1~\text{GeV}$ using leading-log running.
}
\label{tab:pheno:setup:pars}
\end{table}

Ref.~\cite{Bordone:2019guc} provides correlated parameters for the full set of ten hadronic form factors
in $\bar{B}_{(q)}\to D_{(q)}^{(*)}$ transitions in the heavy-quark expansion to order $1/m_c^2$.
This information enables us to evaluate the form factors at arbitrary momentum transfer $q^2$.
However, for our analysis we only require four of these ten form factors. Moreover, we only
need to evaluate them at two distinct values of the momentum transfer; the required
quantities are $f_0^{B_s\to D_s}(q^2 = m_\pi^2)$, $A_0^{B_s\to D_s^*}(q^2 = m_\pi^2)$,
$f_0^{B\to D}(q^2 = m_K^2)$ and $A_0^{B\to D^*}(q^2 = m_K^2)$.
Hence, we produce posterior predictions for these four quantities. Their distribution is a
multivariate Gaussian distribution, which is available in \EOS under the qualified name
\begin{center}
\verb|B_(s)->D_(s)^(*)::FormFactors[f_0(Mpi2),f_0(MK2),A_0(Mpi2),A_0(MK2)]@BGJvD:2019A|\,.
\end{center}
We use this multivariate distribution as a multivariate prior
PDF in our analysis.

\subsection{Methods and results}
\label{sec:pheno:results}

\begin{table}[t]
\renewcommand{\arraystretch}{1.2}
\setlength{\tabcolsep}{25pt}
\centering
\begin{tabular}{@{}cccc@{}}
    \toprule
    Fit model $M$ & Labelled mode & $\chi^2$  & $\log P(D, M)$   \\
    \midrule
    SM            & ---           & $26.69$   &  $9.04 \pm 0.03$ \\
    \midrule
    SM+PC         & ---           & $25.18$   & $ 9.71 \pm 0.10$ \\
    SM+PC'        & ---           &  $1.58$   & $29.12 \pm 0.04$ \\
    \midrule
    WET-1         & A             &  $1.61$   & $21.75 \pm 0.03$ \\
    WET-1         & B             &  $1.67$   & $21.77 \pm 0.03$ \\
    \midrule
    WET-2         & A             &  $1.57$   & $18.40 \pm 0.03$ \\
    WET-2         & B             &  $1.34$   & $16.35 \pm 0.03$ \\
    \midrule
    WET-3         & A             &  $1.62$   & $21.60 \pm 0.03$ \\
    WET-3         & B             &  $1.32$   & $21.50 \pm 0.03$ \\
    \midrule
    WET-4         & A             &  $1.54$   & $18.48 \pm 0.03$ \\
    WET-4         & B             &  $1.69$   & $16.62 \pm 0.03$ \\

   \bottomrule
\end{tabular}
\renewcommand{\arraystretch}{1.0}
\caption{%
    Goodness-of-fit values for the fits conducted as part of this analysis.
    We provide $\chi^2 = -2 \log P(\text{experimental data}\,|\,\vecx^*)$ at the posterior's labelled modes
    $\vecx^*$ and the natural logarithm of the localized evidence $\log P(D, M)$, \ie, the evidence
    obtained by integrating the posterior only in the vicinity of the local mode.
    Providing a $p$-value is not useful in our analyses, since most of the fit models have
    at least as many parameters as they have observations. We remark that neither the SM fit model
    nor the SM+PC fit model
    provide a good fit, since both $\chi^2 / \text{d.o.f.} = 26.69 / 4$ and $\chi^2 / \text{d.o.f.} = 25.18 / 2$ correspond
    to $p$-values well below our a-priori threshold of $3\%$.
    The SM+PC' fit model yields a $p$-value of $45\%$, albeit with power corrections that exceed the theory expectations (see the definition of SM+PC) by two orders of magnitude.
}
\label{tab:pheno:results:gof}
\end{table}

We pursue three objectives with our analysis, which can be summarized
by the following questions:
\begin{enumerate}
    \item[(a)] Can the available data on exclusive \BqToDqL processes be described jointly in the SM?

    \item[(b)] Is a BSM/WET interpretation of the data favoured or disfavoured with respect to the SM hypothesis?

    \item[(c)] How strongly does the available data restrict the parameter space of the $dbcu$ and $sbcu$ sectors of the WET?
\end{enumerate}
To achieve objective (a), we maximize the posterior PDFs $P(\vecx \,|\, D, \text{SM})$.
We observe that, for the models \text{WET-1} through $\text{WET-4}$, the same exercise
yields at least two well-separated modes of the posterior PDF. By mode, we refer to a distinct local
maximum of the PDF. By well-separated, we refer to the fact that any two modes are separated by a region of parameter space that
is so significantly disfavoured that the parameter space effectively splits into two disconnected regions.
We compile our results for the global $\chi^2$ values for the various modes of the posterior
distribution in \cref{tab:pheno:results:gof}. The $\chi^2$ values are obtained in the best-fit points $\vecx^*$.
\\
To achieve objective (b), we sample from the five posterior PDFs and calculate the (local) evidences
$P(D \,|\, M_i)_m$.
Here local evidence refers to integrating the posterior PDF by restricting the integration support to a region that is only connected
to a single mode.
We compile the local evidences in \cref{tab:pheno:results:gof}. Calculating the marginal posteriors enables us
to carry out a Bayesian model comparison as discussed in \cref{sec:pheno:setup}.
\\
To achieve objective (c), we investigate the marginal posteriors for the WET parameters.
The marginal posteriors are discussed in detail in \cref{sec:pheno:results:WET}.

Our analysis uses the \EOS software~\cite{EOSAuthors:2021xpv} in version 1.0.13~\cite{EOS:v1.0.13},
which includes all of the analytical results presented in \cref{sec:class-I,sec:lifetime}.

\subsubsection{Null hypothesis}
\label{sec:pheno:results:SM}

\begin{figure}[t]
    \centering
    \includegraphics[width=0.8\textwidth]{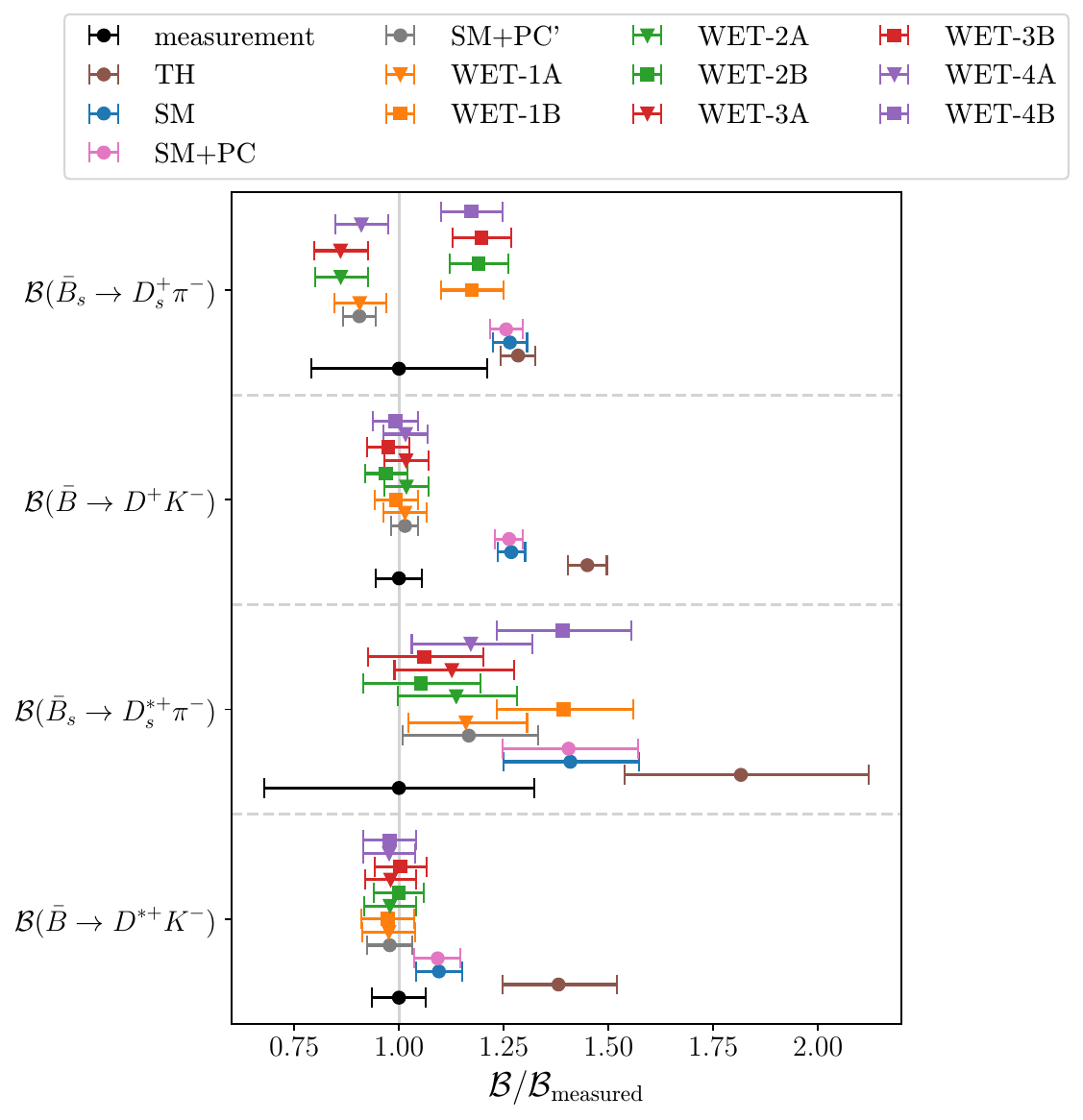}
    \caption{
        Comparison of the branching ratios of the nonleptonic decays in units of the experimentally
        measured branching ratios.
        We show the theory-based prior prediction (TH) and the various posterior postdictions
        for the fit models SM, SM+PC$^{(\prime)}$ and WET-1 to WET-4. In the BSM-like models, the two modes A and B
        are shown separately.
        The tension between the prior predictions and the SM postdiction on the one hand
        and the measurements on the other is clearly visible.
        All intervals shown correspond to the central intervals at $68\%$ probability.
    }
    \label{fig:pheno:results:BRs}
\end{figure}

As anticipated earlier and in line with the results in the literature~\cite{Bordone:2020gao,Cai:2021mlt,Beneke:2021jhp,Endo:2021ifc,Lenz:2022pgw,Panuluh:2024poh}, we find that
the SM fit model does not describe the data well:
we obtain $\chi^2 = 26.69$ when evaluating the experimental likelihood.
Given four degrees of freedom, this corresponds to a $p$-value of $2\cdot 10^{-5}$.
The fit prefers to lower the values of the form factors compared to the prior's central value.
This deviation from the (joint) prior distribution reaches a significance of $3.4\,\sigma$.

Fitting the hadronic nuisance parameters to the experimental data lessens the
tensions with the branching ratios somewhat as can be observed in \cref{fig:pheno:results:BRs}:
all SM postdictions are closer to the measurement than the original theory prediction (marked as TH). The lifetime bound is fulfilled by this fit scenario by construction.

Fitting additionally for potential power corrections as discussed in the definition of the SM+PC fit model
does not alleviate the tension with the experimental data. Accounting for the power corrections using two
flavour-universal parameters (within the estimates produced in Ref.~\cite{Bordone:2020gao}) reduces the $\chi^2$
by less than $2$, yielding a $p$-value of $3\cdot 10^{-6}$.
As expected, loosening the prior for the size of the power corrections beyond the aforementioned estimate
yields an excellent fit with a $p$-value of $45 \%$ and a Bayes factor decisively in favour of the SM fit model.
The preferred size of the power corrections is in the negative $20\%$ range universally for $\delta_P$ and $\delta_V$, exceeding the estimates by roughly two orders of magnitude.

\subsubsection{BSM effects}
\label{sec:pheno:results:WET}

\begin{figure}[t]
    \centering
    \includegraphics[width=0.85\textwidth]{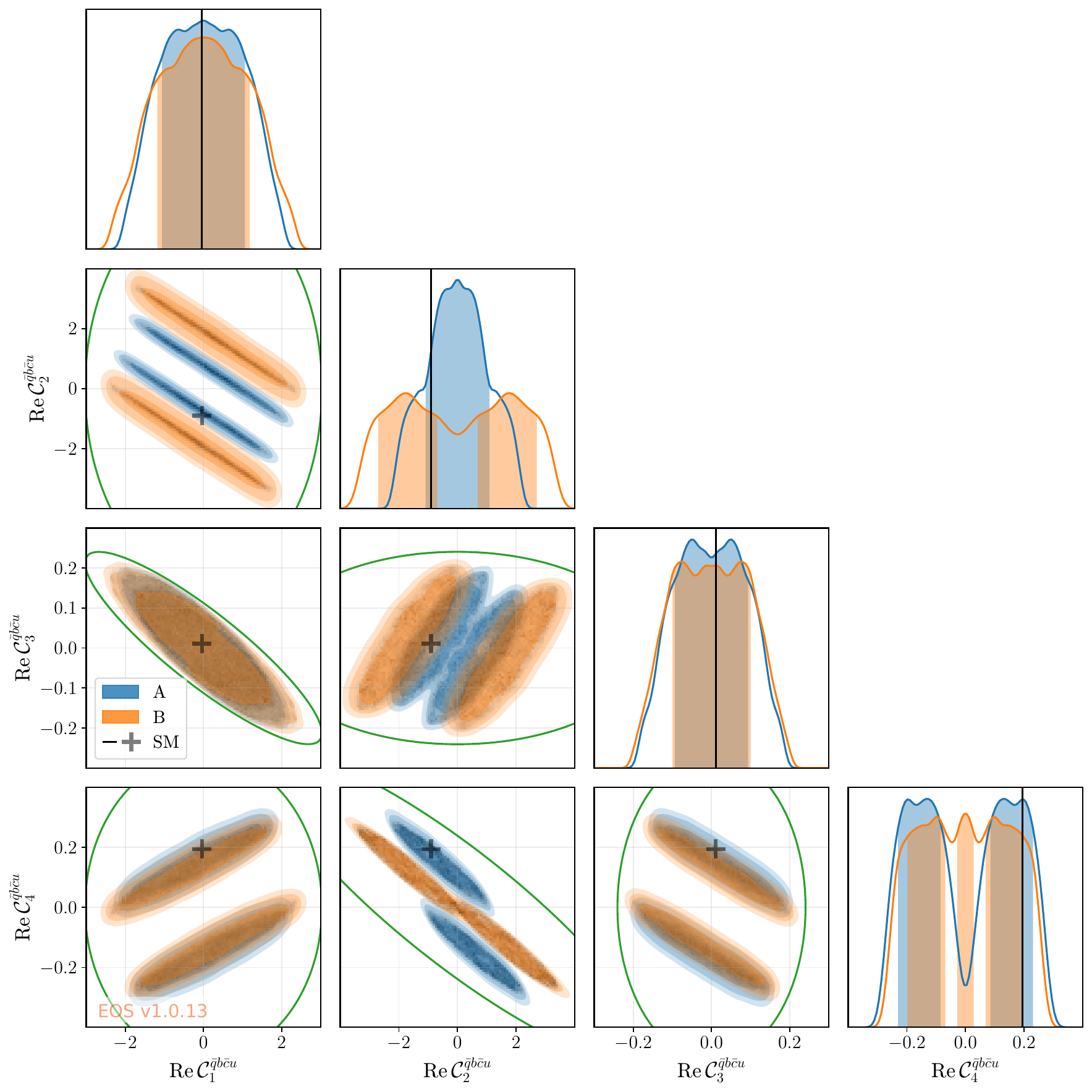}
    \caption{%
        One-dimensional and two-dimensional marginal posterior PDFs for the four parameters of interest of the WET-1 model.
        The ``+'' and the solid black lines show the SM point \cref{eq:preliminaries:matching-conditions-bern}.
        The coloured areas are the central 68\%, 95\%, and 99\% integrated probability contours of the posterior distribution obtained from a kernel density estimation. Blue areas belong to mode A, while orange areas belong to mode B.
        The boundaries due to the lifetime constraint are illustrated as green ellipses.
    }
    \label{fig:WET1}
\end{figure}
\begin{figure}[t]
    \centering
    \includegraphics[width=0.93\textwidth]{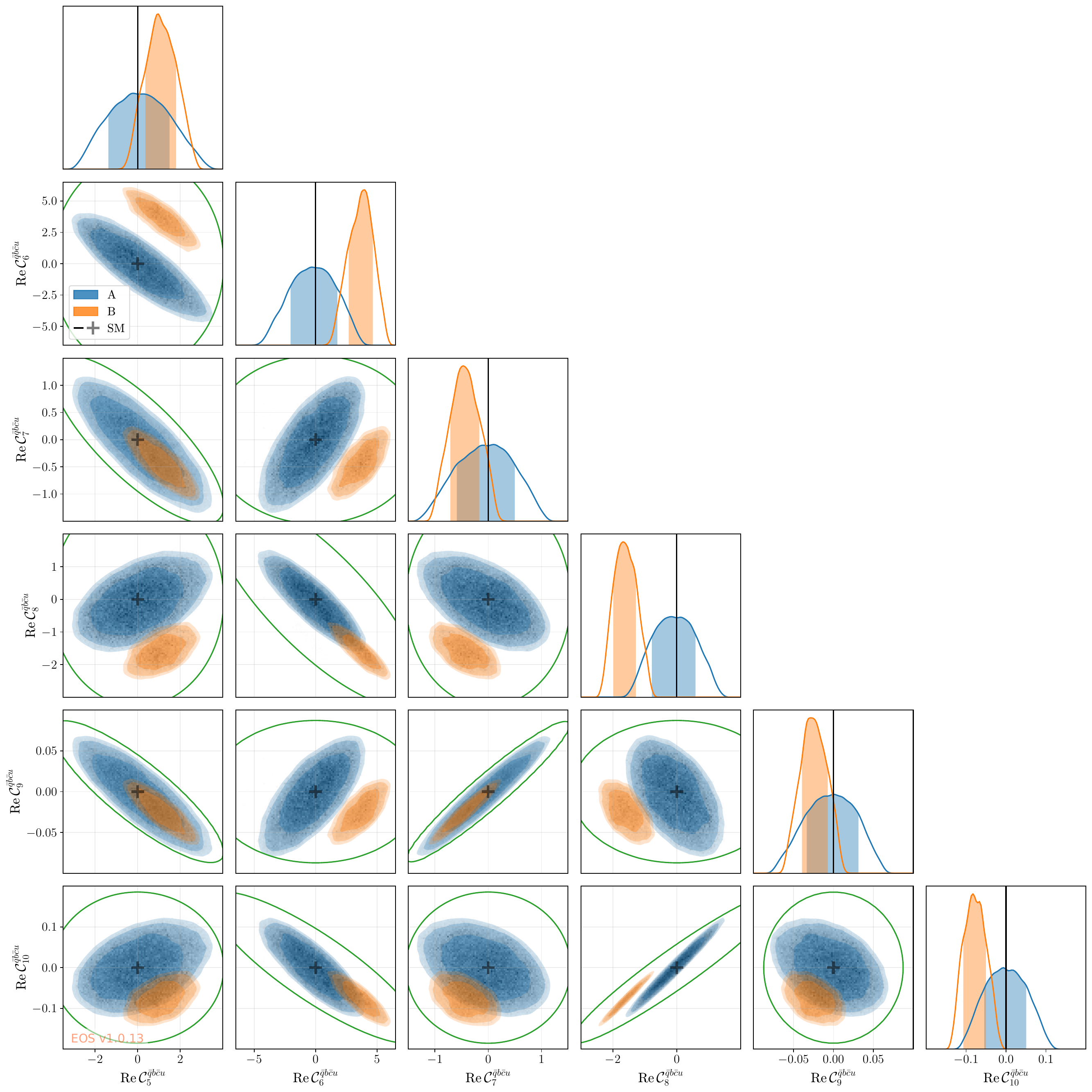}
    \caption{One-dimensional and two-dimensional marginal posterior PDFs for the 6 parameters of interest of the WET-2 model.
    For an explanation of the individual elements, see the caption of \cref{fig:WET1}.}
    \label{fig:WET2}
\end{figure}

The posterior samples for two of the four BSM scenarios are summarized in the corner plots in \cref{fig:WET1,fig:WET2}. Similar corner plots for models WET-3 and WET-4 can be found in the
ancillary material~\cite{EOS-DATA-2024-03}.
The likelihood for the WET-1 model exhibits a discrete symmetry $\wcbern[qbcu]{i} \to -\wcbern[qbcu]{i}$.
This symmetry is broken in models WET-2, WET-3 and WET-4 due to the non-zero
SM contributions to the Wilson coefficients $\wcbern[qbcu]{1\dots 4}$.
Although we confirm this symmetry numerically in our code, our sampling results are
only approximately symmetric. Hence, we symmetrize our results for the WET-1 scenario a posteriori.\\

For each of the BSM fit models, we find two modes with local evidence of at least $2 \%$ of the total evidence.
Details on how we separate the modes in each fit model are discussed in \cref{app:modes}.
In the case of WET-1, these two modes are of course present in duplicate due to the aforementioned symmetry.
We consistently label these modes as ``mode A'' and ``mode B'', based on their relative distance to the
SM point: mode A is always chosen as that mode that is closer to the SM parameter point.
While it might look as if mode A \emph{includes} the SM point in the various projections
shown in \cref{fig:WET1,fig:WET2} and the ancillary figures, we emphasize that this is not
the case. Instead, we find that the samples form a hollow shell around the SM point, which is
excluded at high significance.
In contrast to the SM fit, the BSM fits do not pull the form factors away from the \emph{a-priori} central value.
For all eight modes, we find that they are in perfect agreement with the prior expectation.
From the evidences given in \cref{tab:pheno:results:gof}, we conclude that all BSM models are decisively favoured over the null hypothesis;
their Bayes factors range from $K(\text{WET-2B}, \text{SM}) = 1.5 \cdot 10^3$ to $K(\text{WET-1B}, \text{SM}) = 3.4 \cdot 10^5$.\\

Both of the four-operator models WET-1 and WET-3 feature modes with almost identical evidences ($\log P(D, \text{WET-1})_\text{A} \approx \log P(D, \text{WET-1})_\text{B} \approx \log P(D, \text{WET-3})_\text{A} \approx \log P(D, \text{WET-3})_\text{B} \approx 22$).
Hence, we can neither distinguish between the modes, nor do we see a preference for either of the four-operator models.
The WET-1A and WET-3A modes are close to the SM point and thus would correspond to a relatively small change in the Wilson coefficients. The WET-1B and WET-3B modes on the other hand give room for large modifications of the Wilson coefficients.

The six-operator models  WET-2 and WET-4 each feature a mode A that is substantially preferred over mode B. 
Out of all WET-2 and WET-4 modes, mode WET-4A has the largest evidence. Nevertheless, it --- and therefore 
each of the six-operator fit models --- is strongly disfavoured with respect to the four-operator models,
as evidenced by their Bayes factors:
\begin{align}
    K(\text{WET-1A, WET-4A}) &= 26.31 \, , \quad K(\text{WET-1B, WET-4A}) = 26.84 \, , \notag \\
    K(\text{WET-3A, WET-4A}) &= 22.65 \, , \quad K(\text{WET-3B, WET-4A}) = 20.49 \, .
\end{align}

In order to see the effect of the lifetime constraint,
we overlay the two-dimensional marginalized posterior distributions of the Wilson coefficients
with the two-dimensional projections of the ellipsoid defined by the penalty function  \cref{eq:pheno:penalty-function}. 
We see that for many combinations of Wilson coefficients the lifetime constraint does not have a significant impact (\eg for the $\wcbern[qbcu]{1}$--$\wcbern[qbcu]{4}$ plane in \cref{fig:WET1}).
However, for some combinations
(\eg for the $\wcbern[qbcu]{1}$--$\wcbern[qbcu]{3}$ plane in \cref{fig:WET1}) the lifetime ellipses are almost completely filled.
This indicates that the lifetime gives the dominant constraint on this combination of Wilson coefficients, thereby identifying directions in the parameter space that are presently
either unconstrained or only very mildly constrained by the exclusive decay modes.

\subsection{Outlook}
\label{sec:pheno:outlook}

The analysis presented in this article is of explorative nature and could be improved in a number of ways. One chief improvement would be finding a way to separate between the two types of modes
found in our analysis. Another improvement would be to merge the four-operator and six-operator
scenarios into a ten-operator scenario.
Both types of improvements require data providing further and complementary constraints on the WET coefficients.

Examples for this kind of data include the lifetime ratio $\tau(B^+)/\tau(B^0)$ and the semi-leptonic CP asymmetry $a_{sl}^{d}$. These observables have been recently studied in Ref.~\cite{Lenz:2022pgw} in the context of one-operator and two-operator scenarios,
improving the bounds previously obtained from the exclusive nonleptonic
decays in Ref.~\cite{Cai:2021mlt}.
Other examples include determinations of the inclusive decay widths $\bar{B}_s\to X_{c \bar{s}}$
and $\bar{B} \to X_{c s}$, 
where the final state has total charm $C=+1$ and total strangeness of either $S=+1$ or $S=-1$,
respectively.
Even upper bounds on these inclusive decay widths would help to sharpen the effect
of the penalty function \cref{eq:pheno:penalty-function}.
Furthermore, a more precise measurement of $\mathcal{B}(\bar{B}_s \to D_s^+ \pi^-)$ would help to distinguish between the A and B modes of the different fit models. This is because all posterior predictions for this branching ratio undershoot the experimental central value for the A modes while they overshoot it for the B modes as can be seen in \cref{fig:pheno:results:BRs}. A more precise measurement would discard either all the A modes or all the B modes.

\section{Conclusions}
\label{sec:conclusion}

In this article we have investigated possible BSM interpretations of the puzzle related to the nonleptonic \BqToDqL decays within the Weak Effective Theory (WET). To this end, we have recalculated the two-particle contributions to the hard-scattering kernels up to NLO in QCD for the full set of WET operators up to mass-dimension six. We find agreement with the literature. Additionally, we have calculated the three-particle contributions for the full set of WET operators at LO in QCD. Our results refine the previously published SM result and present for the first time the BSM contributions. 

Based on our analytical results, we have performed a model-independent Bayesian analysis of the available data for our decays of interest. 
Aside from the experimental data on the relevant exclusive branching ratios, we have included a global bound from the $B$-meson lifetime.
To that end, we have calculated the nonleptonic lifetime at LO in QCD and at leading power in the heavy quark expansion for the full set of WET operators in the $qbcu$ sector for the first time.

Our phenomenological analysis begins by fitting only hadronic nuisance parameters and fixing the Wilson coefficients to their SM values. As expected, this fit model cannot adequately describe the experimental data.
Fitting possible power corrections to the data, while constrained by estimates for their size, can only provide a very slight reduction of the tensions. They can only be fully removed by allowing for power corrections that are larger than previous estimates by two orders of magnitude.
Consequently, we tested four different BSM models, two of them simultaneously varying four Wilson coefficients (WET-1 and WET-3) and two of them varying six Wilson coefficients (WET-2 and WET-4).
We find that all BSM models feature two distinct modes, one of them close to the SM point (mode A) and one of them farther away (mode B). All eight BSM modes are decisively favoured over the null hypothesis (the SM hypothesis), but not over the unconstrained power-correction hypothesis (SM+PC'). 
Furthermore, all the modes of the models containing four Wilson coefficients have almost identical evidences and therefore cannot be distinguished. 
However, they are strongly favoured over all four modes featured in the six Wilson coefficient fit models. 
This suggests that among the possible BSM solutions, WET-1 and WET-3 are the most likely. These fit models cover scenarios
with (WET-1) and without (WET-3) mixing between the SM and BSM contributions. 

We point out a number of ways in which this four-fold degeneracy of solutions can be resolved. This includes using additional data on the lifetime ratio $\tau(B^+)/\tau(B^0)$, the semi-leptonic CP asymmetry~$a^d_{sl}$, more precise measurements of $\mathcal{B}(\bar{B}_s \to D_s^+ \pi^-)$, and dedicated inclusive measurements.

\acknowledgments

We thank Xin-Qiang Li for clarifying and useful communication on the choice of evanescent
operators in Ref.~\cite{Cai:2021mlt}.
We thank Alexander Lenz, Maria Laura Piscopo, and Aleksey Rusov for helpful communication on the nonleptonic $B$-meson decay width, and Maria Laura Piscopo for a careful reading of our manuscript.
DvD is grateful to the Mainz Institute for Theoretical Physics (MITP) of the Cluster of Excellence PRISMA$^+$ (Project ID 390831469), for its hospitality and its partial support during the completion of this work.
SM is supported by the Studienstiftung des deutschen Volkes.
DvD~acknowledges support by the UK Science and Technology Facilities Council (grant numbers ST/V003941/1 and ST/X003167/1).
J.V. acknowledges funding from grant 2021-SGR-249 (Generalitat de Catalunya), and from the Spanish
MCIN/AEI/10.13039/501100011033 through the following grants: 
grant CNS2022-135262 funded by the “European Union NextGenerationEU/PRTR”,
grant CEX2019-000918-M through the “Unit of Excellence Mar\'ia de Maeztu 2020-2023” award to the Institute of Cosmos Sciences,
and grant PID2022-136224NB-C21.

\appendix

\section{Details on the operator bases}

\subsection{Evanescent operators}
\label{app:bases:evanescent}

For the one-loop calculations done in this article, a number of evanescent operators are needed. In order to fix these evanescent operators, we follow the standard method of ``Greek projections'' \cite{Tracas:1982gp,Buras:2000if} in the form presented in Ref.~\cite{Dekens:2019ept}. 
For the most part this results in the basis for evanescent operators for $\Delta F = 1$ transitions given in Ref.~\cite{Buras:2000if}. 
However, in the Bern basis, the one-loop corrections to the current-current diagrams generate Dirac structures that cannot be reduced to physical and evanescent operators with the operators given in Ref.~\cite{Buras:2000if}. This is because there are operators with up to four $\gamma$~matrices in each of the currents, which when introduced into one-loop diagrams produce structures with up to six $\gamma$~matrices per current. Using Greek projections, we find the following additional evanescent operators
\begin{align}
    E_{\text{Bern}}^\text{VLL,C} =& [\bar{q}_\alpha \gamma_{\mu \nu \rho \sigma \eta } P_L  b_\beta] [\bar{c}_\beta \gamma^{\mu \nu \rho \sigma \eta } P_L u_\alpha] - 32 \, (8 - 7 \epsilon) \, [\bar{q}_\alpha \gamma_{\mu} P_L  b_\beta] [\bar{c}_\beta \gamma^{\mu } P_L u_\alpha] \, ,
    \\[1mm]
    E_{\text{Bern}}^\text{VLL,S} =& [\bar{q}_\alpha \gamma_{\mu \nu \rho \sigma \eta } P_L  b_\alpha] [\bar{c}_\beta \gamma^{\mu \nu \rho \sigma \eta } P_L u_\beta]- 32 \, (8 - 7 \epsilon) \, [\bar{q}_\alpha \gamma_{\mu} P_L  b_\alpha] [\bar{c}_\beta \gamma^{\mu } P_L u_\beta] \, , 
    \\[1mm]
    E_{\text{Bern}}^\text{VLR,C} =& [\bar{q}_\alpha \gamma_{\mu \nu \rho \sigma \eta } P_L  b_\beta] [\bar{c}_\beta \gamma^{\mu \nu \rho \sigma \eta } P_R u_\alpha] - 16 \, (1 + 8 \epsilon ) \, [\bar{q}_\alpha \gamma_{\mu} P_L  b_\beta] [\bar{c}_\beta \gamma^{\mu } P_R u_\alpha]\, , 
    \\[1mm]
    E_{\text{Bern}}^\text{VLR,S} =& [\bar{q}_\alpha \gamma_{\mu \nu \rho \sigma \eta } P_L  b_\alpha] [\bar{c}_\beta \gamma^{\mu \nu \rho \sigma \eta } P_R u_\beta] - 16 \, (1 + 8 \epsilon ) \, [\bar{q}_\alpha \gamma_{\mu} P_L  b_\alpha] [\bar{c}_\beta \gamma^{\mu } P_R u_\beta] \, , 
    \\[1mm]
    E_{\text{Bern}}^\text{SLR,C} =& [\bar{q}_\alpha \gamma_{\mu \nu \rho \sigma \eta \tau} P_L  b_\beta] [\bar{c}_\beta \gamma^{\mu \nu \rho \sigma \eta \tau} P_R u_\alpha] - 64 \, (1 + 45 \epsilon) \, [\bar{q}_\alpha P_L  b_\beta] [\bar{c}_\beta P_R u_\alpha] \, , 
    \\[1mm]
    E_{\text{Bern}}^\text{SLR,S} =& [\bar{q}_\alpha \gamma_{\mu \nu \rho \sigma \eta \tau} P_L  b_\alpha] [\bar{c}_\beta \gamma^{\mu \nu \rho \sigma \eta \tau} P_R u_\beta] - 64 \, (1 + 45 \epsilon) \, [\bar{q}_\alpha P_L  b_\alpha] [\bar{c}_\beta P_R u_\beta] \, , 
    \\[1mm]
    E_{\text{Bern}}^\text{SLR2,C} =& [\bar{q}_\alpha \gamma_{\mu \nu \rho \sigma} P_L  b_\beta] [\bar{c}_\beta \gamma^{\mu \nu \rho \sigma} P_R u_\alpha] - 16 \, (1 + 8 \epsilon) \, [\bar{q}_\alpha P_L  b_\beta] [\bar{c}_\beta P_R u_\alpha] \, , 
    \\[1mm]
    E_{\text{Bern}}^\text{SLR2,S} =& [\bar{q}_\alpha \gamma_{\mu \nu \rho \sigma} P_L  b_\alpha] [\bar{c}_\beta \gamma^{\mu \nu \rho \sigma} P_R u_\beta] - 16 \, (1 + 8 \epsilon) \, [\bar{q}_\alpha P_L  b_\alpha] [\bar{c}_\beta P_R u_\beta] \, ,
    \\[1mm]
    E_{\text{Bern}}^\text{SLL,C} =& [\bar{q}_\alpha \gamma_{\mu \nu \rho \sigma \eta \tau} P_L  b_\beta] [\bar{c}_\beta \gamma^{\mu \nu \rho \sigma \eta \tau} P_L u_\alpha] + 128 \, (-8 + 23 \epsilon) \, [\bar{q}_\alpha P_L  b_\beta] [\bar{c}_\beta P_L u_\alpha] \notag \\
    &+ 32 \, (8 - 11 \epsilon) \, [\bar{q}_\alpha \sigma_{\mu \nu} P_L  b_\beta] [\bar{c}_\beta \sigma^{\mu \nu} P_L u_\alpha] \, ,
    \\[1mm]
    E_{\text{Bern}}^\text{SLL,S} =& [\bar{q}_\alpha \gamma_{\mu \nu \rho \sigma \eta \tau} P_L  b_\alpha] [\bar{c}_\beta \gamma^{\mu \nu \rho \sigma \eta \tau} P_L u_\beta] + 128 \, (-8 + 23 \epsilon) \, [\bar{q}_\alpha P_L  b_\alpha] [\bar{c}_\beta P_L u_\beta] \notag \\
    &+ 32 \, (8 - 11 \epsilon) \, [\bar{q}_\alpha \sigma_{\mu \nu} P_L  b_\alpha] [\bar{c}_\beta \sigma^{\mu \nu} P_L u_\beta] \, .
\end{align}
The operators of the parity flipped sectors can be obtained by exchanging $P_L \leftrightarrow P_R$. 
With these operators and the ones given in Ref.~\cite{Buras:2000if}, any structure that might appear in our one-loop calculations can be written in terms of physical and evanescent operators.

\subsection{Change of basis between BMU and Bern at NLO}
\label{app:bases:change-of-basis}

The rotation from an operator basis $Q_i$ to an operator basis $Q'_i$ in $d$ dimensions is given by
\begin{align}
    Q_i' = R_{ij} (Q_j + W_{jk} E_k) \ , 
\end{align}
where $E_k$ is the chosen set of evanescent operators needed for the change of basis, $\hat{R}$ is the tree-level transformation matrix (calculable in $d=4$), and $W_{jk}$ depends on the scheme (on the $E_k$).

One considers now matrix elements of the operators and their Wilson coefficients in a loop expansion up to order $\alpha_s^n$,
\begin{align}
    \braket{Q_i} = \sum_{\ell = 0}^n \left(\frac{\alpha_s}{4 \pi}\right)^\ell r_{ij}^{(\ell)} \braket{Q_j}^{\text{tree}} \, ,
    \qquad
    C_i = \sum_{\ell=0}^n \left(\frac{\alpha_s}{4 \pi}\right)^\ell C_i^{(\ell)} \, .
\end{align}
Here $r_{ij}^{(\ell)}$ is the $\ell$-loop amplitude matrix with $r_{ij}^{0} = \delta_{ij}$. 
The corresponding amplitude is given by
\begin{align}
    A = C_i^{(0)} \braket{Q_i}^{\text{tree}} + \frac{\alpha_s}{4 \pi} \left[C_i^{(1)} \braket{Q_i}^{\text{tree}} + C_i^{(0)} \hat{r}_{ij}^{(1)} \braket{Q_j}^{\text{tree}}\right] + \mathcal{O}(\alpha_s^2) \, .
\end{align}
This amplitude must be invariant under a basis transformation,
\begin{align}
    A = C_i \braket{Q_i} = C'_i \braket{Q'_i} = A' \label{eq:bases:change-of-basis} \, ,
\end{align}
which if imposed order by order in $\alpha_s$ leads to the following transformation rules for the Wilson coefficients and the anomalous dimensions,
\begin{align}
C_i'^{(0)} &= C_j^{(0)} R^{-1}_{ji} \, , 
&& \hat{\gamma}'^{(0)} = \hat{R} \hat{\gamma}^{(0)} \hat{R}^{-1} \, , 
    \\
C_i'^{(1)} &= C_j^{(1)} R^{-1}_{ji} + C_j^{(0)} R^{-1}_{jk} \Delta r_{ki} \, ,
&&  \hat{\gamma}'^{(1)} = \hat{R} \hat{\gamma}^{(1)} \hat{R}^{-1} - 2 \beta_0 \Delta \hat{r} - \left[\Delta \hat{r}, \hat{\gamma}'^{(0)}\right] \, .
\end{align}
Here the matrix $\Delta \hat{r}$ is defined as
\begin{align}
    \Delta r_{ij} = (\hat{R} \, \hat{r}^{(1)} \hat{R}^{-1})_{ij} -r'^{(1)}_{ij} \, .
\end{align}
These NLO transformation rules are well known~\cite{Buras:1991jm,Chetyrkin:1997gb,Gorbahn:2004my,Morell:2024aml,Aebischer:2025hsx}.

\bigskip

In order to apply these formulas to the change of basis at hand, it is necessary to compute the tree-level transformation matrix $\hat{R}$, and the one-loop amplitude matrix $\hat{r}^{(1)}$ in both bases. While $\hat{r}^{(1)}$ depends on the IR structure of the amplitude, the shift $\Delta \hat{r}$ does not, and thus we are free to choose conveniently the external states and an IR regulator. We choose all four external quark states massless and with zero momentum and regulate IR divergences by introducing a fictitious gluon mass~$\lambda$. 

We begin by grouping the operators in each basis into the following vectors,
\begin{align}
    \Vec{\mathcal{O}}_{\text{Bern}} =& \left\{\mathcal{O}^{qbcu}_1,\mathcal{O}^{qbcu}_2 ,\mathcal{O}^{qbcu}_3 ,\mathcal{O}^{qbcu}_4 ,\mathcal{O}^{qbcu}_5 ,\mathcal{O}^{qbcu}_6 ,\mathcal{O}^{qbcu}_7 ,\mathcal{O}^{qbcu}_8 ,\mathcal{O}^{qbcu}_9 ,\mathcal{O}^{qbcu}_{10}  \right\} \, , 
    \\[1mm]
    \Vec{\mathcal{O}}_{\text{Bern}}' =& \left\{\mathcal{O}^{qbcu}_{1'},\mathcal{O}^{qbcu}_{2'} ,\mathcal{O}^{qbcu}_{3'} ,\mathcal{O}^{qbcu}_{4'} ,\mathcal{O}^{qbcu}_{5'}, \mathcal{O}^{qbcu}_{6'}, \mathcal{O}^{qbcu}_{7'}, \mathcal{O}^{qbcu}_{8'} ,\mathcal{O}^{qbcu}_{9'} ,\mathcal{O}^{qbcu}_{10'}  \right\} \, , 
    \\[1mm]
    \Vec{\mathcal{Q}}_{\text{BMU}} =& \left\{ \mathcal{Q}_{1}^{VLL}, \mathcal{Q}_{2}^{VLL}, \mathcal{Q}_{1}^{SLR}, \mathcal{Q}_{2}^{SLR}, \mathcal{Q}_{1}^{VRL}, \mathcal{Q}_{2}^{VRL}, \mathcal{Q}_{1}^{SRR}, \mathcal{Q}_{2}^{SRR}, \mathcal{Q}_{3}^{SRR}, \mathcal{Q}_{4}^{SRR}  \right\} \, , 
    \\[1mm]
    \Vec{\mathcal{Q}}_{\text{BMU}}' =& \left\{ \mathcal{Q}_{1}^{VRR}, \mathcal{Q}_{2}^{VRR}, \mathcal{Q}_{1}^{SRL}, \mathcal{Q}_{2}^{SRL}, \mathcal{Q}_{1}^{VLR}, \mathcal{Q}_{2}^{VLR}, \mathcal{Q}_{1}^{SLL}, \mathcal{Q}_{2}^{SLL}, \mathcal{Q}_{3}^{SLL}, \mathcal{Q}_{4}^{SLL}  \right\} \, ,
\end{align}
such that the tree-level transformation matrix~$\hat{R}$ is block diagonal. Primed and unprimed sectors constitute identical independent blocks, and hence it suffices to write down the matrices for the unprimed operators.
The tree-level transformation matrix $\hat{R}^{\text{Bern} \to \text{BMU}}$ is defined as
\begin{align}
     \mathcal{Q}_i^{\text{BMU}}\big|_{d=4} &= R^{\text{Bern} \to \text{BMU}}_{ij} \mathcal{O}_j^{\text{Bern}}\big|_{d=4} \, .
\end{align}
It has the following block-diagonal structure
\begin{align} 
 \hat{R}^{\text{Bern} \to \text{BMU}} &= \left( 
    \begin{array}{cccc}
        \hat{R}^{\text{Bern} \to \text{BMU}}_{1-4}  & 0  \\
        0 & \hat{R}^{\text{Bern} \to \text{BMU}}_{5-10} \\
    \end{array}
    \right) \, ,
\end{align}
with the blocks
\begin{align}
\hat{R}^{\text{Bern} \to \text{BMU}}_{1-4} = \left(
\begin{array}{cccc}
 -\frac{1}{3} & 0            & \frac{1}{12} & 0            \\
 -\frac{1}{9} & -\frac{2}{3} & \frac{1}{36} & \frac{1}{6}  \\
 -\frac{2}{3} & 0            & \frac{1}{24} & 0            \\
 -\frac{2}{9} & -\frac{4}{3} & \frac{1}{72} & \frac{1}{12} \\
\end{array}
\right) \, , \quad
\hat{R}^{\text{Bern} \to \text{BMU}}_{5-10} = \left(
\begin{array}{cccccc}
-\frac{8}{3}  & 0              & \frac{2}{3}   & 0             & \frac{1}{24}   & 0             \\
-\frac{8}{9}  & -\frac{16}{3}  & \frac{2}{9}   & \frac{4}{3}   & \frac{1}{72}   & \frac{1}{12}  \\
\frac{1}{6}   & 0              & -\frac{7}{24} & 0             & -\frac{1}{96}  & 0             \\
\frac{1}{18}  & \frac{1}{3}    & -\frac{7}{72} & -\frac{7}{12} & -\frac{1}{288} & -\frac{1}{48} \\
2             & 0              & -\frac{3}{2}  & 0             & -\frac{1}{8}   & 0             \\
\frac{2}{3} & 4 & -\frac{1}{2} & -3 & -\frac{1}{24} \
& -\frac{1}{4} 
\\
\end{array}
\right) \, .
\end{align}
The one-loop amplitude matrices $\hat{r}$ are also block-diagonal,
\begin{align}
    \hat{r}_{\text{Bern}}^{(1)} = \left( 
    \begin{array}{cc}
        \hat{r}_{\text{Bern}}^{1-4} & 0   \\
        0  & \hat{r}_{\text{Bern}}^{5-10} \\
    \end{array}
    \right) \, , \quad
    \hat{r}_{\text{BMU}}^{(1)} = \left( 
    \begin{array}{cccccccc}
        \hat{r}_{\text{BMU}}^{1-2}  & 0 & 0 & 0 \\
        0 & \hat{r}_{\text{BMU}}^{3-4}  & 0 & 0 \\
        0 & 0 & \hat{r}_{\text{BMU}}^{5-6}  & 0 \\
        0 & 0 & 0 & \hat{r}_{\text{BMU}}^{7-10} \\
    \end{array}
    \right) \, ,
\end{align}
with the sub-blocks
\begin{align}
 \hat{r}_{\text{Bern}}^{1-4} =&
 \left(
\begin{array}{cccc}
 -\frac{4}{3} & 9 & 0 & -\frac{3}{2} \\
 2 & \frac{32}{3} & -\frac{1}{3} & -\frac{5}{8} \\
 -\frac{320}{3} & 64 & \frac{28}{3} & -13 \\
 \frac{128}{9} & 100 & -\frac{26}{9} & -\frac{35}{6} \\
\end{array}
\right)
+
\log \frac{\lambda}{\mu} 
\left(
\begin{array}{cccc}
 -\frac{16}{3}  & -20            & 0             & 2           \\
 -\frac{40}{9}  & -\frac{68}{3}  & \frac{4}{9}   & \frac{5}{6} \\
 0              & -128           & -\frac{16}{3} & 20          \\
 -\frac{256}{9} & -\frac{160}{3} & \frac{40}{9}  & -6
\end{array} 
\right)\, , \\
\notag \\
\hat{r}_\text{Bern}^{5-10} 
=&
\left (
\begin{array}{cccccc}
\frac{16}{3}    & 0             & 0             & \frac{3}{2}    & 0            & 0 \\
0               & \frac{7}{3}   & \frac{1}{3}   & \frac{5}{8}    & 0            & 0 \\
0               & -8            & 0             & 12             & 0            & \frac{3}{2} \\
-\frac{16}{9}   & -\frac{10}{3} & \frac{8}{3}   & 11             & \frac{1}{3}  & \frac{5}{8} \\
-\frac{2560}{3} & -256          & \frac{896}{3} & 64             & \frac{80}{3} & -6 \\
-\frac{512}{9}  & 480           & \frac{128}{9} & -\frac{536}{3} & -\frac{4}{3} & -\frac{89}{6} 
 \end{array}
 \right) 
 +
 \log \frac{\lambda}{\mu} 
\left(
\begin{array}{cccccc}
-\frac{64}{3} & 0 & 0 & -2 & 0 & 0 \\
0 & -\frac{10}{3} & -\frac{4}{9} & -\frac{5}{6} & 0 & 0 \\
0 & 32 & 0 & -32 & 0 & -2 \\
\frac{64}{9} & \frac{40}{3} & -\frac{64}{9} & -\frac{94}{3} & -\frac{4}{9} & -\frac{5}{6} \\
0 & -512 & -\frac{1024}{3} & 384 & -\frac{64}{3} & 32 \\
-\frac{1024}{9} & -\frac{640}{3} & \frac{256}{3} & \frac{1184}{3} & \frac{64}{9} & 10 \\
\end{array} 
\right ) \, ,
\end{align}
\begin{align}
\hat{r}_{\text{BMU}}^{1-2} =& 
\left(
\begin{array}{cc}
 \frac{1}{2} & -\frac{11}{2} \\
 -\frac{11}{2} & \frac{1}{2} 
\end{array}
\right) 
+
\log \frac{\lambda}{\mu} \left(
\begin{array}{cc}
 -\frac{22}{3} & 6 \\
 6 & -\frac{22}{3} \\
\end{array}
\right) \, , \quad 
\hat{r}_{\text{BMU}}^{3-4} =
\left(
\begin{array}{cc}
 -\frac{7}{6} & -\frac{1}{2} \\
 -3           & \frac{19}{3} \\
\end{array}
\right) 
+
\log \frac{\lambda}{\mu} \left(
\begin{array}{cc}
-\frac{10}{3} & -6            \\
 0            & -\frac{64}{3} \\
\end{array}
\right) \, ,
\end{align}
\begin{align}
\hat{r}_{\text{BMU}}^{5-6} =&
\left(
\begin{array}{cc}
\frac{19}{3} & -3           \\
-\frac{1}{2} & -\frac{7}{6} \\
\end{array}
\right) 
+
\log \frac{\lambda}{\mu} \left(
\begin{array}{cc}
-\frac{64}{3} & 0              \\
-6            & -\frac{10}{3}  \\
\end{array}
\right) \, ,
\end{align}
\begin{align}
\hat{r}_{\text{BMU}}^{7-10} =&
\left(
\begin{array}{cccc}
\frac{7}{3} & 1 & \frac{7}{8} & \frac{3}{8} \\
0 & \frac{16}{3} & \frac{3}{4} & -\frac{1}{4} \\
-\frac{14}{3} & -2 & \frac{11}{3} & -3 \\
-4 & \frac{4}{3} & -2 & \frac{2}{3} 
\end{array}
\right) 
+ 
\log \frac{\lambda}{\mu} \left(
\begin{array}{cccc}
-\frac{10}{3} & -6            & -\frac{7}{6} & -\frac{1}{2} \\
0             & -\frac{64}{3} & -1           & \frac{1}{3}  \\
-56           & -24           & -18          & 6            \\
-48           & 16            & 0            & 0 
\end{array}
\right) \, .
\end{align}
With this information one can reproduce the results in~\cref{eq:preliminaries:matching-conditions-bern,eq:preliminaries:adms-bern}.

\section{Light-cone distribution amplitudes}
\label{app:lcda}

\subsection{Two-particle LCDAs}
\label{app:lcda:2pt}

As discussed in Ref.~\cite{Beneke:2000wa}, we can write the vacuum-to-light pseudoscalar meson matrix element of the non-local $\bar q u$ current up to twist-3 as follows,
\begin{align}
    \bra{P(q)} \bar{q}_i(z) u_j(0) \ket{0} = 
    \frac{if_P}{4} \int du \, e^{iuq\cdot z} \left \{\slashed{q} \gamma_5 \phi_{P}(u) - \mu_P \gamma_5 \left(\phi_P^p(u) - \sigma_{\mu \nu} q^\mu z^\nu \frac{\phi_P^{\sigma}(u)}{6} \right) \right\}_{ji} \, , \label{LCDATwist23PseudScalar}
\end{align}
where $P$ denotes the light pseudoscalar meson with momentum $q = E n_+$, $\mu_P = m_P^2/(m_q + m_u)$, and $i, \, j$ are Dirac indices. 
The momentum-space representation projector then reads \cite{Beneke:2000wa,Beneke:2001ev,Beneke:2003zv}
\begin{align}
    M_P =  \frac{if_P}{4} \left \{\slashed{q} \gamma_5 \phi_{P}(u) - \mu_P \gamma_5 \left(\phi_P^p(u)
    - i \sigma_{\mu \nu} n_+^\mu n_-^\nu \frac{\partial_u \phi_P^{\sigma}(u)}{12} 
    + i \sigma_{\mu \nu} q^\mu \frac{\phi_P^{\sigma}(u)}{6} \frac{\partial}{\partial l_{\perp \nu}}  \right) \right\}
     \, .
\end{align}
The twist-2 LCDA $\phi_P$ can then be expanded in Gegenbauer polynomials $C_n^{3/2}$,
\begin{align}
    \phi_{P}(u,\mu) = 6 u \bar{u} \left[1 + \sum_{n=1}^\infty  \alpha_n^P(\mu)\, C_n^{3/2}(2u - 1) \right] \, ,
    \label{eq:LCDAs:2pt:psd:twist2}
\end{align}
with the Gegenbauer moments $\alpha_n^P(\mu)$. We use the LL running for the Gegenbauer moments as presented in Ref.~\cite{Ball:2006wn}. Under the assumption that three-particle contributions can be neglected, the two twist-three LCDAs are not independent, due to equations of motion. This implies
\begin{align}
    \phi_P^p(u) = 1 \, , \qquad \phi_P^\sigma(u) = 6 u (1 - u) \, , \qquad \partial_u \phi_P^\sigma(u) = 6 (1 - 2u) \, .
\end{align}
Thus the projector in momentum space takes the final form
\begin{align}
    M_P =  \frac{if_P}{4} \left \{\slashed{q} \gamma_5 \phi_{P}(u) - \mu_P \gamma_5 \left(\phi_P^p(u)
    - i \sigma_{\mu \nu} n_+^\mu n_-^\nu \frac{(1-2u)}{2} \, \phi_P^p(u)
    + i \sigma_{\mu \nu} q^\mu u \bar{u} \, \phi_P^p(u) \frac{\partial}{\partial l_{\perp \nu}}  \right) \right\}
    \label{eq:PseudoScalarMomentum} \, ,
\end{align}
as presented in~Ref.~\cite{Beneke:2000wa}.

\subsection{Three-particle LCDAs}
\label{app:lcda:3pt}

For the three-particle LCDAs, we follow the conventions of~Ref.~\cite{Ball:2006wn}. We assume $z$ to be almost light-like, $z^2 \simeq 0$. It is convenient to define another light-like vector $p_\mu$ through
\begin{align}
    p_\mu & = q_\mu -\frac{1}{2} z_\mu \frac{m_P^2}{p \cdot z} \, ,
\end{align}
with the light-meson momentum $q^2 = m_P^2$. Note that it would seem like $p$ depends on $z$. However, we are free to choose $z$ to point in the $n_-$ direction. Substituting this back into the definition of $p$ makes it clear that the $z$-dependence cancels. Furthermore, we define the perpendicular part of the metric tensor as
\begin{align}
    g_{\perp \mu \nu} = g_{\mu \nu} - \frac{1}{p \cdot z} (p_\mu z_\nu + p_\nu z_\mu) \, .
\end{align}

Let us start with the twist-3 part for pseudoscalar mesons \cite{Ball:2006wn}: \footnote{%
Our convention for the covariant derivative is $D_\mu = \partial_\mu - ig_s A_\mu \, .$
}
\begin{align}
     \bra{P^-(q)}\bar{q}(0) \sigma_{\mu \nu} \gamma_5 g_s G_{\alpha \beta}(vz)u(0) \ket{0} = i f_{3 P} [(p_\alpha p_\mu g_{\perp \nu \beta} - p_\alpha p_\nu g_{\perp \mu \beta}) - (\alpha \leftrightarrow \beta)] \int \mathcal{D} \underline{u} \, e^{ivu_3 p \cdot z} \Phi_{3; P}(\underline{u}) \, ,
     \label{3ParticleTwist3}
\end{align}
where $\underline{u} = \{u_1, u_2, u_3\}$ denotes the momentum fractions of the quark, antiquark, and the gluon of the light meson momentum $q$ and $\int \mathcal{D} \underline{u} \equiv \int_0^1 du_1 du_2 du_3\, \delta(1-u_1-u_2-u_3)$ defines the integration measure. 
Note that the $g_\perp$ in this definition could readily be exchanged for a $g$ since all the additional terms appearing in $g_\perp$ cancel. For our purposes it is easier to deal with the regular $g$.
The corresponding position space projector is
\begin{align}
      \bra{P^-(q)}\bar{q}_i(0) g_s G_{\alpha \beta}(vz) u_j(0) \ket{0} = \frac{i f_{3 P}}{4} (p_\alpha p_\mu g_{\beta \nu} - p_\beta p_\mu g_{\nu \alpha}) [\sigma^{\mu \nu} \gamma_5 ]_{ji} \int \mathcal{D} \underline{u} e^{ivu_3 p\cdot z} \Phi_{3; P}(\underline{u})  \, , \label{3ParticleTwist3ProjPos}
\end{align}
where $i, \, j$ are Dirac indices. This can be confirmed by performing Dirac traces.
The explicit expression of $\Phi_{3;P}$ reads
\begin{align}
    \Phi_{3;P}(\underline{u}) = 360 u_1 u_2 u_3^2 \left\{1 + \lambda_{3P} (u_1 - u_2) + \omega_{3P} \frac{1}{2} (7 u_3 - 3) \right\} \, ,
    \label{eq:lcda:3pt:tw3}
\end{align}
where the parameters $\lambda_{3P}$ and $\omega_{3P}$ are defined in Ref.~\cite{Ball:2006wn}.

At twist-4, there are two non-vanishing vacuum-to-$P^-$ matrix elements:
\begin{align}
    \bra{P^-(q)}\bar{q}(0)\gamma_\mu \gamma_5 g_s G_{\alpha \beta}(vz)u(0) \ket{0} 
    =& f_P \int \mathcal{D} \underline{u} \ e^{ivu_3 p \cdot z} \Bigg\{\left[p_\beta g_{\alpha \mu} - p_\alpha g_{\beta \mu} \right] \Psi_{4;P}(\underline{u}) \notag \\ 
    &\qquad \qquad \qquad \quad+  \left [\frac{p_\mu p_\alpha z_\beta}{p \cdot z} - \frac{p_\mu p_\beta z_\alpha}{p \cdot z} \right] \left(\Phi_{4;P}(\underline{u}) + \Psi_{4;P}(\underline{u}) \right)  \Bigg\} \, , \label{AltThreeParticleTwist4} \\
    \bra{P^-(q)}\bar{q}(0)\gamma_\mu g_s \tilde{G}_{\alpha \beta}(vz)u(0) \ket{0} 
    =& if_P \int \mathcal{D} \underline{u} \ e^{ivu_3 p \cdot z} \Bigg\{\left[p_\beta g_{\alpha \mu} - p_\alpha g_{\beta \mu} \right] \tilde{\Psi}_{4;P}(\underline{u}) \notag \\ 
    &\qquad \qquad \qquad \quad+  \left [\frac{p_\mu p_\alpha z_\beta}{p \cdot z} - \frac{p_\mu p_\beta z_\alpha}{p \cdot z} \right] \left(\tilde{\Phi}_{4;P}(\underline{u}) + \tilde{\Psi}_{4;P}(\underline{u}) \right)  \Bigg\} \, . 
\end{align}
We are only interested in the first matrix elements since the second one does not appear in our calculations.
The corresponding projector for the first matrix element reads
\begin{align}
    \bra{P^-(q)}\bar{q}_i(0) g_s G_{\alpha \beta}(vz
    )u_j(0) \ket{0} 
    =& -\frac{f_P}{4} [\gamma^\mu \gamma_5]_{ji} \, \int \mathcal{D} \underline{u} \ e^{ivu_3 p \cdot z} \Bigg\{\left[p_\beta g_{\alpha \mu} - p_\alpha g_{\beta \mu} \right] \Psi_{4;P}(\underline{u}) \notag \\ 
    &\qquad \qquad \qquad \quad+  \left [\frac{p_\mu p_\alpha z_\beta}{p \cdot z} - \frac{p_\mu p_\beta z_\alpha}{p \cdot z} \right] \left(\Phi_{4;P}(\underline{u}) + \Psi_{4;P}(\underline{u}) \right)  \Bigg\} \, . \label{ThreeParticleTwist4ProjPos}
\end{align}
It turns out that for us only $\Phi_{4;P}$ is relevant. Its conformal expansion is
\begin{align}
    \Phi_{4;P}(\underline{u}) = 120 u_1 u_2 u_3 \, \left[\phi_0^P + \phi_1^P (u_1 - u_2) + \phi_2^P (3u_3 - 1) \right] \, ,
    \label{eq:lcda:3pt:tw4}
\end{align}
where the parameters $\phi_i^P$, $i = 1, \, 2, \, 3$ are defined in Ref.~\cite{Ball:2006wn}.
%
\section{Parity transforms of matrix elements}
\label{app:parity}

Let us illustrate how to obtain the parity flipped matrix elements of the ones that we provided in \cref{sec:class-I:results:2pt} with the simple example of the tree-level matrix elements of $\mathcal{Q}_2^{VLL}$ and $\mathcal{Q}_2^{VRR}$. The tree-level discussion will carry through without any modifications to the one-loop results. It is easy to see that these operators are connected by a parity transformation:
\begin{align}
    \mathcal{Q}_2^{VLL} \xrightarrow{\mathcal{P}} \mathcal{Q}_2^{VRR} \, .
\end{align}
The tree-level matrix element of $\mathcal{Q}_2^{VLL}$ is given by
\begin{align}
   \bra{D^{(*)+}_{(s)} P^-} \mathcal{Q}_2^{VLL} \ket{\bar{B}_{(s)}} = \frac{i f_P}{4} \left(J_V - J_A \right)\ .
\end{align}
Applying a parity transform on the left side of this equation yields
\begin{align}
      \bra{D^{(*)+}_{(s)} P^-} \mathcal{Q}_2^{VLL} \ket{\bar{B}_{(s)}} \xrightarrow{\mathcal{P}} - \bra{D^{(*)+}_{(s)} P^-} \mathcal{Q}_2^{VRR} \ket{\bar{B}_{(s)}} \, ,
\end{align}
since also the meson states have to be transformed and all three mesons involved are parity odd. Note that also all the momenta go from $p = (p^0, \Vec{p})$ to $\Tilde{p} = (p^0, -\Vec{p})$, which we do not write explicitly. On the right side of the equation, the transformation reads
\begin{align}
    \frac{i f_P}{4} \left(J_V - J_A \right) \xrightarrow{\mathcal{P}} \frac{i f_P}{4} \left(J_V + J_A \right) \, ,
\end{align}
since the time component of the vector current is parity-even while the time component of the axial-vector current is parity-odd. 
Putting everything together we get
\begin{align}
     \bra{D^{(*)+}_{(s)} P^-} \mathcal{Q}_2^{VRR} \ket{\bar{B}_{(s)}} = - \frac{i f_P}{4} \left(J_V + J_A \right) \, ,
\end{align}
which can be easily verified also by direct calculation. The matrix elements of other operators can be transformed in a similar way.

\section{Leading-order two-particle contributions in QCDF}
\label{app:LO}

In this appendix we collect the LO QCDF results. We will present the results for the singlet operators. The corresponding results for the colour-crossed operators can be inferred directly. Using the colour Fierz identity, we rewrite the colour-crossed operators $\mathcal{Q}_1^X$ as \footnote{This of course holds also for the tensor operators $\mathcal{Q}_3^{SLL(RR)}$ and $\mathcal{Q}_4^{SLL(RR)}$ with the appropriate indices.}
\begin{align}
    \mathcal{Q}_1^X = 2 \left[\bar{c} \Gamma_1^X T^A b \right] \left[\bar{q} \Gamma_2^X T^A u \right] + \frac{1}{N} \left[\bar{c} \Gamma_1^X b \right] \left[\bar{q} \Gamma_2^X u \right]\, ,
\end{align}
with the Dirac structures $\Gamma_1^X$ and $\Gamma_2^X$. It is clear that the colour octet does not have an overlap with the meson states and hence we get
\begin{align}
    \bra{D^{(*)+}_{(s)} L^-}\mathcal{Q}_1^X \ket{\bar{B}^0_{(s)}}_{\text{LO}} = \frac{1}{N}\bra{D^{(*)+}_{(s)} L^-}\mathcal{Q}_2^X \ket{\bar{B}^0_{(s)}}_{\text{LO}} \, .
\end{align}

The leading order contributions from the scalar operators read
\begin{align}
    \bra{D^{(*)+}_{(s)} L^-} \mathcal{Q}^{SLL}_2 \ket{\bar{B}^0_{(s)}}_{\text{LO}} &= + \frac{if_L}{4} \mu_L (J_S - J_P) \int_0^1 du\, \phi_{3;L}(u) \, , \\
    \bra{D^{(*)+}_{(s)} L^-} \mathcal{Q}^{SLR}_2 \ket{\bar{B}^0_{(s)}}_{\text{LO}} &= \mp \frac{if_L}{4} \mu_L (J_S - J_P) \int_0^1 du\, \phi_{3;L}(u)\, , \\
    \bra{D^{(*)+}_{(s)} L^-} \mathcal{Q}^{SRL}_2 \ket{\bar{B}^0_{(s)}}_{\text{LO}} &= + \frac{if_L}{4} \mu_L (J_S + J_P) \int_0^1 du\, \phi_{3;L}(u)\, , \\
    \bra{D^{(*)+}_{(s)} L^-} \mathcal{Q}^{SRR}_2 \ket{\bar{B}_{(s)}^0}_{\text{LO}} &= \mp \frac{if_L}{4} \mu_L (J_S + J_P) \int_0^1 du\, \phi_{3;L}(u)\, .
\end{align}
The upper (lower) sign corresponds to a light pseudo scalar (vector) meson in the final state.
Note that in the scalar sector for pseudoscalar light mesons the LCDA is normalized to unity and hence the integral over the LCDA can be trivially computed. In the vector meson case, the twist 3 LCDA integrates to zero and hence the matrix element vanishes at LO for a light vector meson in the final state. 

For the vector operators we get
\begin{align}
    \bra{D^{(*)+}_{(s)} L^-} \mathcal{Q}^{VLL}_2 \ket{\bar{B}_{(s)}^0}_{\text{LO}} &= \pm \frac{if_L}{4} (J_V - J_A) \int_0^1 du \, \phi_L(u) \, , \\
    \bra{D^{(*)+}_{(s)} L^-} \mathcal{Q}^{VLR}_2 \ket{\bar{B}_{(s)}^0}_{\text{LO}} &= - \frac{if_L}{4} (J_V - J_A) \int_0^1 du \, \phi_L(u)\, , \\
    \bra{D^{(*)+}_{(s)} L^-} \mathcal{Q}^{VRL}_2 \ket{\bar{B}_{(s)}^0}_{\text{LO}} &= \pm \frac{if_L}{4} (J_V + J_A) \int_0^1 du \, \phi_L(u)\, , \\
    \bra{D^{(*)+}_{(s)} L^-} \mathcal{Q}^{VRR}_2 \ket{\bar{B}_{(s)}^0}_{\text{LO}} &= - \frac{if_L}{4} (J_V + J_A) \int_0^1 du \, \phi_L(u)\, .
\end{align}
Here, the only contribution comes from the terms proportional to the leading twist LCDA, $\phi_L(u)$, which is normalized to one in the light pseudoscalar as well as vector meson case.

For the tensor operators, there are no leading contributions:
\begin{align}
    \bra{D^{(*)+}_{(s)} L^-} \mathcal{Q}^{SLL}_4 \ket{\bar{B}_{(s)}^0}_{\text{LO}} &= 0\, , \\
    \bra{D^{(*)+}_{(s)} L^-} \mathcal{Q}^{SRR}_4 \ket{\bar{B}_{(s)}^0}_{\text{LO}} &= 0\, . 
\end{align}
This is because the only non-vanishing trace from projecting onto the light meson state would come from the term with a derivative acting on the hard-scattering amplitude.
At leading order, there is no dependence on any momenta and thus the contribution vanishes.

\section{Separation of modes in posterior distributions}
\label{app:modes}

In this appendix we present the functions that define the hyperplanes in the space of parameters of interest that separate the different modes. For the WET-1 model, we need to define two parallel hyperplanes in order to separate the two modes:
\begin{align}
    f_1^{\text{WET-1}}(\vecth) &= \wcbern[qbcu]{1} + 1.17\, \wcbern[qbcu]{2} + 1.6 \, , \\
    f_2^{\text{WET-1}}(\vecth) &= \wcbern[qbcu]{1} + 1.17\, \wcbern[qbcu]{2} - 1.6 \, \, .
\end{align}
The points in mode A fulfil $f_1^{\text{WET-1}}(\vecth) > 0$ and $ f_2^{\text{WET-1}}(\vecth) < 0$, while mode B has the signs of the inequalities swapped.

In the WET-2 model, again two parallel hyperplanes are necessary:
\begin{align}
    f_1^{\text{WET-2}}(\vecth) &= \wcbern[qbcu]{5} + 0.81\, \wcbern[qbcu]{6} - 2.25 \, , \\
    f_2^{\text{WET-2}}(\vecth) &= \wcbern[qbcu]{5} + 0.81\, \wcbern[qbcu]{6} + 2.25 \, \, .
\end{align}
Mode A is identified by $f_1^{\text{WET-2}}(\vecth) < 0$ and $f_2^{\text{WET-2}}(\vecth) > 0$, 
while mode B is given by $f_1^{\text{WET-2}}(\vecth) > 0$ and $f_2^{\text{WET-2}}(\vecth) > 0$. 
There is a third region within the parameter space defined by $f_1^{\text{WET-2}}(\vecth) < 0$ and $f_2^{\text{WET-2}}(\vecth) < 0$. This region, however, contributes less than $0.1 \%$ to the total evidence and thus is not considered.

The modes in the WET-3 model are separated by the following two parallel hyperplanes:
\begin{align}
    f_1^{\text{WET-3}}(\vecth) &= \wcbern[qbcu]{1'} + 1.17\, \wcbern[qbcu]{2'} - 1.15 \, , \\
    f_2^{\text{WET-3}}(\vecth) &= \wcbern[qbcu]{1'} + 1.17\, \wcbern[qbcu]{2'} - 2.45 \, ,
\end{align}
where mode A fulfils $f_1^{\text{WET-3}}(\vecth) < 0$, and mode B $f_1^{\text{WET-3}}(\vecth) > 0$ and $f_2^{\text{WET-3}}(\vecth) < 0$.
The region with $f_2^{\text{WET-3}}(\vecth) > 0$ contributes less than $0.5 \%$ to the total evidence and is therefore not considered.

In the case of WET-4, we need two non-parallel hyperplanes:
\begin{align}
    f_1^{\text{WET-4}}(\vecth) &= \wcbern[qbcu]{5'} + 0.75 \, \wcbern[qbcu]{6'} + 3.25 \, , \\
    f_2^{\text{WET-4}}(\vecth) &= \wcbern[qbcu]{8'} - 11.25 \, \wcbern[qbcu]{10'} - 0.35 \, ,
\end{align}
where the modes are defined by $f_1^{\text{WET-4}}(\vecth) > 0$ for mode A, and $f_1^{\text{WET-4}}(\vecth) < 0$ and $f_2^{\text{WET-4}}(\vecth) > 0$ for mode B. The region with $f_1^{\text{WET-4}}(\vecth) < 0$ and $f_2^{\text{WET-4}}(\vecth) < 0$ contributes less than $2 \%$ to the total evidence and is thus discarded.

\bibliographystyle{JHEP}
\bibliography{references}

\providecommand{\href}[2]{#2}\begingroup\raggedright\begin{thebibliography}{10}

\bibitem{Isidori:2010kg}
G.~Isidori, Y.~Nir and G.~Perez, \emph{{Flavor Physics Constraints for Physics
  Beyond the Standard Model}},
  \href{https://doi.org/10.1146/annurev.nucl.012809.104534}{\emph{Ann. Rev.
  Nucl. Part. Sci.} {\bfseries 60} (2010) 355}
  [\href{https://arxiv.org/abs/1002.0900}{{\ttfamily 1002.0900}}].

\bibitem{Antonelli:2009ws}
M.~Antonelli et~al., \emph{{Flavor Physics in the Quark Sector}},
  \href{https://doi.org/10.1016/j.physrep.2010.05.003}{\emph{Phys. Rept.}
  {\bfseries 494} (2010) 197}
  [\href{https://arxiv.org/abs/0907.5386}{{\ttfamily 0907.5386}}].

\bibitem{Belle-II:2018jsg}
{\scshape Belle-II} collaboration, \emph{{The Belle II Physics Book}},
  \href{https://doi.org/10.1093/ptep/ptz106}{\emph{PTEP} {\bfseries 2019}
  (2019) 123C01} [\href{https://arxiv.org/abs/1808.10567}{{\ttfamily
  1808.10567}}], [Erratum: PTEP 2020, 029201 (2020)].

\bibitem{BaBar:2014omp}
{\scshape BaBar, Belle} collaboration, \emph{{The Physics of the B Factories}},
  \href{https://doi.org/10.1140/epjc/s10052-014-3026-9}{\emph{Eur. Phys. J. C}
  {\bfseries 74} (2014) 3026}
  [\href{https://arxiv.org/abs/1406.6311}{{\ttfamily 1406.6311}}].

\bibitem{Kuhr:2013hd}
T.~Kuhr, \emph{{Flavor physics at the Tevatron}}, vol.~249 (2013),
  \href{https://doi.org/10.1007/978-3-642-10300-1}{10.1007/978-3-642-10300-1}.

\bibitem{LHCb:2012myk}
{\scshape LHCb} collaboration, \emph{{Implications of LHCb measurements and
  future prospects}},
  \href{https://doi.org/10.1140/epjc/s10052-013-2373-2}{\emph{Eur. Phys. J. C}
  {\bfseries 73} (2013) 2373}
  [\href{https://arxiv.org/abs/1208.3355}{{\ttfamily 1208.3355}}].

\bibitem{Aebischer:2017gaw}
J.~Aebischer, M.~Fael, C.~Greub and J.~Virto, \emph{{B physics Beyond the
  Standard Model at One Loop: Complete Renormalization Group Evolution below
  the Electroweak Scale}},
  \href{https://doi.org/10.1007/JHEP09(2017)158}{\emph{JHEP} {\bfseries 09}
  (2017) 158} [\href{https://arxiv.org/abs/1704.06639}{{\ttfamily
  1704.06639}}].

\bibitem{Beneke:1999br}
M.~Beneke, G.~Buchalla, M.~Neubert and C.T.~Sachrajda, \emph{{QCD factorization
  for $B \to \pi \pi$ decays: Strong phases and CP violation in the heavy quark
  limit}}, \href{https://doi.org/10.1103/PhysRevLett.83.1914}{\emph{Phys. Rev.
  Lett.} {\bfseries 83} (1999) 1914}
  [\href{https://arxiv.org/abs/hep-ph/9905312}{{\ttfamily hep-ph/9905312}}].

\bibitem{Bauer:2000yr}
C.W.~Bauer, S.~Fleming, D.~Pirjol and I.W.~Stewart, \emph{{An Effective field
  theory for collinear and soft gluons: Heavy to light decays}},
  \href{https://doi.org/10.1103/PhysRevD.63.114020}{\emph{Phys. Rev. D}
  {\bfseries 63} (2001) 114020}
  [\href{https://arxiv.org/abs/hep-ph/0011336}{{\ttfamily hep-ph/0011336}}].

\bibitem{Beneke:2000ry}
M.~Beneke, G.~Buchalla, M.~Neubert and C.T.~Sachrajda, \emph{{QCD factorization
  for exclusive, nonleptonic B meson decays: General arguments and the case of
  heavy light final states}},
  \href{https://doi.org/10.1016/S0550-3213(00)00559-9}{\emph{Nucl. Phys. B}
  {\bfseries 591} (2000) 313}
  [\href{https://arxiv.org/abs/hep-ph/0006124}{{\ttfamily hep-ph/0006124}}].

\bibitem{Beneke:2001ev}
M.~Beneke, G.~Buchalla, M.~Neubert and C.T.~Sachrajda, \emph{{QCD factorization
  in $B \to \pi K,\, \pi \pi$ decays and extraction of Wolfenstein
  parameters}},
  \href{https://doi.org/10.1016/S0550-3213(01)00251-6}{\emph{Nucl. Phys. B}
  {\bfseries 606} (2001) 245}
  [\href{https://arxiv.org/abs/hep-ph/0104110}{{\ttfamily hep-ph/0104110}}].

\bibitem{Beneke:2003zv}
M.~Beneke and M.~Neubert, \emph{{QCD factorization for $B \to PP$ and $B \to
  PV$ decays}},
  \href{https://doi.org/10.1016/j.nuclphysb.2003.09.026}{\emph{Nucl. Phys. B}
  {\bfseries 675} (2003) 333}
  [\href{https://arxiv.org/abs/hep-ph/0308039}{{\ttfamily hep-ph/0308039}}].

\bibitem{Bauer:2004tj}
C.W.~Bauer, D.~Pirjol, I.Z.~Rothstein and I.W.~Stewart, \emph{{$B \to M_1 M_2$:
  Factorization, charming penguins, strong phases, and polarization}},
  \href{https://doi.org/10.1103/PhysRevD.70.054015}{\emph{Phys. Rev. D}
  {\bfseries 70} (2004) 054015}
  [\href{https://arxiv.org/abs/hep-ph/0401188}{{\ttfamily hep-ph/0401188}}].

\bibitem{Bauer:2001cu}
C.W.~Bauer, D.~Pirjol and I.W.~Stewart, \emph{{A Proof of factorization for $B
  \to D \pi$}},
  \href{https://doi.org/10.1103/PhysRevLett.87.201806}{\emph{Phys. Rev. Lett.}
  {\bfseries 87} (2001) 201806}
  [\href{https://arxiv.org/abs/hep-ph/0107002}{{\ttfamily hep-ph/0107002}}].

\bibitem{Krankl:2015fha}
S.~Kr\"ankl, T.~Mannel and J.~Virto, \emph{{Three-body non-leptonic $B$ decays
  and QCD factorization}},
  \href{https://doi.org/10.1016/j.nuclphysb.2015.08.004}{\emph{Nucl. Phys. B}
  {\bfseries 899} (2015) 247}
  [\href{https://arxiv.org/abs/1505.04111}{{\ttfamily 1505.04111}}].

\bibitem{Huber:2020pqb}
T.~Huber, J.~Virto and K.K.~Vos, \emph{{Three-Body Non-Leptonic Heavy-to-heavy
  $B$ Decays at NNLO in QCD}},
  \href{https://doi.org/10.1007/JHEP11(2020)103}{\emph{JHEP} {\bfseries 11}
  (2020) 103} [\href{https://arxiv.org/abs/2007.08881}{{\ttfamily
  2007.08881}}].

\bibitem{Neubert:1997uc}
M.~Neubert and B.~Stech, \emph{{Nonleptonic weak decays of B mesons}},
  \href{https://doi.org/10.1142/9789812812667_0004}{\emph{Adv. Ser. Direct.
  High Energy Phys.} {\bfseries 15} (1998) 294}
  [\href{https://arxiv.org/abs/hep-ph/9705292}{{\ttfamily hep-ph/9705292}}].

\bibitem{Beneke:2000wa}
M.~Beneke and T.~Feldmann, \emph{{Symmetry breaking corrections to heavy to
  light B meson form-factors at large recoil}},
  \href{https://doi.org/10.1016/S0550-3213(00)00585-X}{\emph{Nucl. Phys. B}
  {\bfseries 592} (2001) 3}
  [\href{https://arxiv.org/abs/hep-ph/0008255}{{\ttfamily hep-ph/0008255}}].

\bibitem{Huber:2016xod}
T.~Huber, S.~Kr\"ankl and X.-Q.~Li, \emph{{Two-body non-leptonic heavy-to-heavy
  decays at NNLO in QCD factorization}},
  \href{https://doi.org/10.1007/JHEP09(2016)112}{\emph{JHEP} {\bfseries 09}
  (2016) 112} [\href{https://arxiv.org/abs/1606.02888}{{\ttfamily
  1606.02888}}].

\bibitem{Beneke:2021jhp}
M.~Beneke, P.~B\"oer, G.~Finauri and K.K.~Vos, \emph{{QED factorization of
  two-body non-leptonic and semi-leptonic B to charm decays}},
  \href{https://doi.org/10.1007/JHEP10(2021)223}{\emph{JHEP} {\bfseries 10}
  (2021) 223} [\href{https://arxiv.org/abs/2107.03819}{{\ttfamily
  2107.03819}}].

\bibitem{Bordone:2020gao}
M.~Bordone, N.~Gubernari, T.~Huber, M.~Jung and D.~van Dyk, \emph{{A puzzle in
  $\bar{B}_{(s)}^0 \to D_{(s)}^{(*)+} \lbrace \pi^-, K^-\rbrace$ decays and
  extraction of the $f_s/f_d$ fragmentation fraction}},
  \href{https://doi.org/10.1140/epjc/s10052-020-08512-8}{\emph{Eur. Phys. J. C}
  {\bfseries 80} (2020) 951}
  [\href{https://arxiv.org/abs/2007.10338}{{\ttfamily 2007.10338}}].

\bibitem{Bordone:2021cca}
M.~Bordone, A.~Greljo and D.~Marzocca, \emph{{Exploiting dijet resonance
  searches for flavor physics}},
  \href{https://doi.org/10.1007/JHEP08(2021)036}{\emph{JHEP} {\bfseries 08}
  (2021) 036} [\href{https://arxiv.org/abs/2103.10332}{{\ttfamily
  2103.10332}}].

\bibitem{Atkinson:2024hqp}
O.~Atkinson, C.~Englert, M.~Kirk and G.~Tetlalmatzi-Xolocotzi,
  \emph{{Collider-Flavour Complementarity from the bottom to the top}},
  \href{https://arxiv.org/abs/2411.00940}{{\ttfamily 2411.00940}}.

\bibitem{Iguro:2020ndk}
S.~Iguro and T.~Kitahara, \emph{{Implications for new physics from a novel
  puzzle in $\bar{B}_{(s)}^0 \to D^{(\ast)+}_{(s)} \lbrace \pi^-, K^- \rbrace$
  decays}}, \href{https://doi.org/10.1103/PhysRevD.102.071701}{\emph{Phys. Rev.
  D} {\bfseries 102} (2020) 071701}
  [\href{https://arxiv.org/abs/2008.01086}{{\ttfamily 2008.01086}}].

\bibitem{Fleischer:2021cct}
R.~Fleischer and E.~Malami, \emph{{Using $B^0_s\to D_s^\mp K^\pm$ Decays as a
  Portal to New Physics}},
  \href{https://doi.org/10.1103/PhysRevD.106.056004}{\emph{Phys. Rev. D}
  {\bfseries 106} (2022) 056004}
  [\href{https://arxiv.org/abs/2109.04950}{{\ttfamily 2109.04950}}].

\bibitem{Fleischer:2021cwb}
R.~Fleischer and E.~Malami, \emph{{Revealing new physics in
  $B^{0}_{s}\rightarrow D_s^{\mp } K^{\pm }$ decays}},
  \href{https://doi.org/10.1140/epjc/s10052-023-11588-7}{\emph{Eur. Phys. J. C}
  {\bfseries 83} (2023) 420}
  [\href{https://arxiv.org/abs/2110.04240}{{\ttfamily 2110.04240}}].

\bibitem{Piscopo:2023opf}
M.L.~Piscopo and A.V.~Rusov, \emph{{Non-factorisable effects in the decays $
  {\overline{B}}_s^0\to {D}_s^{+}{\pi}^{-} $ and $ {\overline{B}}^0\to
  {D}^{+}{K}^{-} $ from LCSR}},
  \href{https://doi.org/10.1007/JHEP10(2023)180}{\emph{JHEP} {\bfseries 10}
  (2023) 180} [\href{https://arxiv.org/abs/2307.07594}{{\ttfamily
  2307.07594}}].

\bibitem{Cai:2021mlt}
F.-M.~Cai, W.-J.~Deng, X.-Q.~Li and Y.-D.~Yang, \emph{{Probing new physics in
  class-I B-meson decays into heavy-light final states}},
  \href{https://doi.org/10.1007/JHEP10(2021)235}{\emph{JHEP} {\bfseries 10}
  (2021) 235} [\href{https://arxiv.org/abs/2103.04138}{{\ttfamily
  2103.04138}}].

\bibitem{Aebischer:2017ugx}
J.~Aebischer et~al., \emph{{WCxf: an exchange format for Wilson coefficients
  beyond the Standard Model}},
  \href{https://doi.org/10.1016/j.cpc.2018.05.022}{\emph{Comput. Phys. Commun.}
  {\bfseries 232} (2018) 71}
  [\href{https://arxiv.org/abs/1712.05298}{{\ttfamily 1712.05298}}].

\bibitem{Buras:2000if}
A.J.~Buras, M.~Misiak and J.~Urban, \emph{{Two loop QCD anomalous dimensions of
  flavor changing four quark operators within and beyond the standard model}},
  \href{https://doi.org/10.1016/S0550-3213(00)00437-5}{\emph{Nucl. Phys. B}
  {\bfseries 586} (2000) 397}
  [\href{https://arxiv.org/abs/hep-ph/0005183}{{\ttfamily hep-ph/0005183}}].

\bibitem{Jenkins:2017jig}
E.E.~Jenkins, A.V.~Manohar and P.~Stoffer, \emph{{Low-Energy Effective Field
  Theory below the Electroweak Scale: Operators and Matching}},
  \href{https://doi.org/10.1007/JHEP03(2018)016}{\emph{JHEP} {\bfseries 03}
  (2018) 016} [\href{https://arxiv.org/abs/1709.04486}{{\ttfamily
  1709.04486}}], [Erratum: JHEP 12, 043 (2023)].

\bibitem{Buras:1989xd}
A.J.~Buras and P.H.~Weisz, \emph{{QCD Nonleading Corrections to Weak Decays in
  Dimensional Regularization and 't Hooft-Veltman Schemes}},
  \href{https://doi.org/10.1016/0550-3213(90)90223-Z}{\emph{Nucl. Phys. B}
  {\bfseries 333} (1990) 66}.

\bibitem{Dugan:1990df}
M.J.~Dugan and B.~Grinstein, \emph{{On the vanishing of evanescent operators}},
  \href{https://doi.org/10.1016/0370-2693(91)90680-O}{\emph{Phys. Lett. B}
  {\bfseries 256} (1991) 239}.

\bibitem{Herrlich:1994kh}
S.~Herrlich and U.~Nierste, \emph{{Evanescent operators, scheme dependences and
  double insertions}},
  \href{https://doi.org/10.1016/0550-3213(95)00474-7}{\emph{Nucl. Phys. B}
  {\bfseries 455} (1995) 39}
  [\href{https://arxiv.org/abs/hep-ph/9412375}{{\ttfamily hep-ph/9412375}}].

\bibitem{Tracas:1982gp}
N.~Tracas and N.~Vlachos, \emph{{Two Loop Calculations in {QCD} and the $\Delta
  I = 1/2$ Rule in Nonleptonic Weak Decays}},
  \href{https://doi.org/10.1016/0370-2693(82)90530-5}{\emph{Phys. Lett. B}
  {\bfseries 115} (1982) 419}.

\bibitem{Dekens:2019ept}
W.~Dekens and P.~Stoffer, \emph{{Low-energy effective field theory below the
  electroweak scale: matching at one loop}},
  \href{https://doi.org/10.1007/JHEP10(2019)197}{\emph{JHEP} {\bfseries 10}
  (2019) 197} [\href{https://arxiv.org/abs/1908.05295}{{\ttfamily
  1908.05295}}], [Erratum: JHEP 11, 148 (2022)].

\bibitem{Gorbahn:2004my}
M.~Gorbahn and U.~Haisch, \emph{{Effective Hamiltonian for non-leptonic
  $|\Delta F| = 1$ decays at NNLO in QCD}},
  \href{https://doi.org/10.1016/j.nuclphysb.2005.01.047}{\emph{Nucl. Phys. B}
  {\bfseries 713} (2005) 291}
  [\href{https://arxiv.org/abs/hep-ph/0411071}{{\ttfamily hep-ph/0411071}}].

\bibitem{Buras:1998raa}
A.J.~Buras, \emph{{Weak Hamiltonian, CP violation and rare decays}},  in
  \emph{{Les Houches Summer School in Theoretical Physics, Session 68: Probing
  the Standard Model of Particle Interactions}}, pp.~281--539, 6, 1998
  [\href{https://arxiv.org/abs/hep-ph/9806471}{{\ttfamily hep-ph/9806471}}].

\bibitem{Egner:2024azu}
M.~Egner, M.~Fael, K.~Sch\"onwald and M.~Steinhauser, \emph{{Nonleptonic
  $B$-meson decays to next-to-next-to-leading order}},
  \href{https://arxiv.org/abs/2406.19456}{{\ttfamily 2406.19456}}.

\bibitem{Descotes-Genon:2002crx}
S.~Descotes-Genon and C.T.~Sachrajda, \emph{{Factorization, the light cone
  distribution amplitude of the B meson and the radiative decay $B \to \gamma
  \ell \nu_\ell$}},
  \href{https://doi.org/10.1016/S0550-3213(02)01066-0}{\emph{Nucl. Phys. B}
  {\bfseries 650} (2003) 356}
  [\href{https://arxiv.org/abs/hep-ph/0209216}{{\ttfamily hep-ph/0209216}}].

\bibitem{Bordone:2019vic}
M.~Bordone, M.~Jung and D.~van Dyk, \emph{{Theory determination of $\bar{B}\to
  D^{(*)}\ell^-\bar\nu$ form factors at $\mathcal{O}(1/m_c^2)$}},
  \href{https://doi.org/10.1140/epjc/s10052-020-7616-4}{\emph{Eur. Phys. J. C}
  {\bfseries 80} (2020) 74} [\href{https://arxiv.org/abs/1908.09398}{{\ttfamily
  1908.09398}}].

\bibitem{Jeffreys:1939xee}
H.~Jeffreys, \emph{{The Theory of Probability}}, Oxford Classic Texts in the
  Physical Sciences, {Oxford University Press} (1939).

\bibitem{Belle:2008ezn}
{\scshape Belle} collaboration, \emph{{Measurement of the Decay $B_s^0 \to
  D_s^- \pi^{+}$ and Evidence for $B_s^0 \to D_s^\pm K^\pm$ in $e^+ e^-$
  Annihilation at $\sqrt{s}$ \textasciitilde{} 10.87-GeV}},
  \href{https://doi.org/10.1103/PhysRevLett.102.021801}{\emph{Phys. Rev. Lett.}
  {\bfseries 102} (2009) 021801}
  [\href{https://arxiv.org/abs/0809.2526}{{\ttfamily 0809.2526}}].

\bibitem{Belle:2001ccu}
{\scshape Belle} collaboration, \emph{{Observation of Cabibbo suppressed $B \to
  D^{(*)}K^-$ decays at BELLE}},
  \href{https://doi.org/10.1103/PhysRevLett.87.111801}{\emph{Phys. Rev. Lett.}
  {\bfseries 87} (2001) 111801}
  [\href{https://arxiv.org/abs/hep-ex/0104051}{{\ttfamily hep-ex/0104051}}].

\bibitem{BaBar:2006rof}
{\scshape BaBar} collaboration, \emph{{Branching fraction measurement of $B^0
  \to D^{(*)+} \pi^{-}$, $B^{-} \to D^{(*)0} \pi^{-}$ and isospin analysis of
  $\bar{B} \to D^{(*)} \pi$ decays}},
  \href{https://doi.org/10.1103/PhysRevD.75.031101}{\emph{Phys. Rev. D}
  {\bfseries 75} (2007) 031101}
  [\href{https://arxiv.org/abs/hep-ex/0610027}{{\ttfamily hep-ex/0610027}}].

\bibitem{CLEO:2002iij}
{\scshape CLEO} collaboration, \emph{{Measurement of $\mathcal{B}(B^- \to D^0
  \pi^-)$ and $\mathcal{B}(\bar{B}^0 \to D^+ \pi^-)$ and isospin analysis of
  $\bar{B} \to D \pi$ decays}},
  \href{https://doi.org/10.1103/PhysRevD.66.031101}{\emph{Phys. Rev. D}
  {\bfseries 66} (2002) 031101}
  [\href{https://arxiv.org/abs/hep-ex/0206030}{{\ttfamily hep-ex/0206030}}].

\bibitem{BaBar:2006zod}
{\scshape BaBar} collaboration, \emph{{Measurement of the absolute branching
  fractions $B \to D\pi, D^*\pi, D^{**}\pi$ with a missing mass method}},
  \href{https://doi.org/10.1103/PhysRevD.74.111102}{\emph{Phys. Rev. D}
  {\bfseries 74} (2006) 111102}
  [\href{https://arxiv.org/abs/hep-ex/0609033}{{\ttfamily hep-ex/0609033}}].

\bibitem{Belle:2010ldr}
{\scshape Belle} collaboration, \emph{{Observation of $B_s^0 \to D_s^{*-}
  \pi^+$, $B_s^0 \to D_s^{(*)-} \rho^+$ Decays and Measurement of $B_s^0 \to
  D_s^{*-} \rho^+$ Polarization}},
  \href{https://doi.org/10.1103/PhysRevLett.104.231801}{\emph{Phys. Rev. Lett.}
  {\bfseries 104} (2010) 231801}
  [\href{https://arxiv.org/abs/1003.5312}{{\ttfamily 1003.5312}}].

\bibitem{CLEO:1997vmd}
{\scshape CLEO} collaboration, \emph{{A New measurement of $B\to D^* \pi$
  branching fractions}},
  \href{https://doi.org/10.1103/PhysRevLett.80.2762}{\emph{Phys. Rev. Lett.}
  {\bfseries 80} (1998) 2762}
  [\href{https://arxiv.org/abs/hep-ex/9706019}{{\ttfamily hep-ex/9706019}}].

\bibitem{CDF:2006hob}
{\scshape CDF} collaboration, \emph{{Measurement of the Ratios of Branching
  Fractions $\mathcal{B}(B^0_s \to D^-_s \pi^+ \pi^+ \pi^-) / \mathcal{B}(B^0
  \to D^- \pi^+ \pi^+ \pi^-)$ and $\mathcal{B}(B^0_s \to D^-_s \pi^+) /
  \mathcal{B}(B^0 \to D^- \pi^+)$}},
  \href{https://doi.org/10.1103/PhysRevLett.98.061802}{\emph{Phys. Rev. Lett.}
  {\bfseries 98} (2007) 061802}
  [\href{https://arxiv.org/abs/hep-ex/0610045}{{\ttfamily hep-ex/0610045}}].

\bibitem{LHCb:2012wdi}
{\scshape LHCb} collaboration, \emph{{Measurements of the branching fractions
  of the decays $B^{0}_{s} \to D^{\mp}_{s} K^{\pm} $ and $B^{0}_{s} \to
  D^{-}_{s} \pi^{+}$}},
  \href{https://doi.org/10.1007/JHEP06(2012)115}{\emph{JHEP} {\bfseries 06}
  (2012) 115} [\href{https://arxiv.org/abs/1204.1237}{{\ttfamily 1204.1237}}].

\bibitem{LHCb:2013vfg}
{\scshape LHCb} collaboration, \emph{{Measurement of the fragmentation fraction
  ratio $f_{s}/f_{d}$ and its dependence on $B$ meson kinematics}},
  \href{https://doi.org/10.1007/JHEP04(2013)001}{\emph{JHEP} {\bfseries 04}
  (2013) 001} [\href{https://arxiv.org/abs/1301.5286}{{\ttfamily 1301.5286}}].

\bibitem{ParticleDataGroup:2018ovx}
{\scshape Particle Data Group} collaboration, \emph{{Review of Particle
  Physics}}, \href{https://doi.org/10.1103/PhysRevD.98.030001}{\emph{Phys. Rev.
  D} {\bfseries 98} (2018) 030001}.

\bibitem{BaBar:2005osj}
{\scshape BaBar} collaboration, \emph{{Measurement of branching fractions and
  resonance contributions for $B^0 \to \bar{D}^0 K^{+} \pi^{-}$ and search for
  $B^0 \to D^0 K^{+} \pi^{-}$ decays}},
  \href{https://doi.org/10.1103/PhysRevLett.96.011803}{\emph{Phys. Rev. Lett.}
  {\bfseries 96} (2006) 011803}
  [\href{https://arxiv.org/abs/hep-ex/0509036}{{\ttfamily hep-ex/0509036}}].

\bibitem{LHCb:2013fqe}
{\scshape LHCb} collaboration, \emph{{Study of $B^0 \to D^{*-} \pi^+ \pi^-
  \pi^+$ and $B^0 \to D^{*-}K^+\pi^-\pi^+$ decays}},
  \href{https://doi.org/10.1103/PhysRevD.87.092001}{\emph{Phys. Rev. D}
  {\bfseries 87} (2013) 092001}
  [\href{https://arxiv.org/abs/1303.6861}{{\ttfamily 1303.6861}}].

\bibitem{Jung:2015yma}
M.~Jung, \emph{{Branching ratio measurements and isospin violation in B-meson
  decays}}, \href{https://doi.org/10.1016/j.physletb.2015.12.024}{\emph{Phys.
  Lett. B} {\bfseries 753} (2016) 187}
  [\href{https://arxiv.org/abs/1510.03423}{{\ttfamily 1510.03423}}].

\bibitem{Belle:2002lms}
{\scshape Belle} collaboration, \emph{{Studies of $B^0$--$\bar{B}^0$ mixing
  properties with inclusive dilepton events}},
  \href{https://doi.org/10.1103/PhysRevD.67.052004}{\emph{Phys. Rev. D}
  {\bfseries 67} (2003) 052004}
  [\href{https://arxiv.org/abs/hep-ex/0212033}{{\ttfamily hep-ex/0212033}}].

\bibitem{BaBar:2005uwr}
{\scshape BaBar} collaboration, \emph{{Measurement of the branching fraction of
  $\Upsilon(4S) \to B^0 \overline{B}^0$}},
  \href{https://doi.org/10.1103/PhysRevLett.95.042001}{\emph{Phys. Rev. Lett.}
  {\bfseries 95} (2005) 042001}
  [\href{https://arxiv.org/abs/hep-ex/0504001}{{\ttfamily hep-ex/0504001}}].

\bibitem{ParticleDataGroup:2022pth}
{\scshape Particle Data Group} collaboration, \emph{{Review of Particle
  Physics}}, \href{https://doi.org/10.1093/ptep/ptac097}{\emph{PTEP} {\bfseries
  2022} (2022) 083C01}.

\bibitem{EOS-DATA-2024-03}
S.~Meiser, D.~van Dyk and J.~Virto, \emph{{\texttt{EOS/DATA-2024-03}:
  Supplementary material for \texttt{EOS/ANALYSIS-2023-09}}},  Nov., 2024.
\newblock
  \href{https://doi.org/10.5281/zenodo.14161470}{10.5281/zenodo.14161470}.

\bibitem{ParticleDataGroup:2024cfk}
{\scshape Particle Data Group} collaboration, \emph{{Review of particle
  physics}}, \href{https://doi.org/10.1103/PhysRevD.110.030001}{\emph{Phys.
  Rev. D} {\bfseries 110} (2024) 030001}.

\bibitem{FlavourLatticeAveragingGroupFLAG:2021npn}
{\scshape Flavour Lattice Averaging Group (FLAG)} collaboration, \emph{{FLAG
  Review 2021}},
  \href{https://doi.org/10.1140/epjc/s10052-022-10536-1}{\emph{Eur. Phys. J. C}
  {\bfseries 82} (2022) 869}
  [\href{https://arxiv.org/abs/2111.09849}{{\ttfamily 2111.09849}}].

\bibitem{RBC:2014ntl}
{\scshape RBC, UKQCD} collaboration, \emph{{Domain wall QCD with physical quark
  masses}}, \href{https://doi.org/10.1103/PhysRevD.93.074505}{\emph{Phys. Rev.
  D} {\bfseries 93} (2016) 074505}
  [\href{https://arxiv.org/abs/1411.7017}{{\ttfamily 1411.7017}}].

\bibitem{Follana:2007uv}
{\scshape HPQCD, UKQCD} collaboration, \emph{{High Precision determination of
  the $\pi$, $K$, $D$ and $D_{(s)}$ decay constants from lattice QCD}},
  \href{https://doi.org/10.1103/PhysRevLett.100.062002}{\emph{Phys. Rev. Lett.}
  {\bfseries 100} (2008) 062002}
  [\href{https://arxiv.org/abs/0706.1726}{{\ttfamily 0706.1726}}].

\bibitem{MILC:2010hzw}
{\scshape MILC} collaboration, \emph{{Results for light pseudoscalar mesons}},
  \href{https://doi.org/10.22323/1.105.0074}{\emph{PoS} {\bfseries LATTICE2010}
  (2010) 074} [\href{https://arxiv.org/abs/1012.0868}{{\ttfamily 1012.0868}}].

\bibitem{RQCD:2019osh}
{\scshape RQCD} collaboration, \emph{{Light-cone distribution amplitudes of
  pseudoscalar mesons from lattice QCD}},
  \href{https://doi.org/10.1007/JHEP08(2019)065}{\emph{JHEP} {\bfseries 08}
  (2019) 065} [\href{https://arxiv.org/abs/1903.08038}{{\ttfamily
  1903.08038}}], [Addendum: JHEP 11, 037 (2020)].

\bibitem{LatticeParton:2022zqc}
{\scshape Lattice Parton} collaboration, \emph{{Pion and Kaon Distribution
  Amplitudes from Lattice QCD}},
  \href{https://doi.org/10.1103/PhysRevLett.129.132001}{\emph{Phys. Rev. Lett.}
  {\bfseries 129} (2022) 132001}
  [\href{https://arxiv.org/abs/2201.09173}{{\ttfamily 2201.09173}}].

\bibitem{Ball:2006wn}
P.~Ball, V.M.~Braun and A.~Lenz, \emph{{Higher-twist distribution amplitudes of
  the K meson in QCD}},
  \href{https://doi.org/10.1088/1126-6708/2006/05/004}{\emph{JHEP} {\bfseries
  05} (2006) 004} [\href{https://arxiv.org/abs/hep-ph/0603063}{{\ttfamily
  hep-ph/0603063}}].

\bibitem{FermilabLattice:2014tsy}
{\scshape Fermilab Lattice, MILC} collaboration, \emph{{Charmed and Light
  Pseudoscalar Meson Decay Constants from Four-Flavor Lattice QCD with Physical
  Light Quarks}}, \href{https://doi.org/10.1103/PhysRevD.90.074509}{\emph{Phys.
  Rev. D} {\bfseries 90} (2014) 074509}
  [\href{https://arxiv.org/abs/1407.3772}{{\ttfamily 1407.3772}}].

\bibitem{Dowdall:2013rya}
R.J.~Dowdall, C.T.H.~Davies, G.P.~Lepage and C.~McNeile, \emph{{$V_{us}$ from
  $\pi$ and $K$ decay constants in full lattice QCD with physical $u$, $d$, $s$
  and $c$ quarks}},
  \href{https://doi.org/10.1103/PhysRevD.88.074504}{\emph{Phys. Rev. D}
  {\bfseries 88} (2013) 074504}
  [\href{https://arxiv.org/abs/1303.1670}{{\ttfamily 1303.1670}}].

\bibitem{Carrasco:2014poa}
N.~Carrasco et~al., \emph{{Leptonic decay constants $f_{K},f_{D},$ and
  $f_{{D}_{s}}$ with $N_{f} = 2+1+1$ twisted-mass lattice QCD}},
  \href{https://doi.org/10.1103/PhysRevD.91.054507}{\emph{Phys. Rev. D}
  {\bfseries 91} (2015) 054507}
  [\href{https://arxiv.org/abs/1411.7908}{{\ttfamily 1411.7908}}].

\bibitem{Bordone:2019guc}
M.~Bordone, N.~Gubernari, D.~van Dyk and M.~Jung, \emph{{Heavy-Quark expansion
  for ${{\bar{B}}_s\rightarrow D^{(*)}_s}$ form factors and unitarity bounds
  beyond the ${SU(3)_F}$ limit}},
  \href{https://doi.org/10.1140/epjc/s10052-020-7850-9}{\emph{Eur. Phys. J. C}
  {\bfseries 80} (2020) 347}
  [\href{https://arxiv.org/abs/1912.09335}{{\ttfamily 1912.09335}}].

\bibitem{EOSAuthors:2021xpv}
{\scshape EOS Authors} collaboration, \emph{{{\EOS}: a software for flavor
  physics phenomenology}},
  \href{https://doi.org/10.1140/epjc/s10052-022-10177-4}{\emph{Eur. Phys. J. C}
  {\bfseries 82} (2022) 569}
  [\href{https://arxiv.org/abs/2111.15428}{{\ttfamily 2111.15428}}].

\bibitem{EOS:v1.0.13}
D.~van Dyk, M.~Reboud, N.~Gubernari, P.~Lüghausen, D.~Leljak, M.~Kirk et~al.,
  \emph{{\EOS} version 1.0.13},  Oct., 2024.
\newblock
  \href{https://doi.org/10.5281/zenodo.13947381}{10.5281/zenodo.13947381}.

\bibitem{Endo:2021ifc}
M.~Endo, S.~Iguro and S.~Mishima, \emph{{Revisiting rescattering contributions
  to $ \overline{B} _{(s)}$ \textrightarrow{} $ {D}_{(s)}^{\left(\ast \right)}M
  $ decays}}, \href{https://doi.org/10.1007/JHEP01(2022)147}{\emph{JHEP}
  {\bfseries 01} (2022) 147}
  [\href{https://arxiv.org/abs/2109.10811}{{\ttfamily 2109.10811}}].

\bibitem{Lenz:2022pgw}
A.~Lenz, J.~M\"uller, M.L.~Piscopo and A.V.~Rusov, \emph{{Taming new physics in
  b \textrightarrow{} c\={u}d(s) with
  \ensuremath{\tau}(B$^{+}$)/\ensuremath{\tau}(B$_{d}$) and $ {a}_{sl}^d $}},
  \href{https://doi.org/10.1007/JHEP09(2023)028}{\emph{JHEP} {\bfseries 09}
  (2023) 028} [\href{https://arxiv.org/abs/2211.02724}{{\ttfamily
  2211.02724}}].

\bibitem{Panuluh:2024poh}
A.H.~Panuluh, S.~Tanaka and H.~Umeeda, \emph{{$B_{(s)}\to D_{(s)}^{(*)}M$
  decays in the presence of final-state interaction}},
  \href{https://arxiv.org/abs/2408.15466}{{\ttfamily 2408.15466}}.

\bibitem{Buras:1991jm}
A.J.~Buras, M.~Jamin, M.E.~Lautenbacher and P.H.~Weisz, \emph{{Effective
  Hamiltonians for $\Delta S = 1$ and $\Delta B = 1$ nonleptonic decays beyond
  the leading logarithmic approximation}},
  \href{https://doi.org/10.1016/0550-3213(92)90345-C}{\emph{Nucl. Phys. B}
  {\bfseries 370} (1992) 69} [Addendum: Nucl.Phys.B 375, 501 (1992)].

\bibitem{Chetyrkin:1997gb}
K.G.~Chetyrkin, M.~Misiak and M.~Munz, \emph{{$|\Delta F| = 1$ nonleptonic
  effective Hamiltonian in a simpler scheme}},
  \href{https://doi.org/10.1016/S0550-3213(98)00131-X}{\emph{Nucl. Phys. B}
  {\bfseries 520} (1998) 279}
  [\href{https://arxiv.org/abs/hep-ph/9711280}{{\ttfamily hep-ph/9711280}}].

\bibitem{Morell:2024aml}
P.~Morell and J.~Virto, \emph{{On the two-loop penguin contributions to the
  Anomalous Dimensions of four-quark operators}},
  \href{https://doi.org/10.1007/JHEP04(2024)105}{\emph{JHEP} {\bfseries 04}
  (2024) 105} [\href{https://arxiv.org/abs/2402.00249}{{\ttfamily
  2402.00249}}].

\bibitem{Aebischer:2025hsx}
J.~Aebischer, P.~Morell, M.~Pesut and J.~Virto, \emph{{Two-Loop Anomalous
  Dimensions in the LEFT: Dimension-Six Four-Fermion Operators in NDR}},
  \href{https://arxiv.org/abs/2501.08384}{{\ttfamily 2501.08384}}.

\end{thebibliography}\endgroup

\end{document}